\documentclass[twocolumn,tighten,astrosymb,trackchanges]{aastex701}
\newcommand{\LJMU}{\affiliation{Astrophysics Research Institute, Liverpool John Moores University, 146 Brownlow Hill, Liverpool L3 5RF, UK}}

\newcommand{\HUBerlin}{\affiliation{Institut f\"ur Physik, Humboldt-Universit\"at zu Berlin, Newtonstr. 15, 12489 Berlin, Germany}}

\newcommand{\PennState}{\affiliation{Department of Astronomy and Astrophysics, The Pennsylvania State University}}

\newcommand{\OKC}{\affiliation{Department of Physics, Oskar Klein Centre, Stockholm University, SE-106 91, Stockholm, Sweden}}
\newcommand{\OKCAstro}{\affiliation{Department of Astronomy, Oskar Klein Center, Stockholm University, SE-106 91 Stockholm, Sweden}}
\newcommand{\CaltechPhys}{\affiliation{Division of Physics, Mathematics and Astronomy, California Institute of Technology, Pasadena, CA 91125, USA}}
\newcommand{\CaltechOO}{\affiliation{Caltech Optical Observatories, California Institute of Technology, Pasadena, CA 91125, USA}}

\newcommand{\Caltech}{\affiliation{Cahill Center for Astronomy and Astrophysics, California Institute of Technology, Mail Code 249-17, Pasadena, CA 91125, USA}}

\newcommand{\UMDastro}{\affiliation{Department of Astronomy, University of Maryland, College Park, MD 20742, USA}}

\newcommand{\IoAKavli}{\affiliation{
Institute of Astronomy and Kavli Institute for Cosmology, University of Cambridge, Madingley Road, Cambridge, CB3 0HA, UK}}

\newcommand{\IPAC}{\affiliation{IPAC, California Institute of Technology, 1200 E. California Blvd, Pasadena, CA 91125, USA}}

\newcommand{\Birmingham}{\affiliation{School of Physics \& Astronomy and Institute for Gravitational Wave Astronomy, University of Birmingham, Birmingham B15 2TT, UK}}

\newcommand{\INAFBol}{\affiliation{INAF - Osservatorio di Astrofisica e Scienza dello Spazio di Bologna, via Piero Gobetti 93/3, I-40129 Bologna, Italy}}
\newcommand{\INAFRoma}{\affiliation{INAF - Osservatorio Astronomico di Roma, via Frascati 33, 00078 Monte Porzio Catone, Italy}}

\newcommand{\mcwilliams}{\affiliation{McWilliams Center for Cosmology and Astrophysics, Department of Physics, Carnegie Mellon University, 5000 Forbes Avenue, Pittsburgh, PA 15213, USA}}
\newcommand{\LMU}{\affiliation{University Observatory, Faculty of Physics, Ludwig-Maximilians-Universität, Scheinerstr. 1, 81679 Munich, Germany}}
\newcommand{\ORIGINS}{\affiliation{Excellence Cluster ORIGINS, Boltzmannstr. 2, 85748 Garching, Germany}}

\newcommand{\Weizmann}{\affiliation{Department of Particle Physics and Astrophysics, Weizmann Institute of Science, Rehovot, Israel}}

\usepackage[dvipsnames]{xcolor}
\usepackage{newtxtext,newtxmath}
\usepackage{graphicx}
\usepackage{amsmath}
\usepackage{bm}
\usepackage[T1]{fontenc}
\usepackage{placeins}
\usepackage{needspace}

\NewDocumentCommand{\companioncite}{o o m}{%
  {\hypersetup{citecolor=red}%
   \IfNoValueTF{#1}
     {\citet{#3}}
     {\IfNoValueTF{#2}
       {\citet[#1]{#3}}
       {\citet[#1][#2]{#3}}}}%
}

\NewDocumentCommand{\companioncitep}{o o m}{%
  {\hypersetup{citecolor=red}%
   \IfNoValueTF{#1}
     {\citep{#3}}
     {\IfNoValueTF{#2}
       {\citep[#1]{#3}}
       {\citep[#1][#2]{#3}}}}%
}

\begin{document}

\title{Follow-up of SN 2025wny IV: Photometric Time-delay Measurements of a Strongly Lensed Superluminous Supernova}

\correspondingauthor{Alice Townsend}


\author[0000-0001-6343-3362]{Alice~Townsend}
\Birmingham
\email[show]{a.townsend@bham.ac.uk}

\author[0000-0002-2376-6979]{Suhail~Dhawan}
\Birmingham
\email{s.dhawan@bham.ac.uk}


\author[0000-0003-3847-0780]{Erin~E.~Hayes}
\IoAKavli
\email{eeh55@cam.ac.uk}

\author[0009-0001-6911-9144]{Maggie L.~Li}
\Caltech
\email{maggieli@caltech.edu}

\author[0000-0001-5975-290X]{Joel~Johansson}
\OKC
\email{joeljo@fysik.su.se}

\author[0000-0002-8380-6143]{Edvard~Mörtsell}
\OKC
\email{edvard@fysik.su.se}

\author[0000-0002-4163-4996]{Ariel~Goobar}
\OKC
\email{ariel@fysik.su.se}

\author[0000-0003-1710-9339]{Lin~Yan}
\Caltech
\email{lyan@caltech.edu}

\author[0000-0002-4557-6682]{Charlotte~Ward}
\PennState
\email{cvw5890@psu.edu}

\author[0009-0000-7317-5256]{Veena~Krishnaraj}
\UMDastro
\email{veenak@umd.edu}

\author[0000-0001-6797-1889]{Steve~Schulze} 
\Weizmann
\email{steve.schulze@weizmann.ac.il}

\author[0009-0009-6243-8300]{Jacob~Osman~Hjortlund}
\OKC
\email{jacob.hjortlund@fysik.su.se}

\author[0000-0003-3658-6026]{Yu-Jing~Qin}
\Caltech
\email{qinyj.astro@gmail.com}

\author[0009-0006-7102-3674]{Hannah~C.~Turner}
\Birmingham
\email{h.c.turner@bham.ac.uk}

\author[0009-0005-8228-0329]{Peter~Massey}
\Birmingham
\email{pxm588@student.bham.ac.uk}

\author[0000-0001-8342-6274]{Jakob~Nordin}
\email{jnordin@physik.hu-berlin.de}
\HUBerlin

\author[0009-0002-6662-4900]{Jule~Augustin}
\LMU
\email{Jule.augustin@campus.lmu.de}

\author[0009-0008-2714-2507]{Aleksandra Bochenek}
\LJMU
\email{A.M.Bochenek@2023.ljmu.ac.uk}

\author[0009-0001-0574-2332]{Malte~Busmann}
\LMU
\ORIGINS
\email{m.busmann@physik.lmu.de}

\author[0000-0002-4223-103X]{Christoffer Fremling}
\CaltechOO
\CaltechPhys
\email{fremling@caltech.edu}

\author[0000-0003-3270-7644]{Daniel~Gruen}
\LMU
\ORIGINS
\email{daniel.gruen@lmu.de}

\author[0000-0002-9364-5419]{Xander~J.~Hall}
\mcwilliams
\email{xhall@cmu.edu}

\author[0000-0002-0129-806X]{K.~-R.~Hinds}
\CaltechPhys
\email{khinds@caltech.edu}

\author[0000-0001-8368-8565]{Ezequiel~J.~Marchesini} 
\INAFBol
\email{ezequiel.marchesini@inaf.it}

\author[0009-0006-0726-1328]{Zoë~McGrath}
\LJMU
\email{z.mcgrath@2024.ljmu.ac.uk}

\author[0000-0002-9646-8710]{Conor~M.~B.~Omand}
\LJMU
\email{C.M.Omand@ljmu.ac.uk}

\author[0000-0002-8691-7666]{Eliana~Palazzi}
\INAFBol
\email{eliana.palazzi@inaf.it}

\author[0000-0001-8472-1996]{Daniel~A.~Perley}
\LJMU
\email{D.A.Perley@ljmu.ac.uk}

\author[0000-0002-8860-6538]{Andrea~Rossi}
\INAFBol
\email{andrea.rossi@inaf.it}

\author[0009-0000-7976-1416]{Killa~Santer}
\LMU
\email{killasanter0@gmail.com}

\author[0000-0003-1546-6615]{Jesper~Sollerman}
\OKCAstro
\email{jesper@astro.su.se}

\author[0009-0009-9751-9215]{Chiara~Ventura}
\INAFRoma
\email{chiara.ventura@inaf.it}

\author[0000-0003-0733-2916]{Jacob~L.~Wise}
\LJMU
\email{J.L.Wise@2022.ljmu.ac.uk}

\author[0000-0001-9152-6224]{Tracy X. Chen}
\IPAC
\email{xchen@ipac.caltech.edu}

\author[0000-0001-5668-3507]{Steven L. Groom}
\IPAC
\email{sgroom@ipac.caltech.edu}

\author[0000-0002-5619-4938]{Mansi~M.~Kasliwal}
\CaltechPhys
\email{mansi@astro.caltech.edu}

\author[0000-0003-1227-3738]{Josiah~Purdum}
\CaltechPhys
\email{mansi@astro.caltech.edu}

\shorttitle{Follow-up of SN~2025wny IV: Photometric Time-delay Measurements}
\shortauthors{A. Townsend et al.}

\begin{abstract}

We present photometric time-delay measurements of SN~2025wny, the first strongly lensed Type I superluminous supernova (SLSN-I), discovered at $z = 2.015$. Time-delay measurements from strongly lensed supernovae provide an independent probe of cosmology and the Hubble constant, $H_0$, without reliance on the local distance ladder. Using multi-facility imaging data, we performed scene-modelling photometry to deblend four of the lensed images (A--D) and construct $grizJ$-band light curves. We modelled the resolved light curves with Gaussian process regression using \texttt{GausSN} \citep{Hayes2024} to infer relative time delays and magnifications between the lensed images. We found that a constant magnification model provides a suboptimal description of the data, motivating a time-dependent sigmoid magnification model to account for evolving relative magnification of image A. We measured time delays of $\Delta t_{AB} = -10.6^{+2.2}_{-2.5}$~days and
$\Delta t_{AC} = 1.2^{+2.7}_{-2.6}$~days (68\% credible intervals), consistent with independent spectroscopic measurements from \companioncite{Johansson2026}. Combining the photometric time delays with the lens model of \companioncite{Mortsell2026} gives $H_{0,\:\rm photo} = 80.5^{+26.4}_{-16.7}\;\rm km\,s^{-1}\,Mpc^{-1}$, while including the spectroscopic time delays as well yields $H_{0,\:\rm comb} = 70.8^{+8.2}_{-6.1}\;{\rm km\,s^{-1}\,Mpc^{-1}}$. Our results further demonstrate the potential of strongly lensed supernovae as independent probes of $H_0$.

\end{abstract}

\section{Introduction}
\label{sec:intro}

Strong lensing of transient astrophysical sources, such as supernovae (SNe), by foreground galaxies or galaxy clusters provides a powerful probe of cosmology. The delay between the arrival times of multiple images is sensitive to the expansion rate of the Universe, quantified by the Hubble constant ($H_0$), an idea first proposed by \citet{Refsdal1964}. Measuring $H_0$ is critical because it sets the absolute distance scale of the Universe. Currently, a significant tension exists between measurements of $H_0$ inferred from early-Universe probes, such as the cosmic microwave background \citep[CMB; e.g.,][]{Planck2020}, and local measurements based on the Cepheid distance ladder \citep[e.g.,][]{Riess2022}. The discrepancy is at the $\sim5\sigma$ level, corresponding to a difference in $H_0$ of approximately 9\% \citep[for a review, see][and references therein]{DiValentino2025,H0DN2026}. Gravitationally lensed supernovae (glSNe) provide an independent avenue to investigate this discrepancy by measuring the time-delay distance ($D_{\Delta {\rm t}}$), which is independent of both local distance calibrations and assumptions about early-Universe physics.

Lensed supernovae, however, are rare, and the first glSN was not discovered until 2013, half a century after the method was originally proposed \citep{Quimby2013}. The current sample of known glSNe now numbers $\sim$17 objects, spanning a range of SN types, redshifts and deflector scales. However, only a subset of these systems have sufficiently long time delays and high-resolution imaging suitable for cosmological inference: SN Refsdal \citep{Kelly2015,Kelly2023a,Kelly2023b}, SN H0pe \citep{Frye2023,Pascale2024,Grayling2026}, and SN Encore \citep{Pierel2024,Pierel2026,Suyu2026}. Ground-based wide-field surveys have begun to uncover a growing population of galaxy-scale glSNe. The first such systems included the multiply imaged Type~Ia SNe iPTF16geu \citep{Goobar2017} and SN~Zwicky \citep{Goobar2023}. Owing to the compact lensing configurations characteristic of galaxy-scale systems, the corresponding time delays were on the order of hours to days \citep{Dhawan2020,Johansson2021}, making them challenging to exploit for precision time-delay cosmography. Additional examples of galaxy-scale glSNe discovered in ground-based surveys include SN~2025mkn \citep{Lemon2026}, SN~2025wny \citep{Johansson2025,Taubenberger2025}, and SN~2026ngr \citep{Johansson2026TNS}.

Lensed quasars have previously been used to constrain $H_0$ \citep[e.g.,][]{Wong2020,Shajib2023,Tdcosmo2025}; however, current measurements remain consistent with both CMB-inferred and local values. Gravitationally lensed supernovae provide several advantages over lensed quasars, particularly because current uncertainties on $H_0$ from lensed quasars are dominated by systematic errors, and resolving the tension at the $5\sigma$ level requires precision at the $\sim$2\% level. Unlike quasars, glSNe exhibit well-characterised light curves, enabling more precise time-delay measurements with significantly fewer follow-up observations. Moreover, because SNe fade over time, post-explosion imaging can be used to validate the lens model. For SNe~Ia, their standardisable luminosities can additionally help break lens model degeneracies, although glSNe of all types can provide time-delay measurements. Beyond their importance for cosmology, glSNe also probe the nature of SNe in the high-$z$ Universe through the `gravitational telescope' effect.

In this paper, we present resolved and unresolved photometry of SN~2025wny \citep{Johansson2025,Taubenberger2025}, also known as `SN~Winny'. SN~2025wny is a spectroscopically classified Type I superluminous supernova (SLSN-I) at $z\simeq2.01$, strongly lensed by two foreground galaxies with the primary deflector at $z\simeq0.375$. Follow-up imaging with higher spatial resolution revealed five images of the SN \citep{Wise2025,Aryan2025}, making SN~2025wny one of the few known glSNe with sufficient image multiplicity and temporal coverage for precision time-delay measurements. It is also the first galaxy-scale glSN for which both photometric and spectroscopic time-delay cosmography are feasible. In this work, we present time-delay measurements inferred from the photometric light curves using semi-parametric methods.

Accurate time-delay measurements from glSNe remain observationally challenging due to irregular cadence, finite signal-to-noise, image blending, and microlensing by stars in the lens galaxy. These effects are particularly important for long-duration transients and high-redshift systems, where the observed light curves may span several months in the observer frame. In particular, microlensing can introduce chromatic and time-dependent distortions to the SN light curves, potentially biasing inferred delays if not properly accounted for \citep[e.g.,][]{Dobler2006,Huber2021,Arendse2025}.

Unlike SNe~Ia, the diversity of SLSN-I light curves and the limited number of well-observed events, especially at $z>2$, motivate approaches that do not rely on spectral templates. SN~2025wny is itself an unusual SLSN-I, exhibiting strong rest-frame ultraviolet emission and spectral properties that are not well represented by existing low-redshift samples \companioncitep{Li2026}, further motivating flexible, data-driven methods for time-delay inference. The analysis presented here for SN~2025wny is particularly timely given the expected increase in glSN discoveries from upcoming wide-field time-domain surveys such as the Vera C. Rubin Observatory's Legacy Survey of Space and Time (LSST) \citep[e.g.,][]{Arendse2024,Sainzdemurieta2024,Ponte2026}.

The structure of this paper is as follows. Section~\ref{sec:overview} provides an overview of SN~2025wny and the companion papers analysing this system. Section~\ref{sec:data} summarises the resolved and unresolved imaging used in this analysis. In Sect.~\ref{sec:phot}, we describe the scene-modelling photometry pipeline used to construct the multi-band light curves, including the photometric corrections and data-quality masking applied. In Sect.~\ref{sec:time_delays}, we describe our methods for inferring time delays using Gaussian process regression with \texttt{GausSN} \citep{Hayes2024}. We present the inferred time delays and best-fitting \texttt{GausSN} models in Sect.~\ref{sec:time_delays_results}, considering both a constant magnification model and a time-dependent magnification model that accounts for the effects of achromatic microlensing. Finally, Sect.~\ref{sec:conclusion} summarises our main results and discusses the cosmological constraints from the combined analysis of SN~2025wny.

\section{Overview of SN~2025wny}
\label{sec:overview}
SN~2025wny (also known as ZTF25abnjznp, GOTO25gqt, and `SN~Winny') was independently discovered in the wide-field transient surveys Gravitational-wave Optical Transient Observer \citep[GOTO;][]{Steeghs2022} and the Zwicky Transient Facility \citep[ZTF;][]{Graham2019,Bellm2019,Dekany2020,Masci2019}. The first report was submitted to the Transient Name Server (TNS) by GOTO \citep{ONeill2025} on 1 September 2025, although the earliest alert was recorded by ZTF on 23 August 2025. Follow-up imaging with the Liverpool Telescope \citep{Wise2025} subsequently resolved the transient into multiple point-like images. Spectroscopic observations classified SN~2025wny as a hydrogen-poor superluminous supernova (SLSN-I) at $z\simeq2.01$, based on its blue continuum and lack of hydrogen features, with a good match to rest-frame ultraviolet SLSN-I spectra obtained with HST \citep{Johansson2025TNS}. High-resolution imaging later revealed a total of five multiply imaged SN components \citep{Aryan2025}.

The SN is spatially coincident with the previously identified strong-lens candidate PS1~J0716+3821 \citep{Canameras2020}, consisting of two massive foreground galaxies. Spectroscopic observations from the first data release of the Dark Energy Spectroscopic Instrument (DESI) measured the redshift of the primary deflector galaxy (G1) to be $z=0.3754\pm0.0001$ \citep{DESI2025}, while subsequent follow-up spectroscopy confirmed a consistent redshift for the secondary lens galaxy \citep{Taubenberger2025}.

SLSNe are a rare class of stellar explosions with peak absolute magnitudes exceeding $M \lesssim -21$~mag, making them approximately an order of magnitude more luminous than normal SNe~Ia \citep{Quimby2011,GalYam2012}. Their rarity is reflected in volumetric rates of $\sim 10^{-4}$ to $10^{-3}$ times those of SNe~Ia \citep{Quimby2013b,Prajs2017,Perley2020}. SLSNe-I lack hydrogen features and often display O~II absorption at early times, a signature of highly ionised ejecta \citep{Quimby2011}. Their light curves typically show rise times of $20-50$~days and slower declines compared to normal stripped-envelope SNe \citep{Chen2023}. Some exhibit undulations or secondary peaks \citep[e.g.,][]{Nicholl2016,Chen2023}.

The physical powering mechanism of SLSNe remains an active area of research. The radioactive decay of large quantities of $^{56}$Ni, which powers the light curves of many normal supernovae, is generally insufficient to explain the luminosities and temporal evolution observed in most SLSNe \citep{GalYam2012,Inserra2013}. Several alternative mechanisms have therefore been proposed, including: (1) spin-down of a rapidly rotating, highly magnetised neutron
star (magnetar), injecting energy into the ejecta \citep{Kasen2010,Woosley2010}; (2) strong interaction between the ejecta and massive CSM, which is hydrogen-poor in SLSN-I and hydrogen-rich in SLSN-II \citep{Chevalier2011}; or (3) more exotic possibilities such as pair-instability explosions in very massive stars ($M \gtrsim 140 ~\mathrm{M_\odot}$), where the conversion of energetic photons into electron-positron pairs reduces pressure support, leading to runaway collapse and explosive oxygen burning \citep{GalYam2009,Schulze2024}.

Observational evidence suggests that multiple mechanisms may
be at play across the population. Owing to the magnification provided by gravitational lensing, SN~2025wny also provides a rare opportunity to study a high-redshift SLSN in detail. As most known SLSNe have been discovered at $z<1$, this provides a unique opportunity to constrain the nature of SLSN progenitors and their powering mechanisms in the early Universe \companioncite{Li2026}.

This paper is one of a series presenting the follow-up campaign of SN~2025wny: \companioncite{Goobar2026} describe
the HST and JWST follow-up imaging and spectroscopy; \companioncite{Li2026} analyse the underlying
physics of the SLSN; \companioncite{Johansson2026} report the spectroscopic time-delay
measurements of images A--E; \companioncite{Mortsell2026} present the lens modelling and the resulting
$H_0$ inference from the system; \companioncite{Hjortlund2026} present simulations of the ZTF survey estimating the expected rate of lensed SLSN-I in six years of operations; and \companioncite{Qin2026} study the host galaxy of SN~2025wny.

\section{Data}
\label{sec:data}
\begin{table*}
    \centering
    \caption{Summary of Imaging from Different Telescopes and Instruments for SN~2025wny.}
    \begin{tabular}{ccccccc}
    \hline
    \hline
    Telescope & Instrument & Filters & Pixel scale & Observation & Num. of & Median \\
    & & & ($\arcsec$\,pixel$^{-1}$) & period & exposures & seeing (\arcsec) \\
    \hline
    \textit{Resolved} \\
    Gemini North    &  GMOS-N & $griz$ & 0.32 & 2025-11-16 -- 2026-05-06 & 16& 0.7\\
    Large Binocular Telescope & LBC & $griz$ & 0.23 & 2025-10-26 -- 2025-12-17 & 8 & 1.1 \\
    Liverpool Telescope & IO:O & $griz$ & 0.30 & 2025-09-01 -- 2026-02-24 & 231 & 1.3\\
    Liverpool Telescope &  LOCI & $griz$ & 0.28 & 2025-11-24 -- 2025-11-25 & 40 & 1.2\\
    Palomar 60-inch & SEDM & $gri$ & 0.38 & 2025-09-01 -- 2025-10-08 & 44 & 1.4 \\
    Fraunhofer Telescope, Wendelstein & 3KK & $grizJ$ & 0.20 & 2025-10-10 -- 2026-05-08 & 182 & 0.8\\
    \hline
    \textit{Unresolved} \\
    Palomar 48-inch & ZTF & $gr$ & 1.01  & 2025-08-23 -- 2026-04-05 & 225 & 3.0\\
    
\hline         
    \end{tabular}
    \label{tab:dataset}
\end{table*}
SN~2025wny was monitored as part of a multi-facility photometric follow-up campaign designed to obtain densely sampled, multi-band light curves of the resolved lensed images. Table~\ref{tab:dataset} summarises the photometric dataset, which spans a baseline of approximately 2025-09-01 to 2026-05-08 and covers both the rise and decline phases of the light curve across multiple filters. We obtained 521 resolved observations across 88 unique epochs from 6 telescopes, corresponding to an average cadence of approximately 3 days over the full dataset, prior to data quality cuts. The unresolved photometry from the ZTF survey is also presented in Table~\ref{tab:dataset}.

\begin{figure*}
    \centering
    \includegraphics[width=0.3\linewidth]{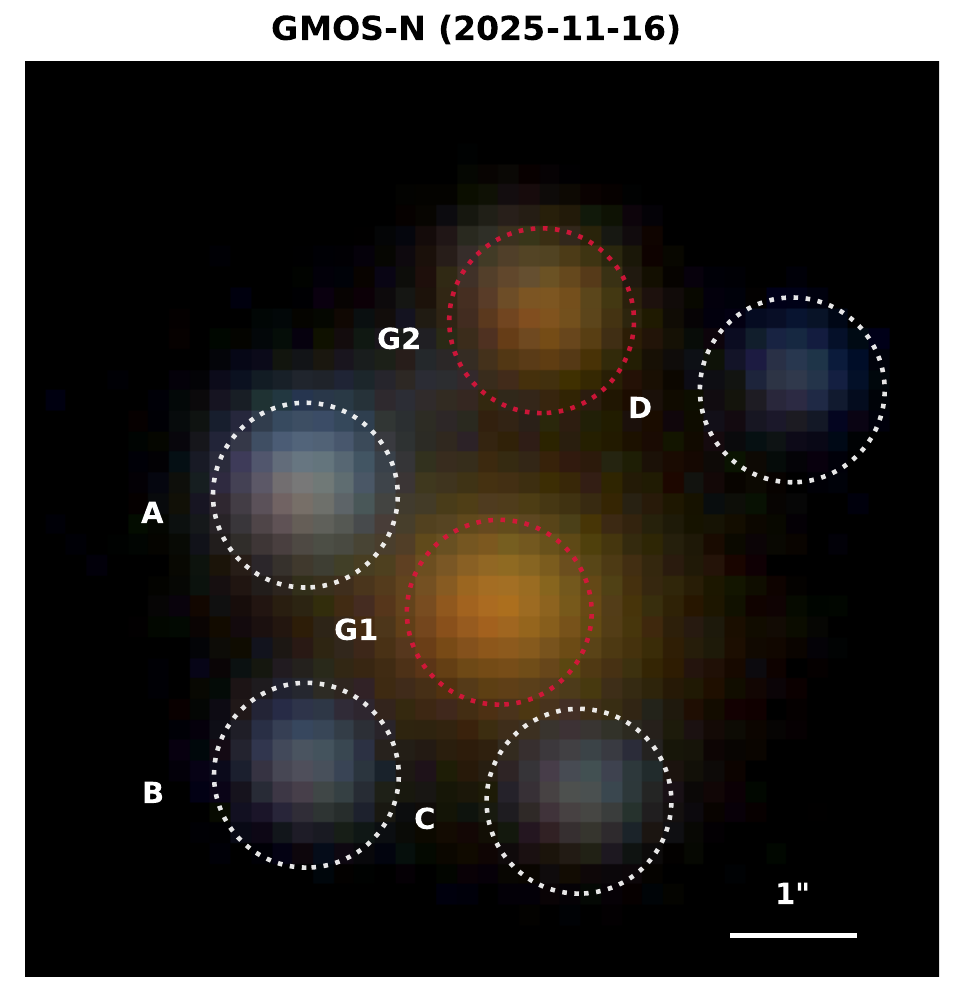}\includegraphics[width=0.3\linewidth]{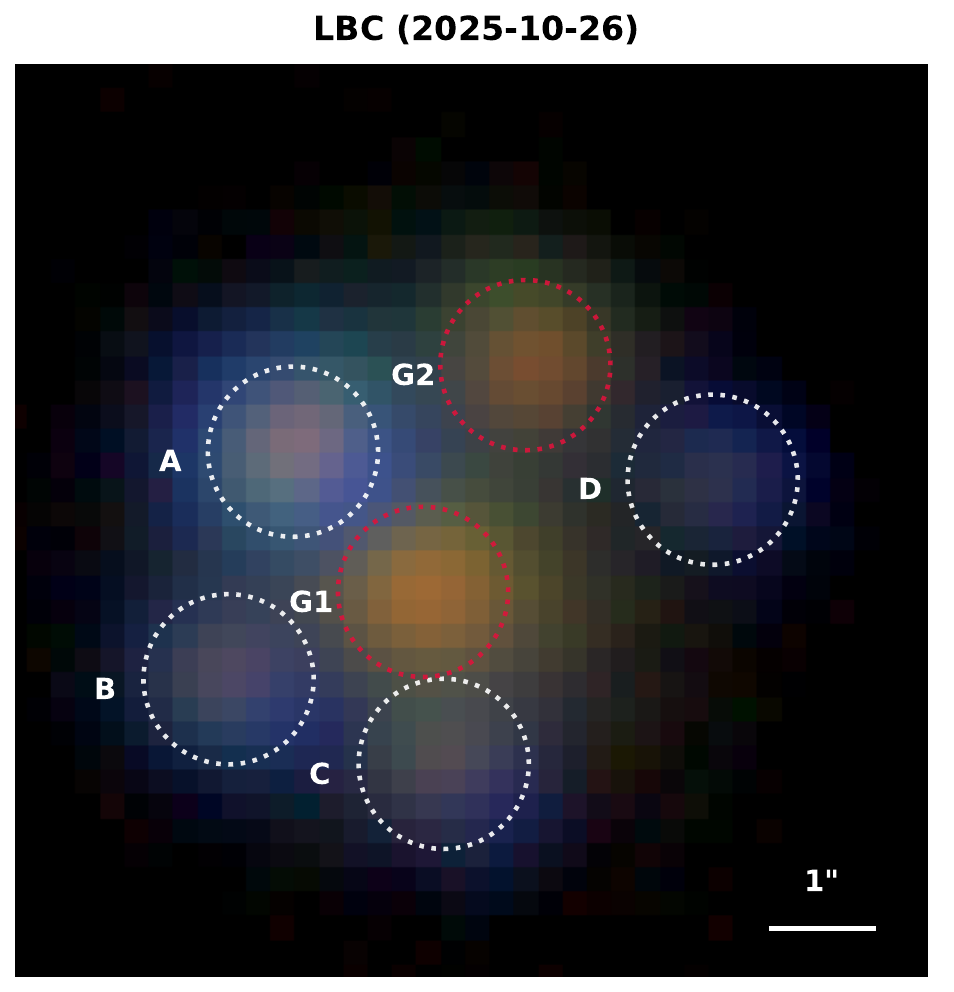}\includegraphics[width=0.3\linewidth]{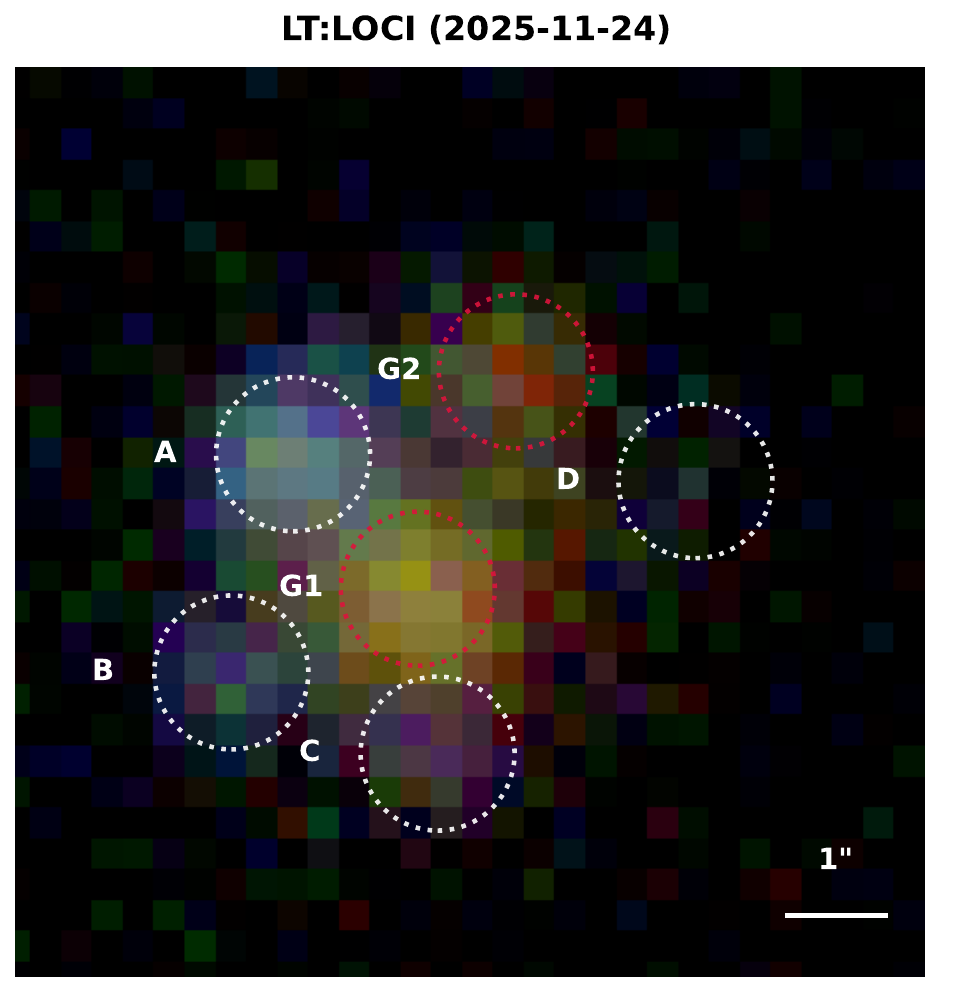}
    \includegraphics[width=0.3\linewidth]{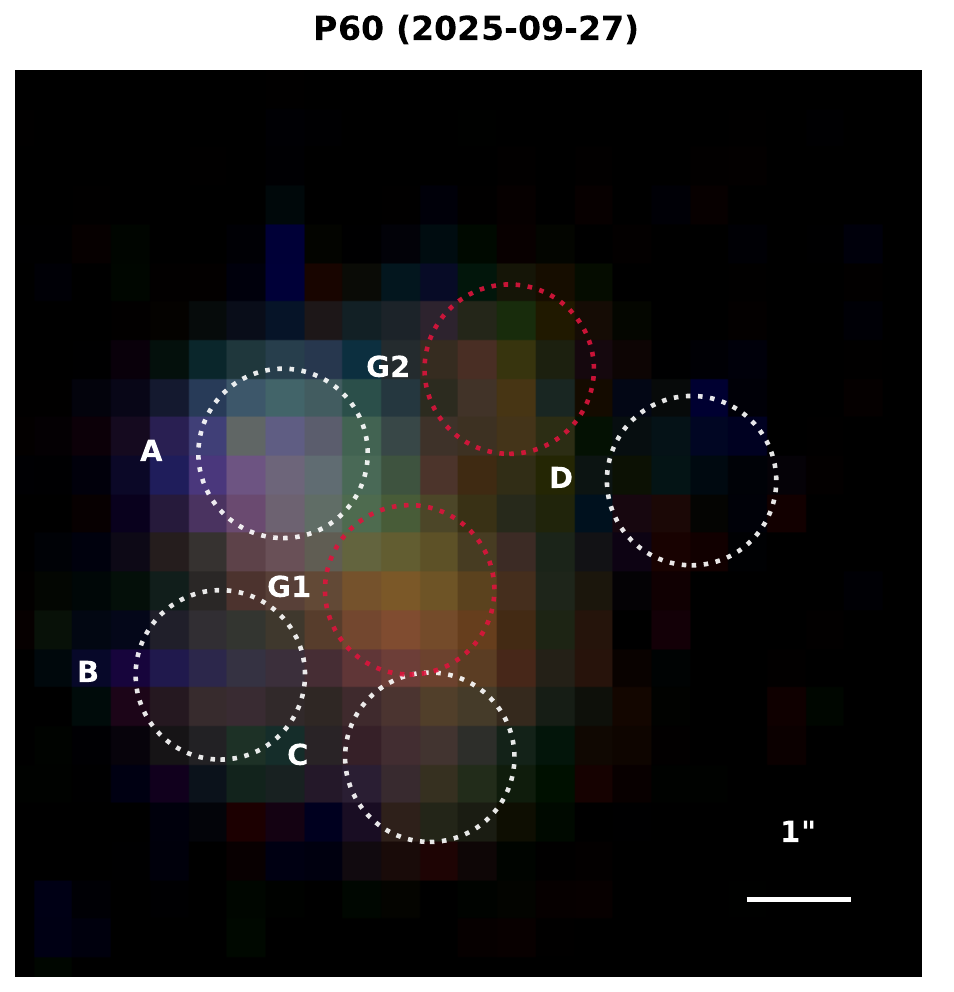}\includegraphics[width=0.3\linewidth]{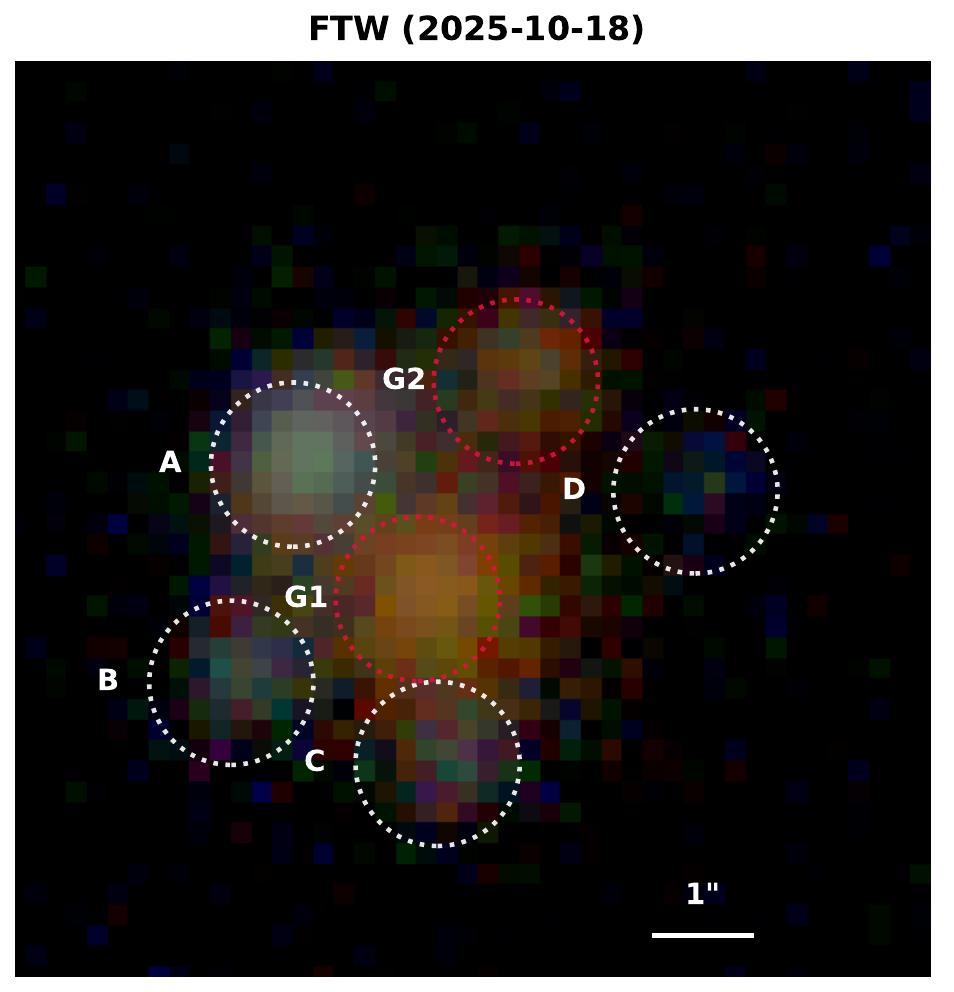}
    \caption{Multi-facility optical RGB composite cutouts of SN~2025wny, reprojected into a North-up, East-left orientation. Dotted circles identify the main lensing galaxies ($\text{G1}$ and $\text{G2}$, crimson) and the four lensed images of the supernova ($\text{A – D}$, white), with a $1\arcsec$ reference scale bar. From left to right, we show cutouts from Gemini/GMOS-N, LBT/LBC, LT/LOCI, P60/RainbowCam, and FTW/3KK.  The panels show the system across variations in telescope aperture, pixel scale, atmospheric seeing, and signal-to-noise ratio.}
    \label{fig:25wny_data}
\end{figure*}

\subsection{Resolved Imaging}
\label{subsec:resolved_data}
The majority of the data collected allowed us to resolve images A, B, C, and D. Image E is heavily blended with the second lens galaxy \companioncitep[for details, see][]{Goobar2026}. Composite RGB images of the $gri$ filters for a particular good-seeing epoch for each telescope are shown in Fig.~\ref{fig:25wny_data}.

\subsubsection{Gemini North Telescope}
\label{subsubsec:gemini}
High spatial resolution imaging was obtained with the 8.1-metre Gemini North telescope using the GMOS instrument in $griz$ bands through the Director's Discretionary Time program GN-2025B-DD-106 (PI Dhawan). Images were bias corrected, flat-fielded and for $z$ band, fringe corrected and co-added using the \texttt{dragons} pipeline \citep{Labrie2023}\footnote{https://dragons.readthedocs.io/}. A total of 5 epochs were acquired between 2025-11-16 and 2026-05-06. The first three epochs were obtained in the $griz$ filters, while the subsequent two epochs were obtained only in the $iz$ filters due to weather conditions during dark time that precluded useful $gr$ observations. Typical exposure times ranged from $4 \times 40$\,s per filter per epoch to $6 \times 60$\,s per filter per epoch, depending on whether the SN was observed in four or two filters, respectively. Observations were obtained under seeing conditions of 0.6--1.1\arcsec, with a median seeing of 0.7\arcsec, providing the spatial resolution required to separate images A, B, C, and D.

\subsubsection{Large Binocular Telescope}
\label{subsubsec:lbt}
We obtained optical imaging in $griz$ with the Large Binocular Cameras \citep[LBCs;][]{Giallongo2008a} at the 8.4\,m Large Binocular Telescope (LBT). Imaging data were processed using the data reduction pipeline developed at INAF, Osservatorio Astronomico di Roma \citep{Fontana2014a}, which includes bias subtraction and flat-fielding, bad-pixel and cosmic-ray masking, astrometric calibration, and co-addition. Field stars from the Gaia DR3 catalogue \citep{Gaia2023} were used for astrometry calibration. A total of 2 epochs were acquired on 2025-10-26 and 2025-12-17 (PI Palazzi), with typical exposure times of $2 \times 180$\,s per filter per epoch. Observations were obtained under seeing conditions of 0.8--1.3\arcsec, with a median seeing of 1.1\arcsec, providing the spatial resolution required to separate images A, B, C, and D.

\subsubsection{Liverpool Telescope}
\label{subsubsec:lt}
Follow-up observations were obtained with the 2-metre Liverpool Telescope \citep[LT;][]{Steele2004} using both the IO:O and the Liverpool Optical Compact Imager (LOCI) instruments in $griz$ bands. Observations obtained between 2025-09-01 and 2025-10-02 were carried out under program ID JL25A09 (PI Perley), while subsequent observations from 2025-10-03 to 2025-10-26 were obtained as part of program JL25B05 (PI Wise). Additional epochs were obtained under program ID JZ24B01 (PI Perley) on nights between 2025-10-06 and 2025-10-14; these observations were conducted in the $gri$ filters. From 2025-10-26 to 2026-02-24, observations were obtained as part of LT reactive time program PQ25B03 (PI Dhawan). Photometric coverage in the $z$ band with IO:O was limited to observations on 2025-12-01 and a second epoch beginning on 2026-02-10. Owing to poor weather conditions, these IO:O $z$-band observations were ultimately discarded.

Images were pre-processed using the automated LT pipeline. This included overscan trimming, bias subtraction, and flat-fielding using daily normalised twilight sky flats. A total of 271 exposures (across 31 unique epochs) were acquired between 2025-09-01 and 2026-02-24, with individual exposure times of 60--250\,s. All exposures were obtained with the IO:O imager, except for two epochs (40 exposures) obtained with the LOCI instrument on 2025-11-24 and 2025-11-25. Exposures exhibiting significant PSF elongation or telescope wobble were flagged and removed during visual inspection. To ensure image quality, we restricted our sample to nights with more than one valid exposure, coadding them into 71 nightly median stacks using \texttt{SWarp} \citep{Bertin2010}. The resulting coadds have total exposure times of 120--840\,s per epoch, with a median of 480\,s. Observations were obtained under seeing conditions of 0.9--2.0\arcsec, with a median seeing of 1.3\arcsec, allowing us to resolve images A, B, and C. Image D was recovered in only a limited number of epochs due to poor signal-to-noise. The LT data provide some of the highest-cadence resolved sampling in the dataset and were essential for tracing the early-time evolution of the light curve.

\subsubsection{Palomar 60-inch Telescope}
\label{subsubsec:p60}
Follow-up observations were obtained with the Palomar 60-inch (1.5-metre) telescope (P60) in the $gri$ bands as part of the ZTF collaboration. The exposures were acquired with the Rainbow Camera (RC), the imaging channel of the Spectral Energy Distribution Machine \citep[SEDM;][]{Blagorodnova2018,Rigault2019}. Images were processed through standard reduction routines following \citet{Blagorodnova2018}, including master bias subtraction, twilight sky flat-fielding across the $g,r,i$ filter quadrants, and night-sky fringe subtraction in $r$ and $i$.

A total of 44 exposures (across 14 unique epochs) were obtained between 2025-09-01 and 2025-10-08, each with an exposure time of 180\,s. Observations were carried out under seeing conditions of 1.2--1.8\arcsec, with a median seeing of 1.4\arcsec, allowing us to resolve images A, B, and C. Although image D was resolved in a limited number of epochs, its photometry was too faint to be measured with a significant signal-to-noise. Together with the LT data, the P60 data were essential for sampling the early-time evolution of the light curve.

\subsubsection{Fraunhofer Telescope at Wendelstein Observatory}
\label{subsubsec:wendelstein}
Follow-up observations were obtained with the Three Channel Imager \citep[3KK;][]{LangBardl2016} on the 2.1-metre Fraunhofer Telescope at Wendelstein Observatory \citep[FTW;][]{Hopp2014} with the $grizJ$ bands. A total of 182 co-added exposures (47 unique epochs) were acquired between 2025-10-10 and 2026-05-08, with total exposure times in the range of 180--7200\,s. Observations were carried out under seeing conditions of 0.5--1.7\arcsec, with a median seeing of 0.8\arcsec, allowing us to resolve images A, B, C, and D. The FTW data were reduced with a custom pipeline \citep{Goessl2002} for bias and dark subtraction, flat-fielding, and cosmic ray masking. For further details, particularly regarding near-infrared detrending, we refer the reader to \citet{Busmann2025}. The FTW data were essential for sampling the light curve evolution at all phases, providing most of the photometry for the $z$-band and all of the $J$-band.

\subsection{Unresolved Imaging}
\label{subsec:unresolved_data}

\begin{figure}[t!]
    \centering
    \includegraphics[width=\linewidth]{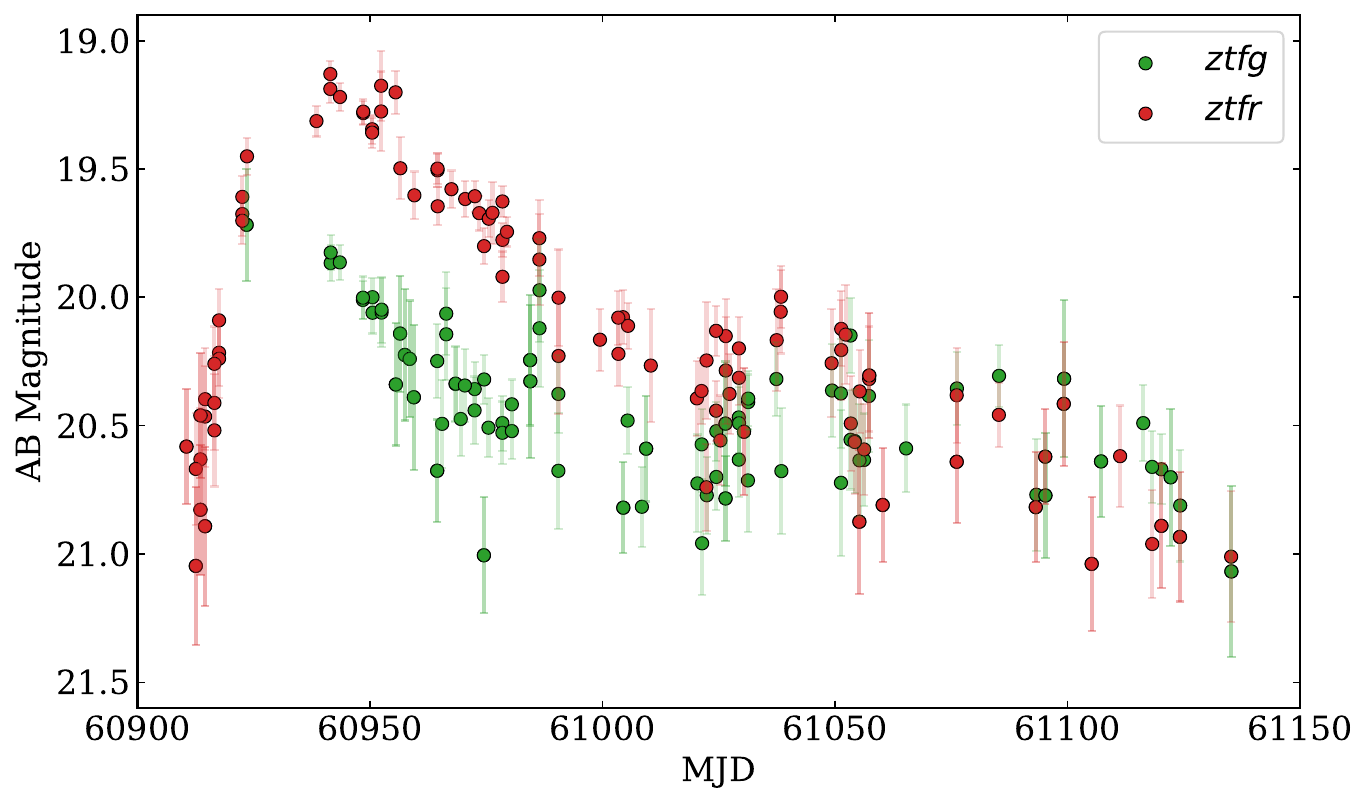}
    \caption{Unresolved forced photometry measurements from ZTF (green: $g$ band; red: $r$ band), showing the blended flux from the unresolved system.}
    \label{fig:ztf_forced}
\end{figure}

\subsubsection{Zwicky Transient Facility (Palomar 48-inch Telescope)}
\label{subsubsec:ztf}
ZTF $g$- and $r$-band photometry was obtained from the public survey and provides coverage of the early rise of the event, spanning approximately 2025-08-23 to 2026-04-05. Owing to the angular resolution of ZTF ($1.0\arcsec$\,pixel$^{-1}$) and the typical seeing at Palomar ($>2\arcsec$), the individual lensed images of SN~2025wny are not spatially resolved. Consequently, the reported photometry reflects a blended flux primarily driven by Image A and nearby components.

Forced photometry data from the IPAC ZTF forced photometry service \citep{Masci2023}, reduced by \companioncite{Li2026}, are presented in Fig.~\ref{fig:ztf_forced}. The flux measurements are obtained through PSF fitting on reference-subtracted images, using a fixed position from the ZTF science platform \texttt{Fritz} \citep{vanderWalt2019,Coughlin2023} corresponding to image A.

\section{Photometry}
\label{sec:phot}
\subsection{Scene-modelling Photometry}
\label{subsec:smp}
SN~2025wny was observed with multiple instruments with different detector characteristics and photometric filter systems. In several cases, the available pre-explosion reference images from wide-field surveys such as Pan-STARRS1 \citep[PS1;][]{Chambers2016}, the Canada-France-Hawaii Telescope Legacy Survey \citep[CFHTLS;][]{cfhtls2012}, and the Legacy Survey \citep[LS;][]{Dey2018} are shallower than the follow-up imaging. Consequently, standard difference imaging analysis (DIA) techniques can perform poorly, potentially introducing systematic and correlated residuals into the recovered photometry. Furthermore, combining observations from multiple telescopes requires consistent treatment of the photometric zeropoints and relative flux calibration to construct a homogeneous set of multi-band light curves for the lensed images of SN~2025wny.

To address these challenges, we adopted scene-modelling photometry (SMP). Scene modelling avoids the need for a reference image by fitting a model of the observed scene simultaneously to all available exposures \citep[e.g.,][]{Astier2006,Holtzman2008}. In this approach, the scene is decomposed into point-source (e.g. transient events) and extended components (e.g. background lens/host galaxy), with the point-source fluxes  allowed to vary between epochs while the extended emission is assumed to remain constant. The underlying scene is forward-modelled by convolving it with the appropriate PSF for each exposure to reproduce the observed images, and the model parameters are optimised by comparing these predictions directly with the observed pixel values. This approach is particularly well suited to our data because it allows the relative contributions of the four lensed images and the underlying host and lens galaxy light to be constrained simultaneously, even when the individual components are partially blended.

More generally, for an exposure $k$, the observed pixel values can be modelled as
\begin{equation}
\begin{aligned}
D_k(x,y) ={}&
\left[
\sum_{i=1}^{N_{\rm PS}}
F_{i,k}\,\delta(x-x_i,y-y_i)
+ B(x,y)
\right]
\otimes K_k(x,y) \\
&+ \epsilon_k(x,y).
\end{aligned}
\label{eq:scene-model}
\end{equation}
where $D_k(x,y)$ is the observed image for exposure $k$, $F_{i,k}$ is the flux of point source $i$ in that exposure, $\delta(x-x_i,y-y_i)$ represents a point source at position $(x_i,y_i)$, $B(x,y)$ represents the time-independent extended emission from the background, $K_k(x,y)$ is the PSF of exposure $k$, and $\epsilon_k(x,y)$ denotes the observational noise.

To do this, we applied the SMP pipeline \texttt{lightcurver}\footnote{The \texttt{lightcurver} pipeline was originally developed for PSF photometry of lensed quasars in the absence of a reference image, where the quasar emission varies continuously and therefore cannot be cleanly separated from the host galaxy emission through conventional image subtraction.} \citep{Dux2024,Dux2025}, which is built on the \texttt{STARRED} forward-modelling formalism \citep{Michalewicz2023,Millon2024}. In this framework, the extended host and lens galaxy emission is modelled on a higher-resolution grid, with the pixel scale subsampled by a factor of two relative to the original data. The extended component is regularised using the starlet transform, an isotropic wavelet transform widely used in astronomical image reconstruction \citep{Starck1994,Starck2015}. Regularisation constrains the reconstructed image to favour significant spatial structure over noise fluctuations, to preserve astrophysically meaningful features such as the lens galaxy emission.

\begin{figure*}    
\centering
\includegraphics[width=.8\textwidth]{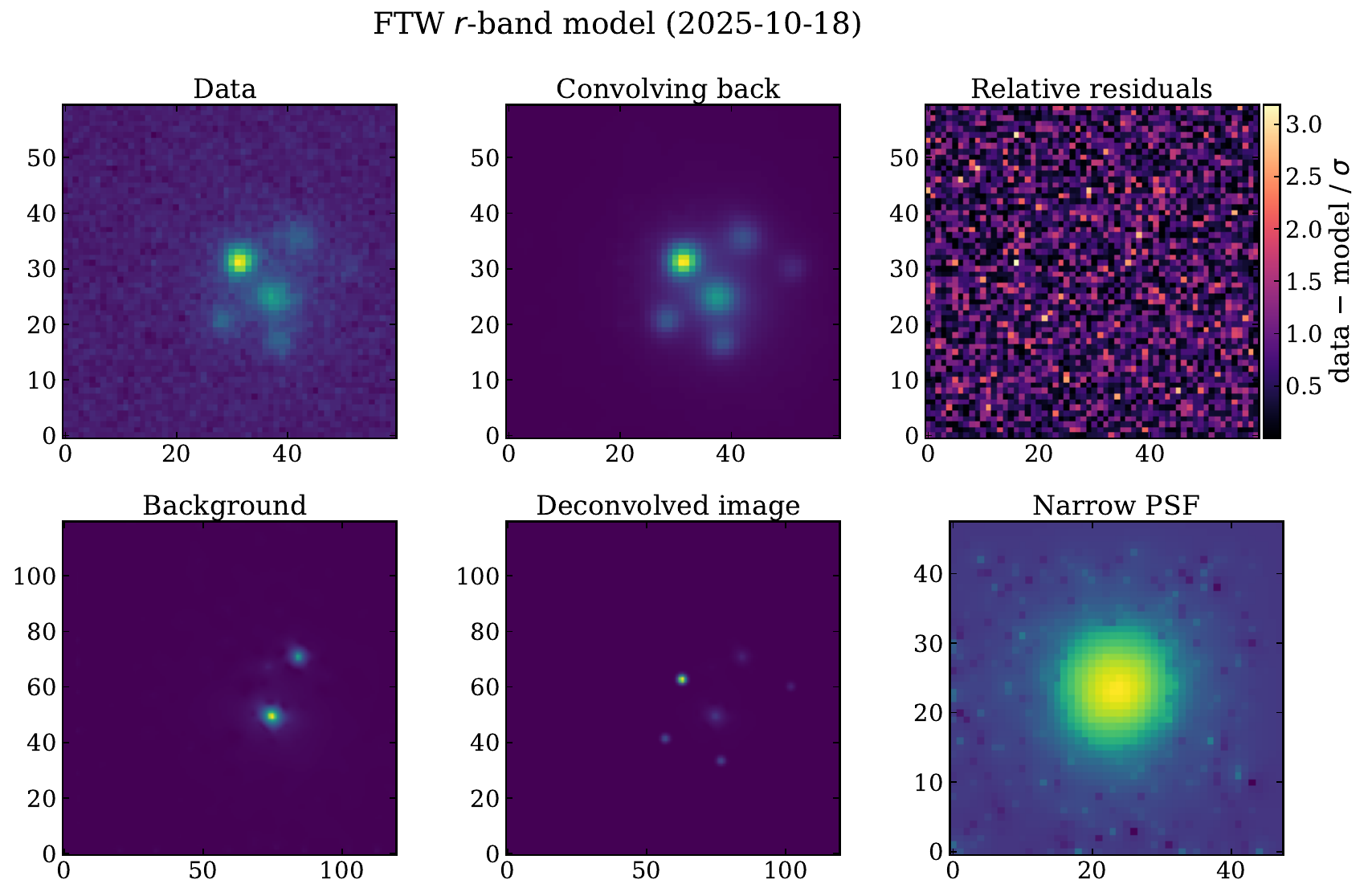}
\caption{Example of the scene reconstruction for a representative epoch from FTW in the $r$-band on 2025-10-18. From left to right and top to bottom: observed image, best-fitting model convolved with the observational PSF, relative residual map, fitted background component, recovered deconvolved source model, and narrow PSF. The narrow PSF is the target PSF to which the source model is deconvolved, with a Gaussian FWHM of 2 pixels \citep{Michalewicz2023}.}
\label{fig:smp_model}
\end{figure*}

Each telescope and filter combination was modelled independently to account for differences in image depth and pixel scale. First, the \texttt{lightcurver} image-processing pipeline uses \texttt{sep} \citep{Bertin1996,Barbary2016} for initial background estimation and source extraction. The resulting source detections are used to obtain an initial astrometric solution, performed with \texttt{Astrometry.net} \citep{Lang2010}. Following the astrometric solution, the pipeline uses \texttt{astroquery} \citep{Ginsburg2019} to identify Gaia calibration stars \citep{Gaia2016} within the field. We typically select stars with $g$-band magnitudes between 14.0 and 21.0~mag \footnote{We also tested an alternative selection with $g$-band magnitudes restricted to 14.0--18.0~mag and found no significant change in the resulting photometry, indicating that the measurements are robust to the choice of reference stars.} and require a minimum mean signal-to-noise ratio (SNR) of 10. For the point-spread function (PSF) reconstruction, we initially select the 10 nearest suitable stars to the target. The final number of stars may differ from this nominal selection after excluding sources affected by nearby contaminants, detector saturation, low signal-to-noise, or astrometric inconsistencies.

Following the initial calibration, the pipeline extracts cutouts of the region of interest (ROI; a small region around the target) and reference stars using \texttt{Astropy} \citep{Astropy2022}, which are subsequently masked and cleaned of cosmic-ray artifacts. For each individual frame, \texttt{STARRED} is used to reconstruct the PSF from the previously identified Gaia calibration stars. PSF photometry of these stars is then used to determine the relative photometric zeropoint and normalisation factor for each exposure. The reduced chi-squared statistic for each fit is recorded in a database, enabling the automated exclusion of frames with poor PSF fits, typically associated with unfavourable atmospheric conditions or inadequate PSF reconstruction.

Finally, we use \texttt{STARRED} to model the scene using an extended component representing the host and lens galaxies, together with point sources representing the multiple lensed images. The positions of images A--D are anchored to those measured from high-resolution HST imaging \companioncitep{Goobar2026}, while small per-exposure astrometric shifts are fitted to account for residual errors in the WCS solutions.

\begin{figure*}    
\centering
\includegraphics[width=.9\textwidth]{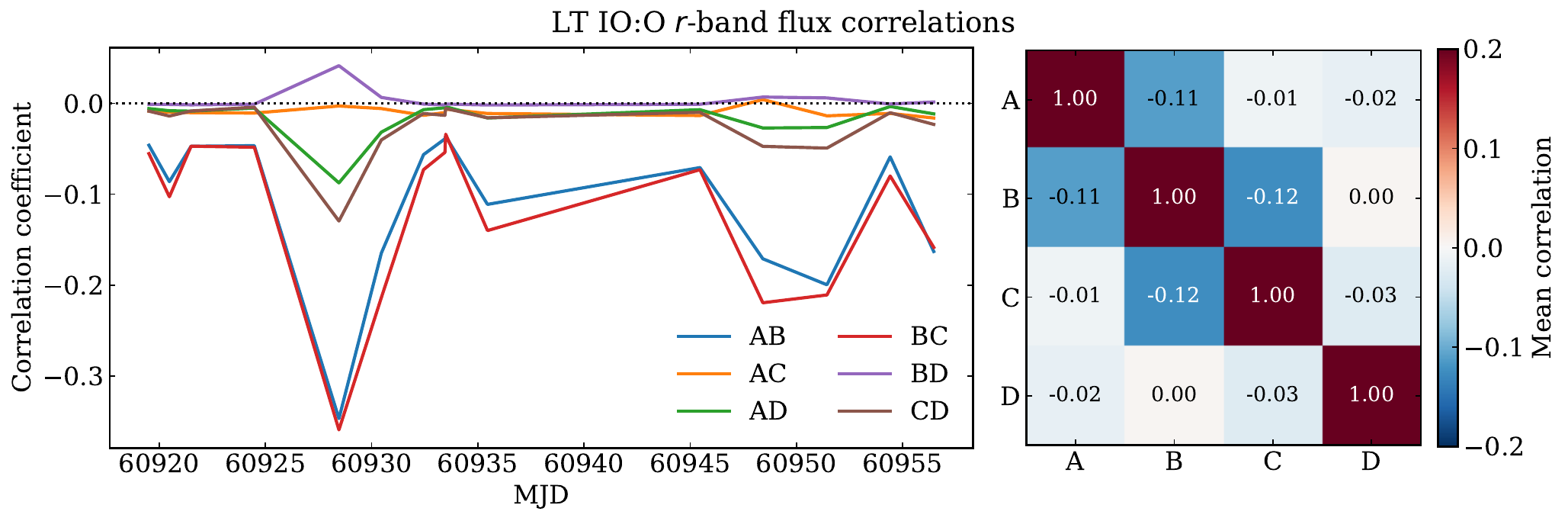}
\caption{Example of the Fisher covariance between the recovered fluxes of the four lensed images for the LT IO:O $r$-band dataset. Left: pairwise correlation coefficients, $\rho_{ij}$, between the fitted fluxes of images A, B, C, and D as a function of epoch. Right: mean correlation matrix averaged over all epochs. The correlations are generally weak ($|\rho| \lesssim 0.1$), indicating that the deconvolution successfully separates the lensed images with minimal flux cross-talk. Epochs exhibiting stronger correlations ($|\rho| > 0.2$) were subsequently excluded by the quality masks described in Sect.~\ref{subsec:masks}.}
\label{fig:flux_covariance}
\end{figure*}

The reconstruction is regularised through hyperparameters that control the strength of the constraints on the extended and point-source components. For the extended emission, these terms regulate the reconstructed structure across spatial scales, suppressing noise-driven or high-frequency features while preserving significant astrophysical structure, and discourage non-physical negative surface brightness. A separate, weaker regularisation is applied to the point-source component to prevent overfitting while preserving the measured SN fluxes \citep{Millon2024}. The default regularisation parameters provided by \texttt{lightcurver} were generally well-suited for high cadence datasets such as the LT data, but were insufficient for datasets with 2--4 epochs such as Gemini/GMOS and LBT/LBC. In these cases, stronger starlet regularisation was required to stabilise the background reconstruction and prevent flux leakage between the extended and point-source components.\footnote{An example of scene modelling and regularisation tuning in \texttt{lightcurver} is provided here \citep{Dux2024}: \url{https://github.com/duxfrederic/lightcurver/blob/main/docs/example_starred_notebooks/example_roi_modelling.ipynb}}.

The \texttt{lightcurver} outputs include reconstructed backgrounds, final scene models, and inferred photometry with associated zeropoints and uncertainties. We visually inspected the data-minus-model residuals for each fit, accepting models with smooth residuals and no significant correlated structure. Small residuals in the lens galaxy cores were considered acceptable (observed for LBT/LBC, e.g., Fig.~\ref{fig:smp_model_lbc}), but photometry was discarded when significant residuals around images A--D indicated potential biases in the point-source reconstruction. This was particularly relevant for datasets with few epochs and sharper PSFs, such as Gemini/GMOS-N and LBT/LBC, where small PSF mismatches are more apparent and the scene model is less well constrained. All LBT/LBC $i$- and $z$-band photometry was discarded due to significant residuals around the lensed images.

An example of the scene reconstruction for a representative epoch is shown for an FTW $r$-band observation (2025-10-18) in Fig.~\ref{fig:smp_model}, using the plotting tools provided by \texttt{STARRED}. The figure shows the observed image, best-fitting model convolved with the observational PSF, normalised residual map, fitted background component, recovered source model, and the narrow PSF used in the reconstruction. The residual map shows the data-minus-model residuals normalised by the pixel uncertainties; the absence of significant systematic structure indicates a good overall description of the observed data. Additional example reconstructions for data from other telescopes are presented in Appendix~\ref{sec:appendix-lightcurver-models}.

Photometric uncertainties were estimated using the Fisher-information formalism \citep{Fisher1922} implemented in \texttt{STARRED}. Following the optimisation of the full scene model, all parameters other than the point-source flux amplitudes were fixed at their best-fitting values, and the fluxes were re-optimised. The Fisher information was then evaluated for the free flux parameters around this optimum, with the resulting diagonal covariance terms used to estimate the $1\sigma$ flux uncertainties. This procedure propagates the pixel-level noise through the forward model, including the effects of the PSF and source blending, while treating the remaining scene-model parameters as fixed. To account for frame-to-frame photometric calibration uncertainties, the Fisher-derived flux uncertainties were combined in quadrature with the relative normalisation uncertainties of each epoch. This normalisation uncertainty represents the uncertainty in the relative scaling applied to bring the photometry from individual exposures onto a common flux scale. The reported photometric uncertainties therefore include both the propagated pixel-noise contribution from the scene model and the uncertainty associated with the relative photometric calibration of each frame.

However, we found that the formal photometric uncertainties were often underestimated, with the observed scatter in the measurements exceeding that expected from the reported uncertainties. We therefore implemented an error floor, adopting a fiducial value of $0.1$~mag. The motivation for this choice and its impact on the analysis are discussed further in Sect.~\ref{sec:time_delays} and Appendix~\ref{sec:appendix_indep_GP}, where an independent Gaussian process analysis of the light curves is used to estimate the additional photometric uncertainty required.

We additionally computed the full Fisher covariance matrix to quantify correlations between the lensed image fluxes arising from imperfect deblending and PSF overlap. The correlation coefficient between images $i$ and $j$ is defined as
\begin{equation}
    \rho_{ij} = \frac{\mathrm{Cov}(F_i,F_j)}{\sigma_{F_i}\sigma_{F_j}},
\end{equation}
where $\mathrm{Cov}(F_i,F_j)$ is the covariance between their fitted fluxes and $\sigma_{F_i}$ and $\sigma_{F_j}$ are their corresponding $1\sigma$ uncertainties. These correlations were generally small ($|\rho| \lesssim 0.1$), indicating minimal coupling between the recovered image fluxes. Epochs in which any pair of images exhibited stronger correlations ($|\rho| > 0.2$) were masked in subsequent analysis (see Sect.~\ref{subsec:masks}). Figure~\ref{fig:flux_covariance} shows the Fisher covariance properties for a representative LT IO:O $r$-band dataset, where the pairwise correlations remain close to zero for most epochs, and the mean correlation matrix confirms the weak coupling between the four image fluxes.

We note that the \texttt{lightcurver} SMP pipeline cannot resolve the individual images in the ZTF P48 data because of the insufficient spatial resolution ($1\arcsec$\,pixel$^{-1}$), which is further limited by the typical seeing at Palomar Observatory ($>2\arcsec$). However, we applied it to the ZTF P48 data at the mean position of the four images to measure their total blended flux, as described in Appendix~\ref{sec:appendix-unres-lightcurver}.

\subsection{Difference Imaging Photometry}
\label{subsec:dia}
In addition to our primary scene-modelling framework, we performed independent DIA photometry across all bands to evaluate photometric systematics and provide a cross-check for the SMP light curves. Science frames were subtracted against CFHTLS ($gr$) and PS1 ($iz$) reference templates, followed by simultaneous forced PSF photometry at the \textit{HST} positions of images A--D \companioncitep{Goobar2026}. While scene modelling remains our primary analytical method due to its superior treatment of blended sources, a complete description of the DIA pipeline is provided in Appendix~\ref{sec:appendix-dia}.

\subsection{Extinction Correction}
\label{subsec:ext_corr}
We corrected all photometric measurements for Milky Way (MW) foreground extinction using $E(B-V)_{\rm MW}=0.0529$~mag from the recalibrated \citet{Schlafly2011} dust maps, and assumed the extinction law of \citet{Fitzpatrick1999} with $R_V=3.1$.

We investigated the possibility of additional extinction associated with the lensing galaxy by comparing the colours of the multiple SN images from resolved HST/WFC3 photometry \companioncitep[described in detail in][]{Goobar2026}. Images A, B, and C have small relative time delays \companioncitep[verified by][]{Johansson2026}, allowing their observed colour differences to provide a first-order probe of differential extinction, while minimising the impact of intrinsic colour evolution. 

The HST/WFC3 photometry provides the strongest constraint on differential extinction because, at the redshift of SN~2025wny, the observed optical filters probe the rest-frame ultraviolet ($\lambda_{\rm rest}\approx1600\text{--}2700$~\AA), where the wavelength dependence of dust extinction is strong. We compare the HST colours of the lensed images at the two available epochs (MJD $\approx$ 60969 and 61010). Taking image A as the reference, images B and C are consistently bluer in the HST bands, with the largest differences occurring towards the bluest filters. For the first HST epoch, the $F475W\text{--}F814W$ colours are $0.85\pm0.02$, $0.66\pm0.02$, and $0.66\pm0.03\:\mathrm{mag}$ for images A, B, and C, respectively, corresponding to colour differences of $0.19\pm0.03\:\mathrm{mag}$ between image A and both images B and C. At the second HST epoch, the same colour differences are preserved: $F475W-F814W=0.56\pm0.02$, $0.34\pm0.03$, and $0.37\pm0.03\:\mathrm{mag}$ for images A, B, and C, corresponding to differences of $0.22\pm0.04$ and $0.19\pm0.04\:\mathrm{mag}$, respectively.

Assuming a Milky Way-like extinction curve at the lens redshift ($z_{\rm lens}=0.375$), these colour offsets correspond to a differential reddening of approximately $\Delta E(B-V)\sim0.07-0.09\:\mathrm{mag}$ between image A and images B/C \companioncitep{Goobar2026}. The consistency of the colour differences between the two HST epochs suggests that the observed trends are not solely due to transient colour evolution. However, the magnitude of the effect remains small.

The inferred reddening should be interpreted with caution because intrinsic SLSN-I colour evolution, small phase differences between images, and residual host galaxy contamination can contribute to the observed colour differences. Furthermore, \companioncite{Li2026} find that the spectra of SN~2025wny do not exhibit strong narrow Na\,\textsc{i}\,D absorption features at either the host or lens redshift, providing no evidence for substantial dust along the line of sight, although we note that Na\,\textsc{i}\,D is not a definitive tracer of extinction \citep[e.g.,][]{Poznanski2012}.

The HST colour differences could be explained by modest differential reddening of $\Delta E(B-V)\lesssim0.1$~mag between image A and images B/C, but do not provide sufficiently strong evidence to justify a lens galaxy extinction correction. Interpreting the observed colour differences purely as differential extinction requires several assumptions: that dust extinction is the dominant origin of the observed colour differences; that the dust in the lensing galaxy follows a Milky-Way-like extinction law; that residual contamination from the lens and host galaxies has a negligible effect on the measured SN photometry; and that chromatic microlensing does not contribute. We therefore do not apply a lens extinction correction in our fiducial analysis, applying only the MW correction. Instead, we treat possible lens galaxy extinction as a systematic uncertainty and explore its impact by applying a correction of $\Delta E(B-V)=0.1$~mag in an alternative analysis presented in Sect.~\ref{subsec:sys_test}.

\subsection{S-correction}
\label{subsec:s_corr}
S-corrections map observed photometry from different telescope and instrument filter sets onto a common, standardised photometric system \citep[e.g.,][]{Stritzinger2002}. They compensate for slight bandpass mismatches by integrating an observer-frame spectrum across the throughput curves of both the observing instrument and the reference standard. While S-corrections are standard practice for homogenising multi-instrument heterogeneous photometry, we explicitly choose not to apply phase-dependent S-corrections to light curves for the following reasons.

\begin{figure*}[t!]    
\centering
\includegraphics[width=.75\textwidth]{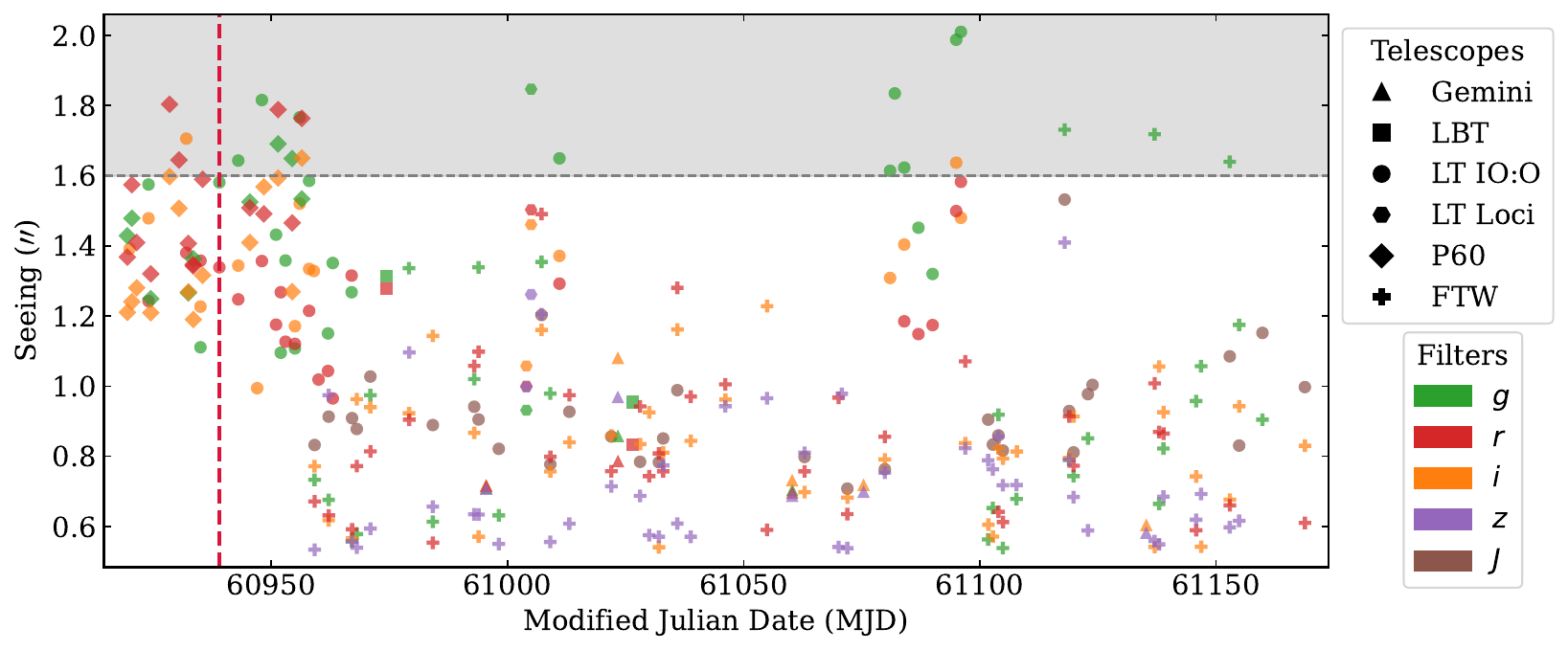}
\caption{Seeing in arcseconds as a function of MJD for the different telescopes and filters used in the campaign. The vertical red dashed line indicates the epoch of peak flux for image A in the $r$ band. The shaded region shows epochs with seeing above the adopted $1.6\arcsec$ threshold.}
\label{fig:seeing_mjd}
\end{figure*}

First, calculating S-corrections requires a representative spectral energy distribution (SED) at each observation epoch. Because no continuous, multi-epoch empirical SED template exists for this object, S-corrections for images B and C must rely on the flux-calibrated spectra obtained from image A, the brightest image. However, these corrections depend non-linearly on rest-frame phase $t_{\text{rest}} = (t_{\text{obs}} - t_0) / (1+z)$. To map a spectrum from image A at epoch $t_A$ to evaluate the correction for image B or C at epoch $t_{B,C}$, one must know the relative time delays a priori to align the rest-frame phases. Incorporating phase-dependent S-corrections would therefore necessitate an iterative fitting scheme, where time delays are assumed, light curves are corrected, and delays are re-measured. Such iterative feedback loops risk introducing subtle circular biases into the final time-delay likelihood surface, anchoring the solution near the initial guess.

Second, calculating accurate S-corrections requires precise, end-to-end optical throughput functions $T(\lambda)$ (including detector quantum efficiency, optical coatings, and atmospheric transmission). Because exact throughput profiles are unavailable for almost all instruments in our dataset\footnote{While the transmission profile for LT IO:O retrieved from the SVO Filter Profile Service \citep{Rodrigo2020} includes the detector quantum efficiency and cryostat window transmission, the profiles available for the remaining instruments represent filter-only transmission curves.}, applying S-corrections using nominal or idealised transmission curves introduces additional, unquantifiable systematic uncertainties into the light curves.

Finally, \companioncite{Li2026} computed S-corrections for image A using flux-calibrated spectra. The corrections are negligible compared with our observational uncertainties, particularly given the adopted $0.1\:\mathrm{mag}$ error floor for the SMP. The median S-correction across all bands is $0.0001$~mag. The $i$- and $z$-band corrections have very small dispersions ($\mathrm{RMS}=0.008$ and $0.010$~mag, respectively), while the larger dispersions in $g$ and $r$ ($\mathrm{RMS}\sim0.033$--$0.037$~mag) are driven by a small number of epochs and remain well below the adopted error floor. Given that the potential systematic errors introduced by circular phase assumptions and imperfect throughput profiles outweigh these sub-percent level offsets, we treat bandpass mismatches as a sub-dominant systematic error rather than applying explicit corrections.

\subsection{Data Quality Mask}
\label{subsec:masks}
To mitigate the impact of erroneous data points in the light curves, we constructed a comprehensive quality mask to identify and exclude problematic epochs based on the following criteria:

\begin{enumerate}
\item Inter-image correlation: Epochs exhibiting strong spatial cross-correlations ($|\rho_{XY}| > 0.2$) were masked for the affected image components $X$ and $Y$ (see Sect.~\ref{subsec:smp}). This removed 6.9\% of the total detections.
\item Independent photometric consistency: We compared the SMP and DIA measurements as an independent check on the recovered photometry. Epochs for which the two measurements differed by more than $1\sigma$ were masked, where $\sigma$ is the combined uncertainty from the SMP and DIA measurements (see Fig.~\ref{fig:25wny_SMphot_residual} for the normalised residuals). The DIA uncertainties include contributions from flux scaling, zeropoint calibration, PSF matching, and potential flux leakage, providing a relatively conservative tolerance for discrepancies between the two methods (see Sect.~\ref{sec:appendix-dia}). The SMP photometry is used for the final analysis, while the DIA measurements provide an independent check against substantially discrepant measurements. When no DIA measurement was available because of quality flags from the DIA pipeline, the SMP measurement was retained. This criterion removed 9.0\% of the total detections.
\item Manual quality control: A subset of highly discrepant epochs was identified via visual inspection and subsequently masked. In particular, we removed the $i$- and $z-$band detections from Gemini due to a significant offset from the other telescope data. This offset was also observed in the DIA light curve (see Fig.~\ref{fig:25wny_DIA_res}), suggesting that the problem is intrinsic to the data, and not an SMP pipeline fault. This removed 5.3\% of the total detections.
\end{enumerate}

These independent flags were combined into a final rejection mask applied to the light curve, resulting in the removal of 19.5\% of the total detections. 

We then applied further quality criteria, requiring an SNR $>5$ and imposing a seeing cut of $1.6\arcsec$. This threshold was chosen conservatively to remove epochs with particularly poor image quality while retaining a significant number of measurements around the peak. Figure~\ref{fig:seeing_mjd} shows the seeing in arcseconds as a function of time in Modified Julian Date (MJD) for the different telescopes and filters used in the analysis. The vertical red dashed line marks the epoch corresponding to the peak flux of image A in the $r$ band. The observations obtained around this epoch were affected by significantly poorer seeing conditions, with many measurements approaching or exceeding the adopted seeing threshold. These observations were predominantly obtained with the LT and P60.

\subsection{Results}
\label{subsec:phot_results}
\begin{figure*}
    \centering
    \includegraphics[width=0.89\linewidth]{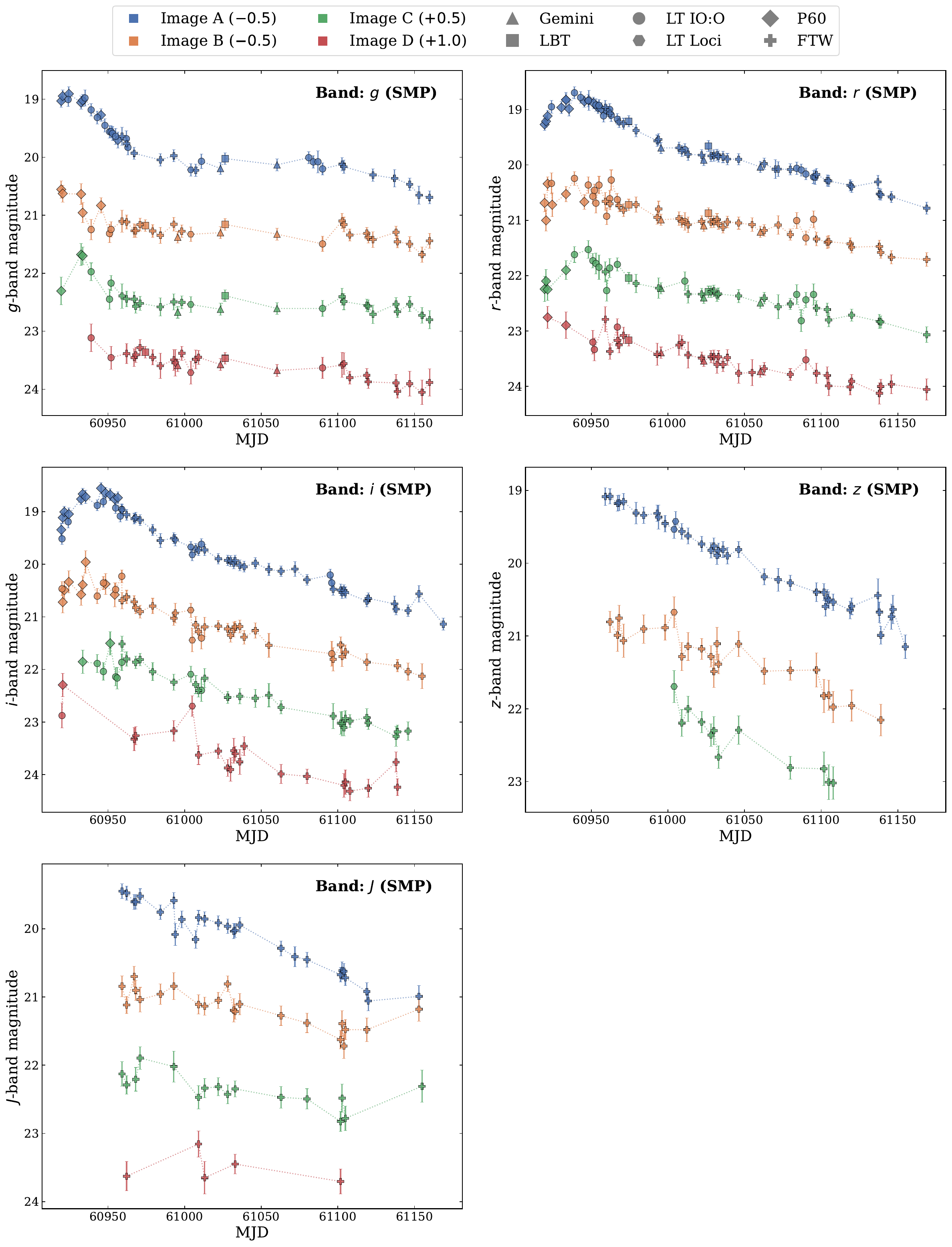}
    \caption{
    Multi-band $grizJ$ light curves of SN~2025wny obtained using the \texttt{lightcurver} SMP framework from observations with Gemini (triangle), LBT (square), LT IO:O (circle), LT LOCI (hexagon), P60 (diamond), and FTW (cross). The resolved images A (blue), B (orange), C (green), and D (red) are shown with vertical offsets of $-0.5$, $-0.5$, $+0.5$, and $+1.0$ mag, respectively, for visual clarity. Only measurements passing the quality cuts described in Sect.~\ref{subsec:masks} are included. The magnitudes are corrected for MW extinction.}
    \label{fig:25wny_SMphot_res}
\end{figure*}

The SMP multi-band $grizJ$ light curve for the resolved data is shown in Fig.~\ref{fig:25wny_SMphot_res}, corrected for MW extinction and with the masking described in Sect.~\ref{subsec:masks} applied. We additionally adopt an error floor of $0.1$~mag to account for the reported photometric uncertainties underestimating the observed scatter. This choice is motivated by the Gaussian process analysis presented in Appendix~\ref{sec:appendix_indep_GP}, which independently inferred error floors of $\sim0.1$~mag in the $g$, $r$, and $i$ bands.

The DIA photometry for the resolved data is presented in Fig.~\ref{fig:25wny_DIA_res} in Appendix~\ref{sec:appendix-dia}, also corrected for MW extinction. A less stringent cut of $\mathrm{SNR}>1$ is adopted for the DIA measurements to account for their larger photometric uncertainties. Compared to the SMP photometry, the DIA light curves exhibit substantially larger scatter, particularly in the $g$- and $i$-band, as well as larger per-epoch uncertainties. These uncertainties account for contributions from flux scaling, zeropoint calibration, PSF matching, and potential flux leakage between the closely separated images, reflecting the additional sources of uncertainty associated with the DIA measurements. Given the larger scatter and uncertainties, we do not use the DIA photometry for time-delay inference.

A comparison of the two photometry methods is shown in Fig.~\ref{fig:25wny_SMphot_residual} in Appendix~\ref{sec:appendix-dia}, where we plot the normalised residuals, defined as the magnitude difference divided by the combined uncertainty. The two methods show good overall agreement, with the majority of measurements consistent within $1\sigma$. Only a small number of points lie between $1\sigma$ and $3\sigma$, with a handful of epochs exceeding the $3\sigma$ threshold. No significant systematic trends or phase-dependent biases are observed across the baseline. This level of agreement is remarkable considering the blended nature of the system and the lack of reference images utilised in the SMP approach, demonstrating the feasibility of using SMP for blended glSN systems.

The raw photometry is listed in Table~\ref{tab:photometry} without extinction corrections or systematic error floors applied; bad epochs identified by our quality cuts are indicated in the mask column.

\section{Time-delay Inference}
\label{sec:time_delays}
In this section, we describe the procedure used to infer time delays from the multi-band light curves of SN~2025wny. Given the absence of a well-established SED template for SLSNe-I, and the possibility that SN~2025wny deviates from the known population \companioncitep[for more details, see][]{Li2026}, we adopt a data-driven approach for modelling the photometry\footnote{We additionally performed template-based light curve fitting using SLSN-I models, which did not provide satisfactory fits to the data; the results are presented in Appendix~\ref{sec:appendix-template}.}. Therefore, we model the intrinsic light curve using a non-parametric Gaussian process model \citep[GPs;][]{Rasmussen2006}, while incorporating the relative magnifications of the lensed images through a parametric lensing model. The combination allows us to infer the photometric time delays without assuming a template for the intrinsic supernova light curve. The time delays of images B and C relative to image A, denoted $\Delta t_{AB} = t_{A}-t_{B}$ and $\Delta t_{AC}= t_{A}-t_{C}$, are inferred in this work. Owing to the low SNR of image D in the majority of the imaging data, and the lack of data around the peak, we do not attempt to measure $\Delta t_{AD}$ here. Instead, this delay was investigated using spectroscopy by \companioncite{Johansson2026}.

We used two approaches to infer the time delays. The primary time-delay measurement was performed using \texttt{GausSN} \citep{Hayes2024}, which provides the main results presented in this work. Additionally, we developed an independent GP analysis, described in Appendix~\ref{sec:appendix_indep_GP}, as a cross-check of the time-delay measurements and to fit an additional photometric error floor. This analysis also provided initial estimates of the time delays and other nuisance parameters used in the \texttt{GausSN} analysis.

\texttt{GausSN} models the intrinsic variability of the source as a draw from a GP. The observed flux of each image is described as a time-shifted and multiplicatively scaled version of this single latent GP, so all images jointly constrain the light curve evolution. The relative time delays, $\Delta t_{AB}$ and $\Delta t_{AC}$, and magnification ratios, $\beta_{B/A}$ and $\beta_{C/A}$, are inferred while marginalising over the shape of the latent light curve.

In the simplest implementation, \texttt{GausSN} assumes constant magnification for each image. To account for physical effects such as microlensing, the model can instead parametrise the magnification as a time-dependent function. Following \citet{Hayes2024}, we adopt a sigmoid magnification model to capture this variation. Physical microlensing simulations \citep[e.g.,][]{FoxleyMarrable2018} show that as an expanding photosphere crosses an individual microcaustic, it can produce a single step-like magnification transition. A sigmoid function provides a suitable functional form to capture this step-like evolution, parametrised as:
\begin{equation}
\label{equ:sigmoid}
    \beta(t)=\beta_0+\frac{\beta_1}{1+e^{-r(t-t_0)}},
\end{equation}
where $\beta_0$ is the macrolensing effect, $\beta_1$ is the scale of the microlensing effect, $r$ is the rate of change in the microlensing effect, and $t_0$ is the time at which the microlensing transition occurs.

In the default implementation of \texttt{GausSN}, the magnification of each image is defined relative to the light curve of an arbitrarily chosen reference image, with the time-dependent sigmoid applied to the other images. Fitting with image A as the reference revealed nearly identical $\beta(t)$ profiles for images B and C, offset only by their initial magnification ratios. As shown in Fig.~\ref{fig:micro_func_default} in Appendix~\ref{sec:appendix-sigmoidgaussn}, both images undergo a concurrent shift toward higher magnifications at later epochs. Since the two images are expected to experience independent microlensing along their spatially distinct lines of sight, the similar time-dependent behaviour of B and C is unlikely to represent two independent microlensing events occurring simultaneously. Instead, this behaviour is more naturally interpreted as evidence that the time-dependent magnification is associated primarily with the reference image A, which is relatively brighter at the earlier epochs. Therefore, we modified the \texttt{GausSN} model such that images B and C are assigned constant magnifications, while image A is assigned a time-dependent sigmoid magnification.

Because \texttt{GausSN} does not use information about the absolute magnitude of the transient, the model is only sensitive to relative magnifications of the images. Instead, the physically relevant quantities are the relative magnifications, $\beta_{B/A} = \beta_B/\beta_{A}(t)$ and $\beta_{C/A} = \beta_C/\beta_{A}(t)$, where $\beta_{A}(t)$ is the time-dependent magnification of image A defined by Equation~\ref{equ:sigmoid}.

The adopted priors on the constant and sigmoid magnification hyperparameters are summarised in Table~\ref{tab:gaussn-priors}. As the light curves are normalised such that the maximum flux is unity, the priors on the kernel amplitude are expressed in these normalised flux units.

\begin{table}[t!]
    \centering
    \caption{Priors on the \texttt{GausSN} Model Parameters.}
    \begin{tabular}{|c|c|}
        \hline
        \textbf{Parameter} & \textbf{Prior} \\
        \hline
        Kernel amplitude, $A$ & $\mathcal{U}(0,5)$ \\
        Kernel timescale, $\tau$ [days] & $\mathcal{U}(20,50)$ \\
        Mean, $c$ & $\mathcal{U}(-0.5,0.5)$ \\
        $\Delta t_{AB}, \Delta t_{AC}$ [days] & $\mathcal{U}(-35,35)$ \\
        \hline
        \multicolumn{2}{|c|}{\textit{Constant magnification}} \\
        \hline
        $\beta_{B/A}$, $\beta_{C/A}$ & $\mathcal{U}(0.01,1.5)$ \\
        \hline
        \multicolumn{2}{|c|}{\textit{Sigmoid magnification}} \\
        \hline
        $\beta_{0,A}$ & $\mathcal{U}(0.01,1.5)$ \\
        $\beta_{1,A}$ & $\mathcal{N}(0,0.5^2)$ \\
        $r_{A}$ [days$^{-1}$] & $\mathcal{N}(0,0.5^2)$ \\
        $t_{0,A}$ [days] & $\mathcal{U}(60900,61200)$\\
        $\beta_{B}$, $\beta_{C}$ & $\mathcal{U}(0.01,1.5)$ \\
        
        \hline
    \end{tabular}
    \label{tab:gaussn-priors}
    \begin{minipage}{\linewidth}
\vspace{1ex}
\footnotesize
\textbf{Notes.} Units are given in square brackets; parameters without units are dimensionless.
\end{minipage}
\end{table}

Following previous applications of GPs to SN light curves \citep[e.g.,][and references therein]{Hayes2024}, we modelled the intrinsic SN variability using a squared-exponential kernel,
\begin{equation}
\label{equ:squared_exp_kernel}
k(t,t') = A^2 \exp\left[-\frac{(t-t')^2}{2\tau^2}\right],
\end{equation}
where $A$ is the characteristic variability amplitude and $\tau$ is the GP kernel correlation timescale. We assume a constant mean function with a single parameter, $c$.

The GP kernel timescale, $\tau$, was assigned a uniform prior over $[20,50]$ days. The time-delay parameters were assigned uniform priors over $[-35,35]$ days, motivated by visual inspection of the light curves, which show a clearly identifiable peak in all three images. The adopted priors on the GP hyperparameters are summarised in Table~\ref{tab:gaussn-priors}.

We sample the posterior using nested sampling with \texttt{dynesty} \citep{Speagle2020}. We used 500 live points and the random-slice sampling method (\texttt{rslice}), with a stopping criterion of $\Delta\ln Z=0.5$.\footnote{We note that we repeated some analyses using 1000 live points and a more stringent $\Delta\ln Z=0.1$ stopping criterion, which resulted in consistent posterior constraints.} Here, $Z$ is the Bayesian evidence and $\Delta\ln Z$ represents the estimated remaining contribution to the evidence from the unexplored parameter space. The nested-sampling output was converted to equal-weight posterior samples using the posterior weights, from which we derived the marginalised posterior distributions for each parameter. We report the posterior median as the central value and the 16th and 84th percentiles as the lower and upper bounds, respectively.

\section{Results and Discussion}
\label{sec:time_delays_results}
The fiducial dataset used in this analysis is the masked, MW-corrected $griz$-band photometry with a $0.1$~mag error floor applied, shown in Fig.~\ref{fig:25wny_SMphot_res}. We restrict the analysis to the $griz$ bands because the $J$-band data are based on observations from only a single telescope (FTW) and therefore cannot be cross-validated against observations from other facilities. Additionally, the observed scatter is greater than that in the other bands, particularly for image A, so the model may attempt to interpret features in the apparent ``wiggles'' in the light curve that are likely to be artefacts of weather.

\begin{figure*}
    \centering
    \includegraphics[width=0.85\linewidth]{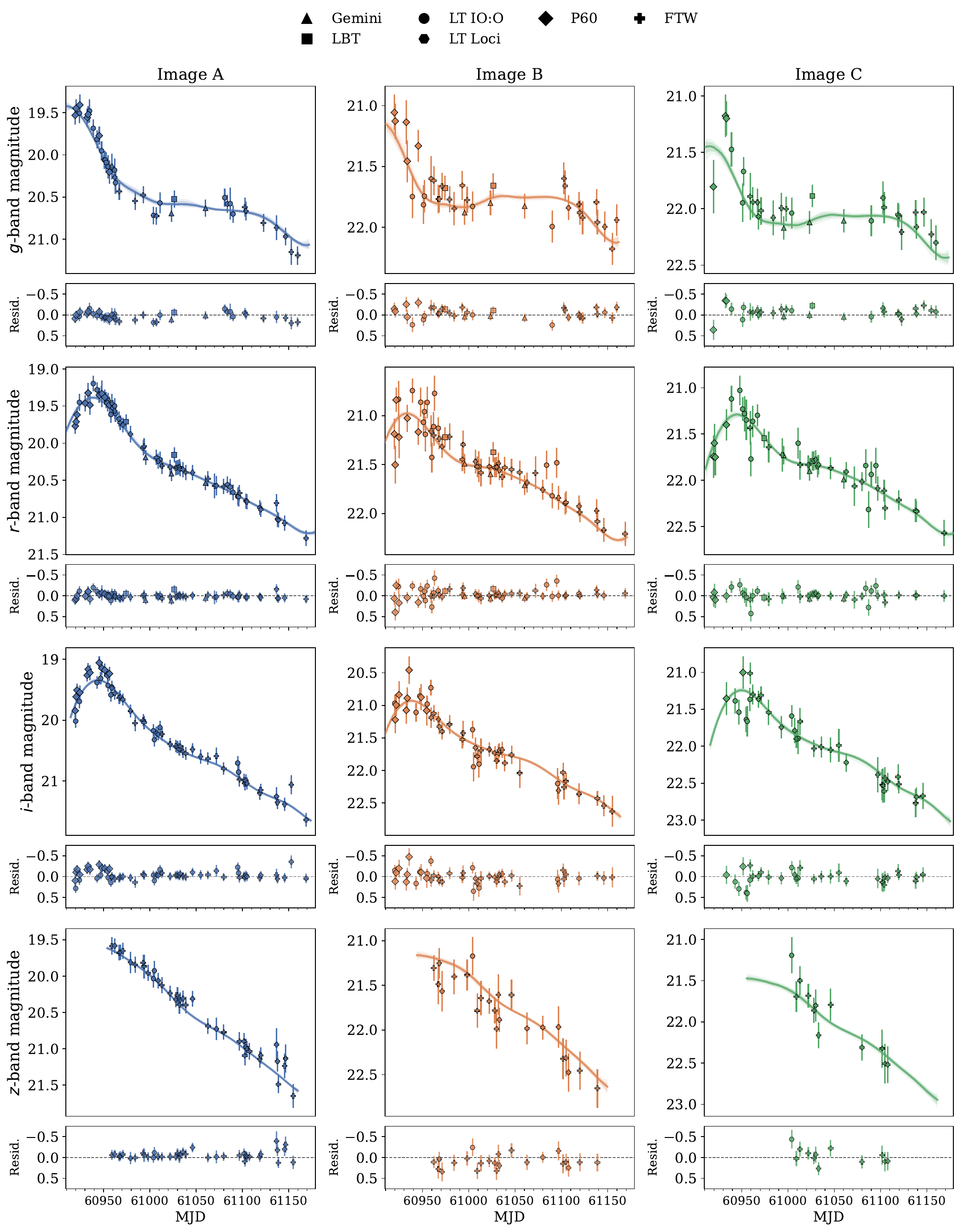}
    \caption{
    \texttt{GausSN} sigmoid magnification model fit to the masked, MW-corrected $griz$-band photometry. The blue, orange, and green points show the observed photometry for images A, B, and C, respectively, with different marker shapes denoting the different telescopes. The solid lines show the median model prediction from the posterior, while the darker and lighter shaded regions indicate the 68\% and 90\% credible intervals, respectively. These intervals are calculated from the distribution of model predictions obtained by evaluating 100 equally spaced samples from the equal-weight posterior, including the uncertainties in the GP and time-dependent magnification parameters. The lower panels show the residuals between the observed magnitudes and the posterior-median model. The inferred relative time delays are $\Delta t_{AB} = -10.6^{+2.2}_{-2.5}$~days and $\Delta t_{AC} = 1.2^{+2.7}_{-2.6}$~days (68\% credible intervals).
    }
    \label{fig:micro_fit}
\end{figure*}

We present the constant magnification \texttt{GausSN} fits to the full dataset and a peak-only subset ($\mathrm{MJD}<61000$) in Appendix~\ref{sec:appendix-constantgaussn}. The fit to the full light curve shows tension around the peak of image A and during the decline phase, with the inferred time delays changing substantially when the analysis is restricted to the early-time data. In contrast, the delay between images B and C remains consistent between the two fits, suggesting that the tension is primarily associated with image A. The inferred relative magnifications also change significantly between the two fits, indicating that a single constant magnification may not adequately describe the observed light curves. These results motivate the inclusion of a time-dependent magnification component, as discussed in Sect.~\ref{sec:time_delays}.

We present the \texttt{GausSN} sigmoid magnification model fit to the fiducial dataset in Fig.~\ref{fig:micro_fit}, with the corresponding parameter estimates given in Table~\ref{tab:micro_time_delay_results}. The inferred time delays are $\Delta t_{AB} = -10.6^{+2.2}_{-2.5}$~days and
$\Delta t_{AC} = 1.2^{+2.7}_{-2.6}$~days (68\% credible intervals). For completeness, we also report $\Delta t_{BC} = \Delta t_{AC} - \Delta t_{AB}=11.3^{+2.6}_{-2.2}$~days, calculated from the posterior samples.

\begin{table}[ht]
\centering
\caption{\texttt{GausSN} Sigmoid Magnification Model Parameter Estimates for the Fiducial Photometry.}
\label{tab:micro_time_delay_results}
\begin{tabular}{lc}
\hline
Parameter & Value \\
\hline
Number of data points, $N$ & 483 \\
Log evidence, $\log Z$ & 1155 \\
\hline
Kernel amplitude, $A$ & $0.18^{+0.07}_{-0.05}$ \\
Kernel timescale, $\tau$ & $35.8^{+5.5}_{-5.1}$ \\
Mean, $c$ & $0.34^{+0.10}_{-0.09}$ \\
\hline
$\Delta t_{AB}$ [days] & $-10.6^{+2.2}_{-2.5}$ \\
$\Delta t_{AC}$ [days] & $1.2^{+2.7}_{-2.6}$ \\
$\Delta t_{BC}$ [days] & $11.3^{+2.6}_{-2.2}$ \\
\hline
$\beta_{0,A}$ & $0.66^{+0.23}_{-0.16}$ \\
$\beta_{1,A}$ & $0.77^{+0.26}_{-0.22}$ \\
$r_{A}$ [days$^{-1}$] & $-0.025^{+0.005}_{-0.007}$ \\
$t_{0,A}$ [days] & $ 60953^{+17}_{-18}$ \\
$\beta_{B}$ & $0.25^{+0.07}_{-0.06}$ \\
$\beta_{C}$ & $0.19^{+0.06}_{-0.04}$ \\
\hline
\end{tabular}
\begin{minipage}{\linewidth}
\vspace{1ex}
\footnotesize
\textbf{Notes.} The reported uncertainties correspond to the $+1\sigma$ and $-1\sigma$ credible intervals. Units are given in square brackets; parameters without units are dimensionless.
\end{minipage}
\end{table}

This model has the highest log-evidence value of all the configurations considered, with $\log Z=1155$, compared with $\log Z=1023$ for the constant magnification model (Appendix~\ref{sec:appendix-constantgaussn}). The corresponding difference in log evidence, $\Delta\log Z=132$, provides strong Bayesian preference for the adopted sigmoid magnification model over the constant magnification model. This indicates the improvement in fit compensates for the additional model complexity.

A second comparison is with the default \texttt{GausSN} sigmoid magnification model, which applies the sigmoid magnification to both B and C relative to image A. This model has $\log Z=1138$, giving $\Delta\log Z=17$ relative to the adopted model. Thus, while both models favour a sigmoid description, the adopted configuration is still preferred by the data. In the adopted model, the sigmoid magnification is applied only to image A, making it less complex than the default \texttt{GausSN} configuration. The higher evidence therefore indicates that this simpler treatment of the magnification is better supported by the data. Importantly, the resulting time-delay estimates are consistent between the two models, indicating that this difference in model complexity does not lead to a significant change in the inferred time delays.

From visual inspection, the sigmoid magnification model provides a significantly improved description of the light curves, particularly around the peak. In comparison with the constant magnification model, the sigmoid magnification model is able to reproduce the shape of the peak in image A more closely. The model also provides a better description of the relative brightness of image A versus B and C over the duration of the light curve.

\begin{figure*}[t!]
    \centering
    \textbf{Sigmoid Magnification \texttt{GausSN} Model}\par\vspace{0.5em}
    \includegraphics[width=0.48\linewidth]{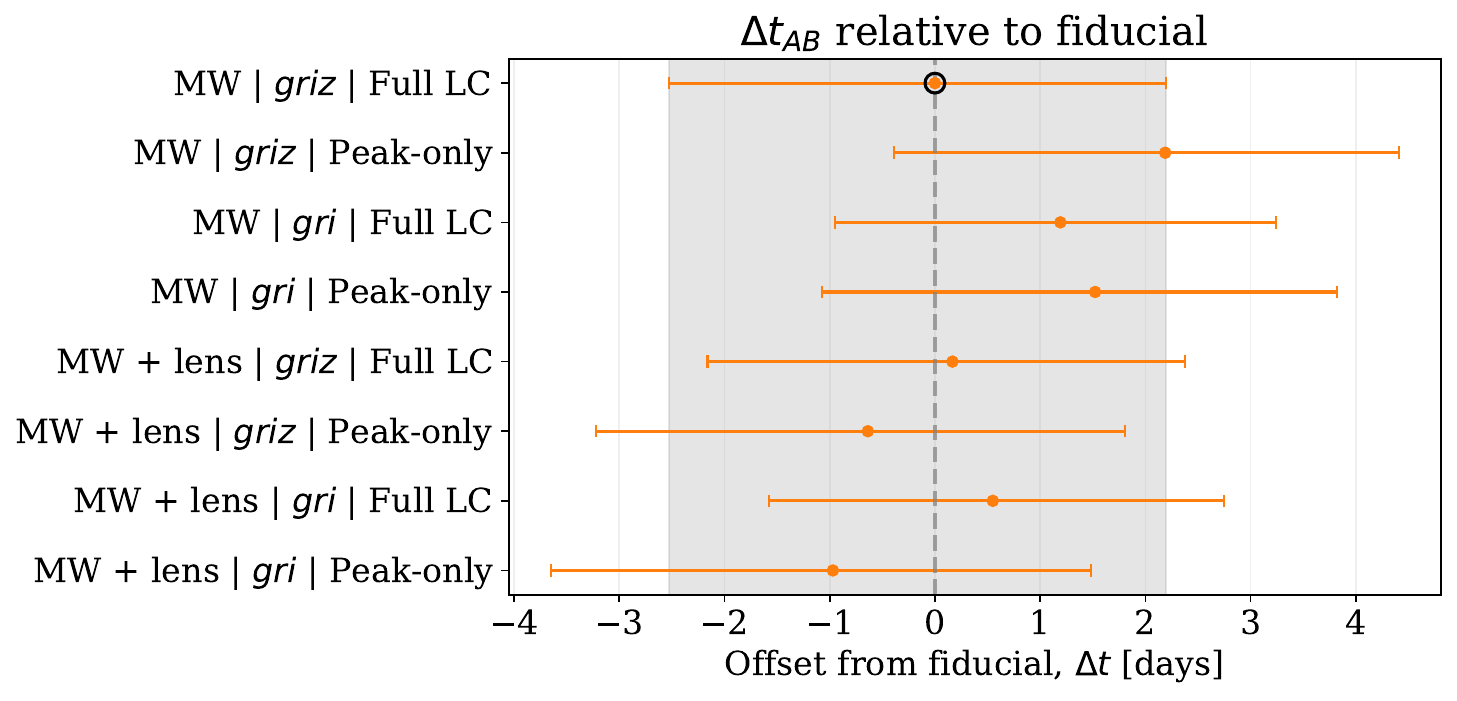}
    \includegraphics[width=0.48\linewidth]{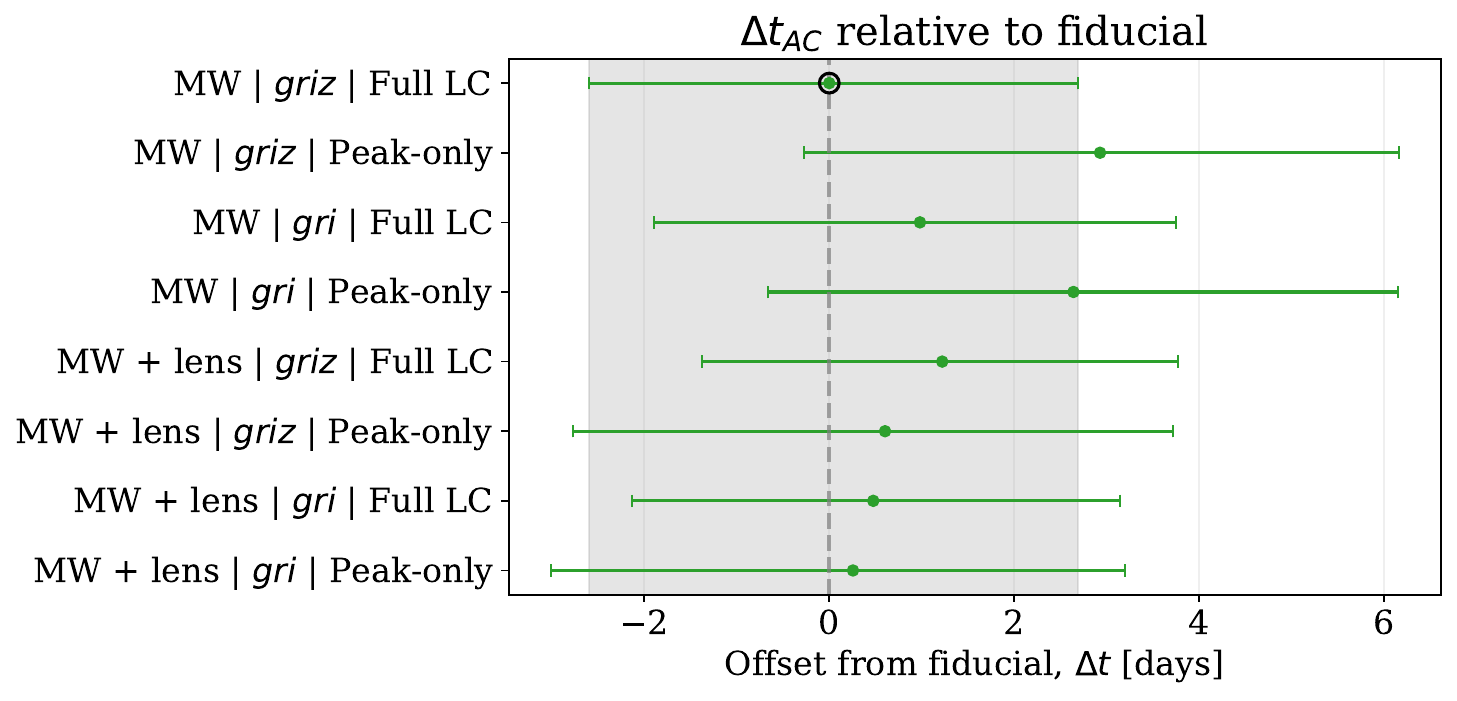}
    \caption{Forest plot showing the systematic tests of the inferred photometric time delays from the \texttt{GausSN} sigmoid magnification model (left: $\Delta t_{AB}$; right: $\Delta t_{AC}$). The horizontal axis shows the offset in the inferred time delay relative to the fiducial analysis, defined as the MW-corrected, $griz$-band, masked, full light curve fit. Each point corresponds to one of the 8 combinations of lens galaxy extinction treatment (MW only or MW and lens correction), temporal range (full light curve or peak-only), and photometric bands ($gri$ or $griz$). Error bars show the statistical uncertainties from the corresponding individual fits. The dashed vertical line marks the fiducial result ($\Delta t=0$), and the open marker identifies the fiducial configuration. The grey shaded region indicates the statistical uncertainty of the fiducial measurement.}
    \label{fig:gaussn_sys_sig}
\end{figure*}

Although the sigmoid magnification model provides a better overall fit, a small systematic discrepancy remains around the peak, with the observations tending to be slightly brighter than the model in the $r$- and $i$-band. One possible explanation is residual differential extinction that is not fully captured by the current corrections, for example extinction associated with the lens or host galaxy. However, when including an additional lens extinction correction for image A at the $\Delta E(B-V)=0.1\:\mathrm{mag}$ level, this did not resolve this discrepancy, with the resulting time-delay measurements remaining consistent to within $\sim1.2\:\mathrm{days}$ (see Sect.~\ref{subsec:sys_test} and Fig.~\ref{fig:gaussn_sys_sig}).

\begin{figure}[t!]
    \centering
    \includegraphics[width=\linewidth]{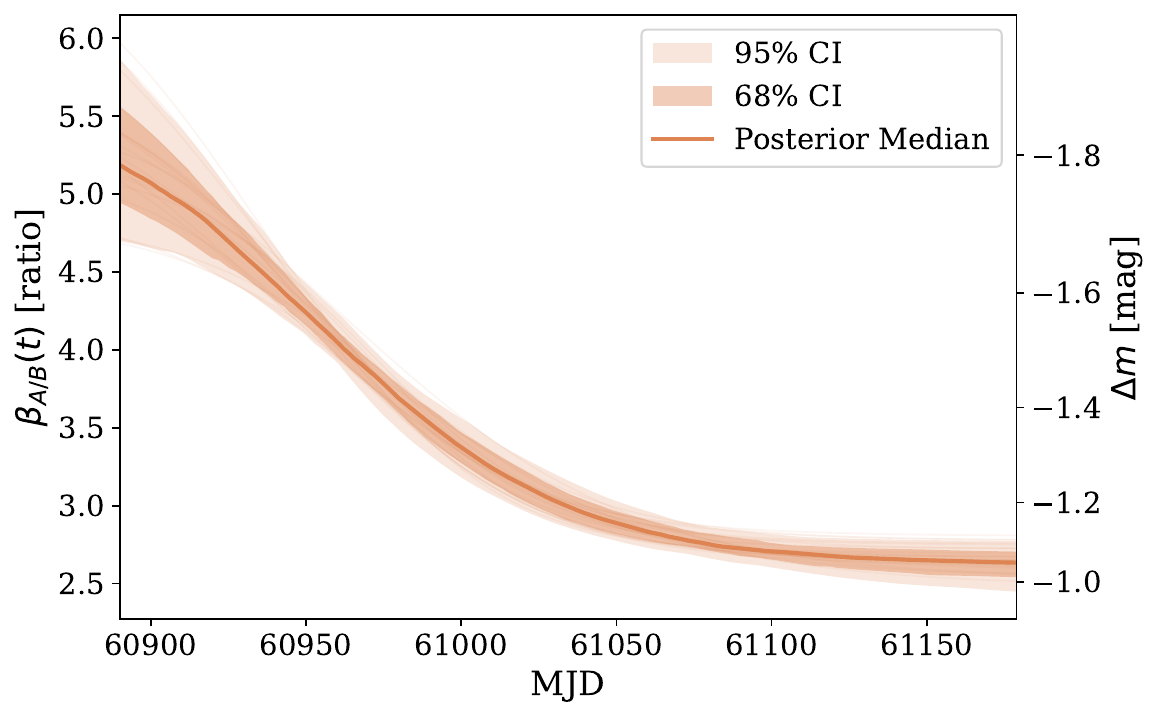}
    \caption{Time-dependent relative magnification inferred from the \texttt{GausSN} sigmoid magnification model for image A relative to image B. The solid line shows the posterior-median relative magnification, $\beta_{A/B}(t)$, while the darker and lighter shaded regions indicate the 68\% and 95\% credible intervals, respectively. The uncertainty regions are calculated from 100 equally spaced samples from the equal-weight posterior, evaluating the sigmoid magnification function for each posterior sample and taking the corresponding percentiles at each time. The right-hand axis displays the corresponding magnitude difference between images A and B.
}
    \label{fig:micro_func}
\end{figure}

Alternatively, the remaining structure could indicate that the sigmoid function is only an approximate description of the underlying microlensing evolution. There could be an additional chromatic component that has not been accounted for. We investigated the possibility of chromatic microlensing by modifying the sigmoid magnification model such that the parameters $\beta_{0,A}$, $\beta_{1,A}$, $r_A$, and $t_{0,A}$ were allowed to vary independently between bands. This did not resolve the tension at the peak of image A and resulted in a slightly lower log-evidence value of $\log Z=1152$, as well as substantially wider credible intervals for the inferred time delays. This indicates that the additional complexity introduced by the chromatic magnification model is not justified by the data, and we therefore did not adopt this model.

Figure~\ref{fig:micro_func} shows the evolution of the relative brightness of image A with respect to image B, by plotting Equation~\ref{equ:sigmoid} using the posterior parameters listed in Table~\ref{tab:micro_time_delay_results}. This demonstrates that the relative magnification cannot be modelled by a fixed constant over the observed time span. We interpret the inferred time-dependent magnification as likely arising from microlensing and estimate that the relative magnification of image A increased by an additional $0.56\pm0.04$~mag during its $r$-band peak ($\mathrm{MJD}\simeq60940$) relative to a later declining phase ($\mathrm{MJD}\simeq61150$).

The posterior distributions of the sigmoid magnification model parameters are presented in the corner plot shown in Fig.~\ref{fig:corner1} (Appendix~\ref{sec:appendix-sigmoidgaussn}). The posterior distributions show a degeneracy between $\beta_{1,A}$ and $r_A$, with two distinct modes corresponding to solutions related by a simultaneous change in the signs of these parameters. These solutions produce equivalent magnification curves and therefore cannot be distinguished by the data. A strong correlation is also evident between $\beta_{0,A}$, $\beta_B$, and $\beta_C$. This arises from an overall scale degeneracy between the magnification parameters and the amplitude of the underlying GP model.

\subsection{Systematic Tests}
\label{subsec:sys_test}

To assess the sensitivity of our time-delay measurements to analysis choices, we performed a full-factorial set of 8 analyses spanning three binary choices: lens galaxy extinction correction (MW extinction correction alone or MW correction supplemented by a correction for extinction from the lens galaxy; see Sect.~\ref{subsec:ext_corr}), temporal range (full light curve or peak-only, with $\mathrm{MJD}<61000$), and photometric bands ($gri$ or $griz$). This approach is useful because some analysis choices are not independent in practice.

Our fiducial analysis uses the \texttt{GausSN} sigmoid magnification model with the masked, MW-corrected $griz$-band photometry. Figure~\ref{fig:gaussn_sys_sig} shows the results of this systematic analysis for the two measured time delays as a forest plot, with each point showing the offset from the fiducial time delay. The error bars represent the statistical uncertainties from the corresponding individual fits.

The systematic variations produce generally modest shifts relative to the statistical uncertainties. The largest shift in $\Delta t_{AB}$ is $+2.2$~days, corresponding to $0.7\sigma$ of the combined statistical uncertainty. For $\Delta t_{AC}$, the largest shift is $+2.9$~days, corresponding to $0.7\sigma$ of the combined statistical uncertainty. Both of these shifts occur for the peak-only analysis with a phase cut of $\mathrm{MJD}<61000$, while retaining the $griz$ bands and the MW extinction correction alone. These shifts are substantially smaller than those determined for the constant magnification model fits in Appendix~\ref{sec:appendix-constantgaussn}.

The factorial analysis also illustrates which analysis choices have the greatest influence on the inferred delays. In particular, restricting the temporal range to the peak-only region produces noticeable shifts, whereas including a lens galaxy extinction correction or changing the photometric bands from $griz$ to $gri$ has a comparatively smaller effect when the other fiducial choices are retained. This is likely caused by microlensing effects not fully captured by the sigmoid magnification model.

A corresponding systematic analysis was performed using the \texttt{GausSN} constant magnification model, shown in Fig.~\ref{fig:gaussn_sys_const} in Appendix~\ref{sec:appendix-constantgaussn}. For this analysis, we also adopted the masked, MW-corrected, $griz$-band photometry as the fiducial case. The inferred time delays show a larger sensitivity to the analysis choices than those obtained with the sigmoid magnification model, with the maximum shifts in both $\Delta t_{AB}$ and $\Delta t_{AC}$ exceeding those found for the sigmoid magnification model. This greater sensitivity suggests that the sigmoid magnification model provides a more stable description of the data with respect to the analysis choices explored here.

The values of $\Delta t_{AB}$ and $\Delta t_{AC}$ from these tests are presented in Table~\ref{tab:sys_time_delays} of Appendix~\ref{sec:appendix-systest}.

\subsection{Comparison with Spectroscopic Time Delays and $H_0$ Inference}

Our photometric time-delay measurements show remarkable agreement with independent spectroscopic time-delay constraints for SN~2025wny from \companioncite{Johansson2026}. Using the evolution of broad absorption-line minima and maxima across several months of observations, they find $\Delta t_{AB}=-10.3\pm2.3$ and $\Delta t_{AC}=0.1\pm3.6$~days. Both delays are consistent within $1\,\sigma$ uncertainties of our photometrically derived values.  This strong agreement between two distinct observational methods provides an important cross-validation of the inferred time delays and suggests that large method-specific systematic biases are unlikely to dominate the measurements.

\companioncite{Mortsell2026} found that combining the lens model with spectroscopically and photometrically measured time delays yields a filter-marginalised constraint of
\[
H_0 = 66.7^{+7.6}_{-6.3}\;
{\rm km\,s^{-1}\,Mpc^{-1}},
\]
for a fiducial model with isothermal mass profiles. Allowing the density slopes of the lens galaxies to vary over a broad range results in
\[
H_0 = 70.8^{+8.2}_{-6.1}\;
{\rm km\,s^{-1}\,Mpc^{-1}}.
\]
We note that these constraints are dominated by the time-delay measurement between images A and D, $\Delta t_{AD}$, from \companioncite{Johansson2026}, which is measured with $\sim5\%$ precision. Using only the photometric time delays and the free-slope lens model yields $H_{0,\:\rm photo} = 80.5^{+26.4}_{-16.7}\;\rm km\,s^{-1}\,Mpc^{-1}$. Nevertheless, our independent measurements of $\Delta t_{AB}$ and $\Delta t_{AC}$ provide important cross-validation of the methodology used by \companioncite{Johansson2026}.

\section{Conclusions}
\label{sec:conclusion}
We have presented photometric time-delay measurements of the strongly lensed SLSN-I SN~2025wny. Combining multi-facility imaging with scene-modelling photometry and Gaussian process regression, we obtained resolved multi-band light curves and measured the time delays between the lensed images. Our findings are as follows:

\begin{itemize}
\item We performed scene-modelling photometry with \texttt{lightcurver} \citep{Dux2024} on imaging obtained from the Gemini North Telescope, Large Binocular Telescope, Liverpool Telescope, Palomar 60-inch Telescope, and the Fraunhofer Telescope at Wendelstein Observatory to deblend the four lensed images of SN~2025wny and construct resolved $grizJ$-band light curves. The smallest angular separation between the images is $\sim2\arcsec$ \companioncitep{Goobar2026}, yet scene-modelling photometry enabled reconstruction of the individual light curves despite the limited angular resolution of the data. This demonstrates that resolved light curves of strongly lensed supernovae can be recovered with 2-metre-class telescopes when the image separations are modest.
\item We considered both a constant magnification model, in which the relative brightness of the images is fixed with time, and a time-dependent sigmoid magnification model motivated by the possibility of microlensing. We implemented these models using \texttt{GausSN} \citep{Hayes2024}.
\item We found that a constant magnification model provides a suboptimal description of the data, motivating a time-dependent sigmoid magnification model to account for evolving relative magnification of image A. The sigmoid magnification model provides an improved description of the observed light curves, while simultaneously accounting for the changing relative brightness of the images over the duration of the observations. The resulting time delays are $\Delta t_{AB} = -10.6^{+2.2}_{-2.5}$~days and
$\Delta t_{AC} = 1.2^{+2.7}_{-2.6}$~days (68\% credible intervals). The evolution of $\beta_{A/B}(t)$ demonstrates that the relative magnification cannot be modelled by a fixed constant over the observed time span. We interpret the inferred time-dependent magnification as likely arising from microlensing and estimate that the relative magnification of image A increased by an additional $0.56\pm0.04$~mag during its $r$-band peak ($\mathrm{MJD}\simeq60940$) relative to a later declining phase ($\mathrm{MJD}\simeq61150$).
\item The constant magnification model provides a useful baseline but shows systematic tension with the observations. In particular, the model underpredicts the brightness of image A around the peak, and there is additional tension during the decline phase. Furthermore, restricting the fit to the rising and peak phases produces shifts of approximately $+7.6$ and $+7.4$~days in $\Delta t_{AB}$ and $\Delta t_{AC}$, respectively. The inferred $\Delta t_{BC}$, however, remains consistent between the two fits to within $0.1$~days. The relative magnifications also differ between the full light curve and peak-only fits at $6\,\sigma$ significance. Together, these results suggest that a constant relative magnification does not fully capture the observed evolution of the three images.
\item Furthermore, our photometrically derived time delays are consistent with independent spectroscopic time-delay measurements from \companioncite{Johansson2026}, providing robust cross-methodological validation of our results prior to cosmological inference.
\item Combining the photometric time delays with the fiducial lens model of \companioncite{Mortsell2026} gives $H_{0,\:\rm photo} = 80.5^{+26.4}_{-16.7}\;\rm km\,s^{-1}\,Mpc^{-1}$, while including the spectroscopic time delays as well yields $H_{0,\:\rm comb} = 70.8^{+8.2}_{-6.1}\;{\rm km\,s^{-1}\,Mpc^{-1}}$.
\end{itemize}

This work demonstrates the potential of scene-modelling photometry for recovering deblended light curves of strongly lensed supernovae, and the utility of Gaussian process regression methods such as \texttt{GausSN} for robustly inferring time delays in supernovae that lack reliable spectral templates. Our results also highlight the importance of accounting for microlensing when measuring photometric time delays in strongly lensed supernovae. These measurements provide a key ingredient for the cosmological analysis of SN~2025wny and further demonstrate the potential of strongly lensed supernovae as independent probes of the Hubble constant.

\begin{acknowledgments}
A.T., S.D., and H.C.T.\ acknowledge support from UK Research and Innovation (UKRI) under the UK government’s Horizon Europe funding Guarantee EP/Z000475/1.

E.E.H.\ is supported by a Gates Cambridge Scholarship (\#OPP1144).

A.G.\ acknowledges financial support from the research project grant “Understanding the Dynamic Universe” funded by the Knut and Alice Wallenberg under Dnr KAW 2018.0067, {\em Vetenskapsr\aa det}, the Swedish Research Council through grants projects Dnr 2020-03444 and 2025-03692 the G.R.E.A.T research environment, Dnr 2016-06012, the EDUCATE excellence center funded by the Swedish Research Council through grant Dnr 2022-06627 and the Swedish National Space Agency, Dnr 2023-00226.

E.M.\ acknowledges support from the Swedish Research Council under Dnr VR 2024-03927. 

C.W.\ acknowledges support from LSST-DA through grant \# 2025-SFF-LFI12-Ward. LINCC Frameworks is supported by Schmidt Sciences, a philanthropic initiative founded by Eric and Wendy Schmidt, as part of the Virtual Institute of Astrophysics (VIA).

A.R., E.P., C.V., E.J.M.\ acknowledge INAF project Supporto Arizona \& Italia.

K.H.\ would like to thank the Caltech Presidential Postdoctoral Fellowship program and President Rosenbaum.

S.D. acknowledges his pet dog Thundercat who passed away during the time the analysis was carried out in the paper (and was a great moral support).

Based on observations obtained with the Samuel Oschin Telescope 48-inch and the 60-inch Telescope at the Palomar Observatory as part of the Zwicky Transient Facility project. ZTF is supported by the National Science Foundation under Award \#2407588 and a partnership including Caltech, USA; Caltech/IPAC, USA; University of Maryland, USA; University of California, Berkeley, USA; Cornell University, USA; Drexel University, USA; University of North Carolina at Chapel Hill, USA; Institute of Science and Technology, Austria; National Central University, Taiwan, and the German Center for Astrophysics (DZA), Germany. Operations are conducted by Caltech's Optical Observatory (COO), Caltech/IPAC, and the University of Washington at Seattle, USA.

The ZTF forced-photometry service was funded under the Heising-Simons Foundation grant No. 12540303 (PI: Graham).

SED Machine is based on work supported by the National Science Foundation under grant No. 1106171.

The LBT is an international collaboration among institutions in the United States and Europe. At the time data were acquired for this research, LBT Corporation Members were the University of Arizona on behalf of the Arizona Board of Regents; Istituto Nazionale di Astrofisica, Italy; LBT Beteiligungsgesellschaft, Germany, representing the Max-Planck Society, the Leibniz Institute for Astrophysics Potsdam, and Heidelberg University; and The Ohio State University, representing The Ohio State University, University of Notre Dame, University of Minnesota, and University of Virginia.

Based on observations obtained at the international Gemini Observatory, a program of NSF NOIRLab, which is managed by the Association of Universities for Research in Astronomy (AURA) under a cooperative agreement with the U.S. National Science Foundation on behalf of the Gemini Observatory partnership: the U.S. National Science Foundation (United States), National Research Council (Canada), Agencia Nacional de Investigación y Desarrollo (Chile), Ministerio de Ciencia, Tecnología e Innovación (Argentina), Ministério da Ciência, Tecnologia, Inovações e Comunicações (Brazil), and Korea Astronomy and Space Science Institute (Republic of Korea).

The Liverpool Telescope is operated on the island of La Palma by Liverpool John Moores University in the Spanish Observatorio del Roque de los Muchachos of the Instituto de Astrofisica de Canarias with financial support from the UK Science and Technology Facilities Council.

This paper contains data obtained with the 2.1-m Fraunhofer Telescope of the Wendelstein Observatory of the Ludwig-Maximilians University Munich. We thank the staff of the Wendelstein observatory and associated students of LMU Munich for their technical help and strong support. In particular, we would like to thank Julius Gassert, Ziyuan Zhu, Hanna Kellermann, Rasika Deshpande, Mitra Maleki, Pablo Vega, Asu Uenver, Xiaoxiong Zuo, Christoph Ries, Michael Schmidt, and Silona Wilke. Funded in part by the Deutsche Forschungsgemeinschaft (DFG, German Research Foundation) under Germany's Excellence Strategy – EXC-2094/2 – 390783311.

We thank Fr\'ed\'eric Dux for his assistance and helpful advice in using the \texttt{lightcurver} package.

\end{acknowledgments}

\begin{contribution}
A.T.\ led the analysis and wrote the majority of the manuscript.
S.D.\ provided supervision and guidance throughout the analysis, led the difference image analysis, some of observing programs and contributed to the manuscript.
E.E.H.\ developed the \texttt{GausSN} software used for the time-delay analysis and contributed to that aspect of the analysis and manuscript.
M.L.\ supported the photometry extraction.
J.J.\ supported the difference image analysis. J.J.\ and E.M.\ contributed to the independent GP time-delay measurement analysis.
A.G., L.Y., S.S., J.O.H., Y.Q., H.C.T., P.M., and J.N.\ were collaborators on the project and key members of the follow-up campaign.
C.W.\ and V.K.\ provided expertise in scene-modelling photometry.
S.D., S.S., J.A., A.B., M.B., C.F., D.G., X.J.H., K.-R.H., E.J.M., C.M.B.O., E.P., D.A.P., A.R., K.S., J.S., C.V., and J.L.W. contributed to the observations, data reduction, and/or leadership of the photometric follow-up campaign.
T.X.C., S.L.G., M.M.K., and J.P. contributed to the construction and development of ZTF collaboration.
All authors reviewed and provided comments on the manuscript.

\end{contribution}

\facilities{Gemini (GMOS-N), FTW (3KK), LBT (LBC), LT (IO:O and LOCI), P60 (SEDM, RC), P48 (ZTF).}

\software{\texttt{lightcurver} \citep{Dux2024},  
          \texttt{STARRED} \citep{Michalewicz2023,Millon2024}, \texttt{GausSN} \citep{Hayes2024}}

\bibliography{2025wny}{}
\bibliographystyle{aasjournalv7}

\appendix

\section{Application of lightcurver to the ZTF P48 Unresolved Photometry}
\label{sec:appendix-unres-lightcurver}

We also analysed the unresolved ZTF forced photometry using \texttt{lightcurver} to test whether the pipeline can independently recover the forced photometry measurements presented in Sect.~\ref{subsec:unresolved_data}. Rather than performing a full deblending analysis (which was explored but found to be unsuccessful for these data), we used \texttt{lightcurver} solely as a photometric extraction tool to measure the total unresolved flux of SN~2025wny. The ZTF $g$- and $r$-band forced photometry was processed independently in \texttt{lightcurver}, without the inclusion of a background image. We also tested an alternative approach in which a background model was generated using historical science images of the field with low seeing; however, this produced comparable results and was not adopted for the final photometry.

\begin{figure*}[h!]
    \centering
    \includegraphics[width=0.7\linewidth]{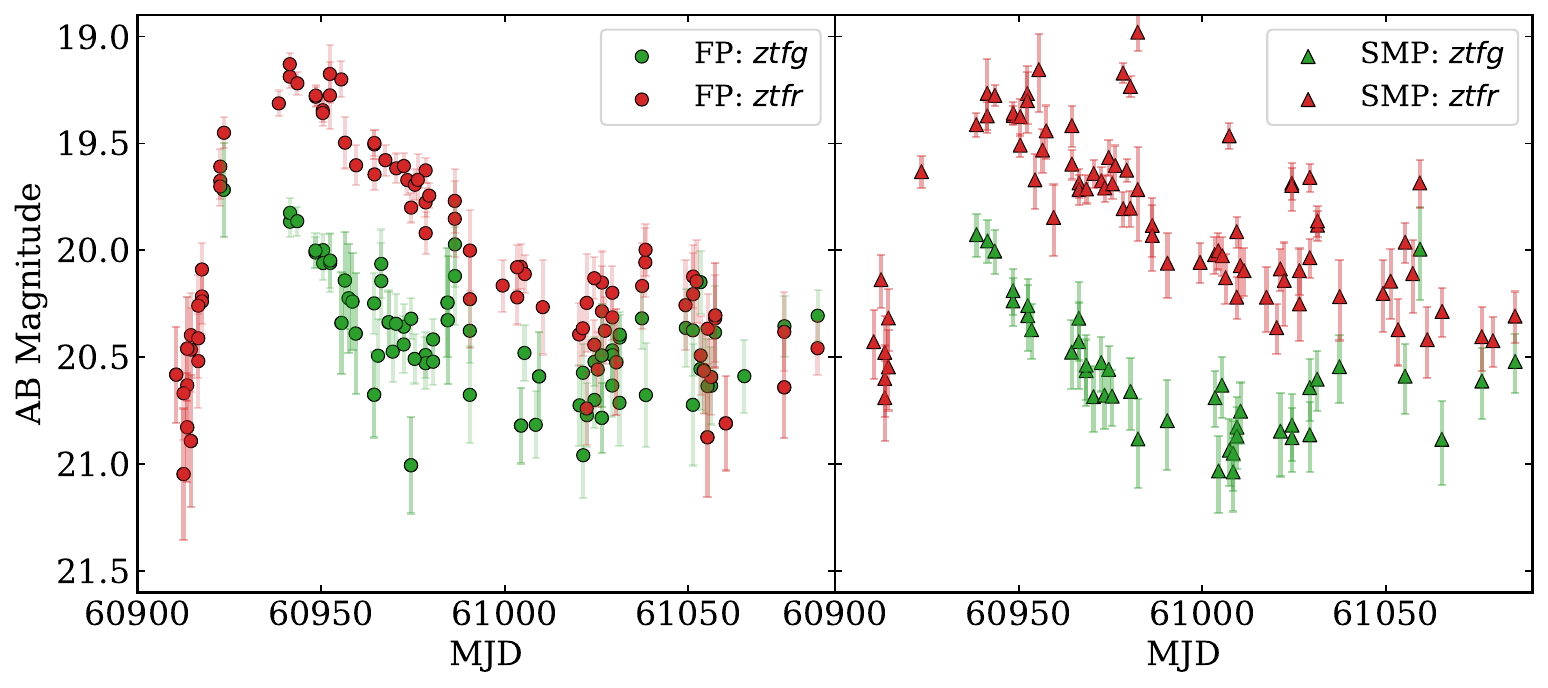}
    \caption{Comparison of unresolved ZTF forced photometry measurements (green: $g$ band; red: $r$ band) from the ZTF forced photometry service (left) and the \texttt{lightcurver} SMP extraction with SNR greater than five (right).}
    \label{fig:ztf_lightcurver}
\end{figure*}

The resulting \texttt{lightcurver} photometry is compared with the ZTF forced photometry measurements in Fig.~\ref{fig:ztf_lightcurver}. The two extractions show consistent evolution across the observed baseline, providing a consistency check on the unresolved light curve. The uncertainties and scatter from \texttt{lightcurver} are noticeably larger than those from the ZTF forced photometry, likely due to the poorer spatial resolution of the photometry and the increased uncertainty in separating the transient flux from unresolved lens and host galaxy contributions without a reference. 

\FloatBarrier

\clearpage

\section{Example scene reconstructions from lightcurver}
\label{sec:appendix-lightcurver-models}

\begin{figure*}[h!]    
\centering
\includegraphics[width=.65\textwidth]{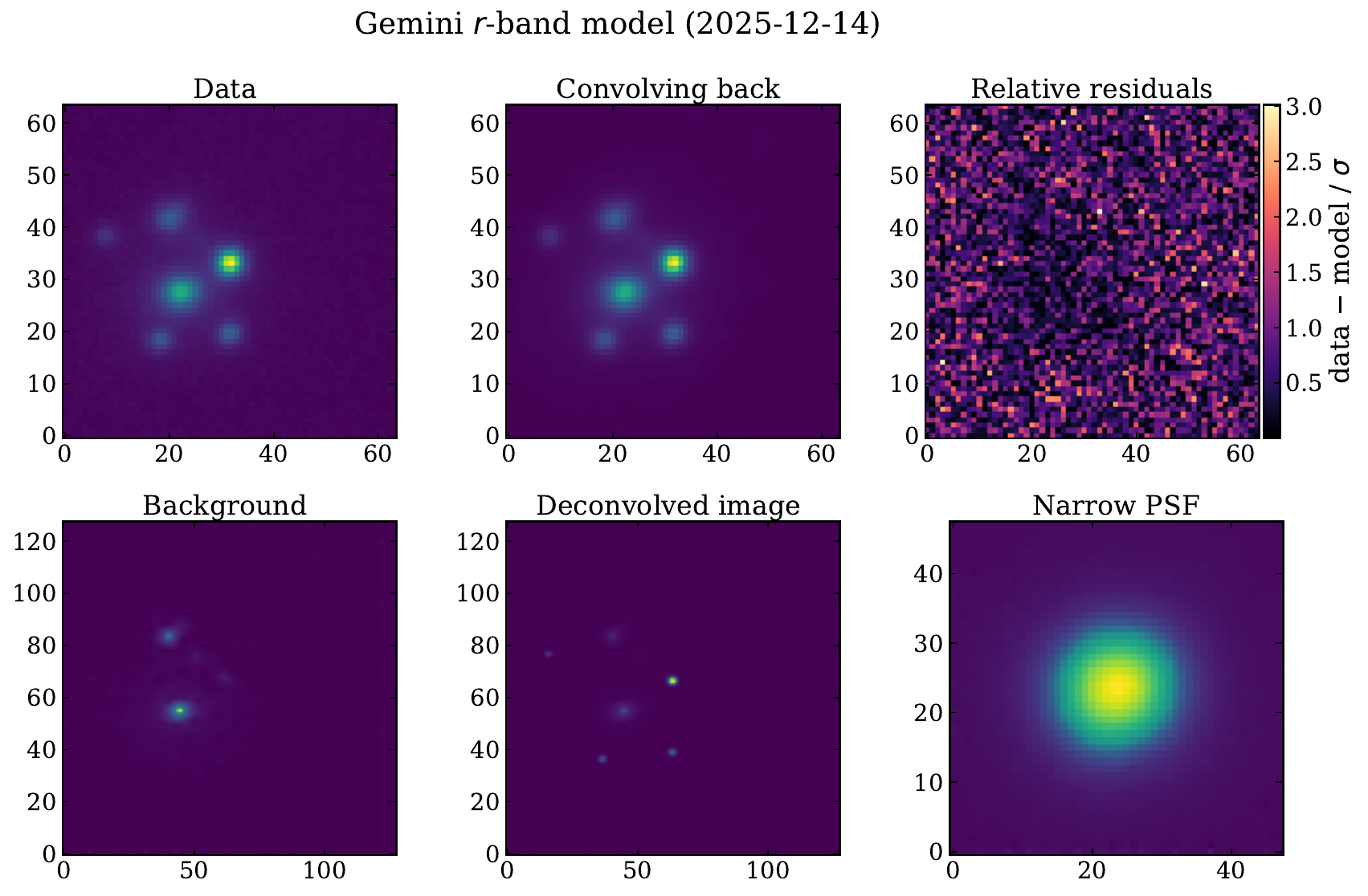}
\caption{Example of the scene reconstruction for a representative Gemini $r$-band epoch on 2025-12-14. From left to right and top to bottom: observed image, best-fitting model convolved with the observational PSF, normalised residual map, fitted background component, recovered source model, and narrow PSF. The residual map is shown in units of the pixel noise, demonstrating that the model provides an adequate description of the observed data. Weak residual structure is visible around the lens system, likely arising from imperfect modelling of the PSF and background; however, these residuals are consistent with the expected noise level and do not significantly impact the reconstruction.}
\label{fig:smp_model_gemini}
\end{figure*}

\begin{figure*}    
\centering
\includegraphics[width=.65\textwidth]{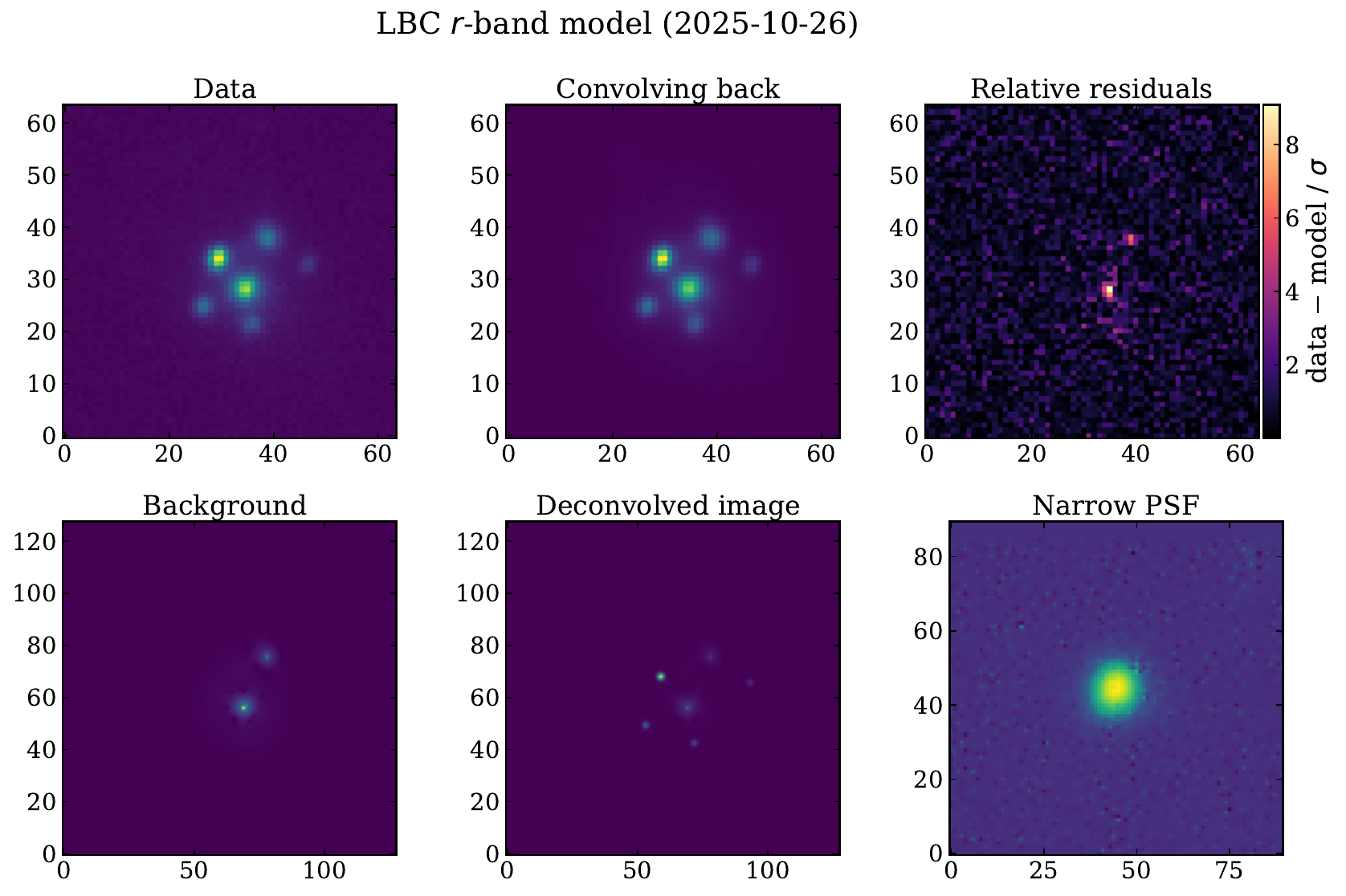}
\caption{Example of the scene reconstruction for a representative LBC $r$-band epoch on 2025-10-26. The panels show the observed image, best-fitting model convolved with the observational PSF, normalised residual map, fitted background component, recovered source model, and narrow PSF. The residual map is shown in units of the pixel noise. Some correlated residual structure is visible around the centres of the two lensing galaxies; however, these residuals are not expected to have a substantial impact on the recovered photometry (see Sect.~\ref{subsec:smp} for further discussion).}
\label{fig:smp_model_lbc}
\end{figure*}

\begin{figure*}    
\centering
\includegraphics[width=.65\textwidth]{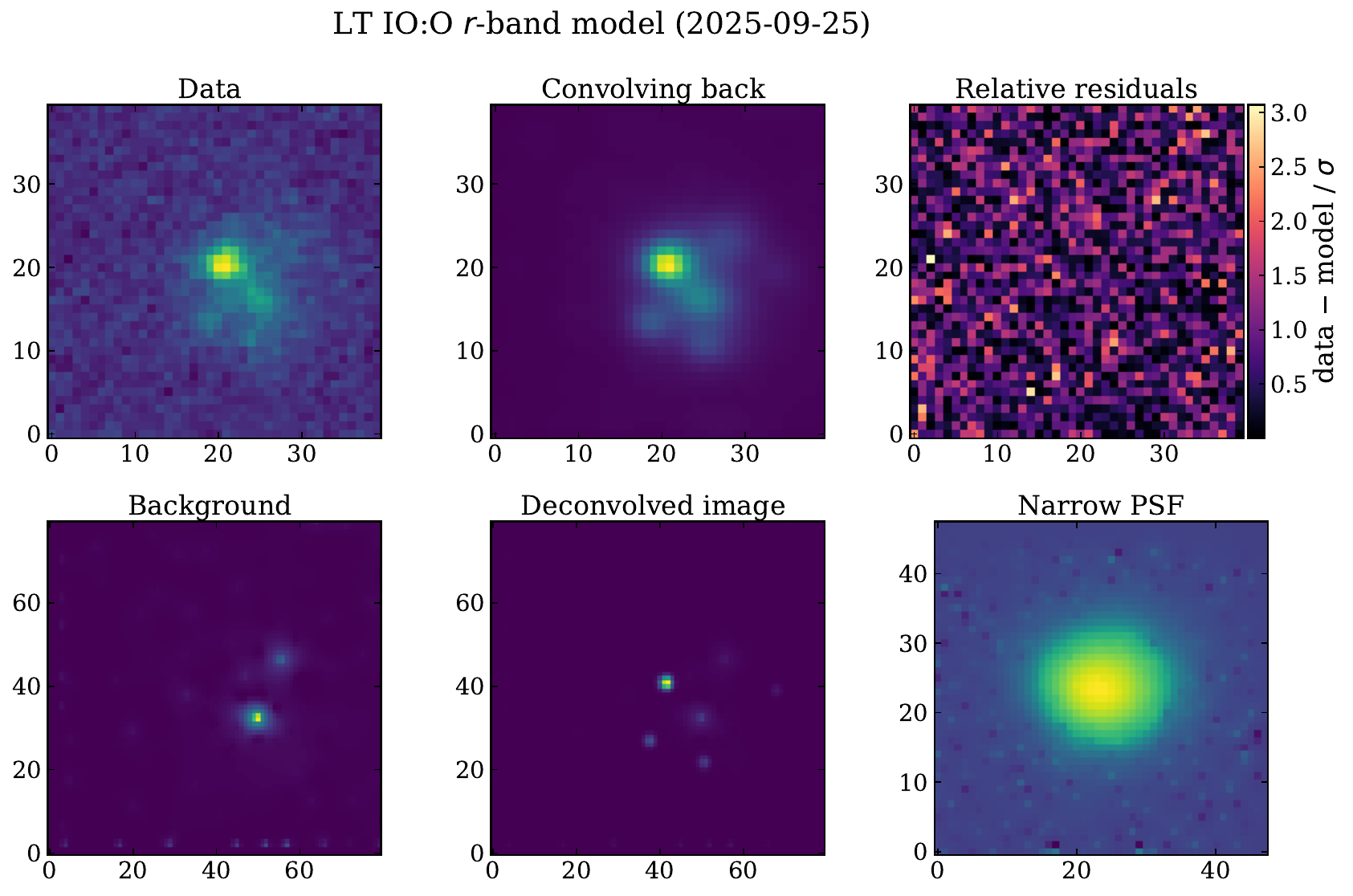}
\caption{Example of the scene reconstruction for a representative LT IO:O $r$-band epoch on 2025-09-25. The panels show the observed image, best-fitting model convolved with the observational PSF, normalised residual map, fitted background component, recovered source model, and narrow PSF. The residual map is shown in units of the pixel noise, demonstrating that the model provides an adequate description of the observed data.}
\label{fig:smp_model_lt}
\end{figure*}

\begin{figure*}    
\centering
\includegraphics[width=.65\textwidth]{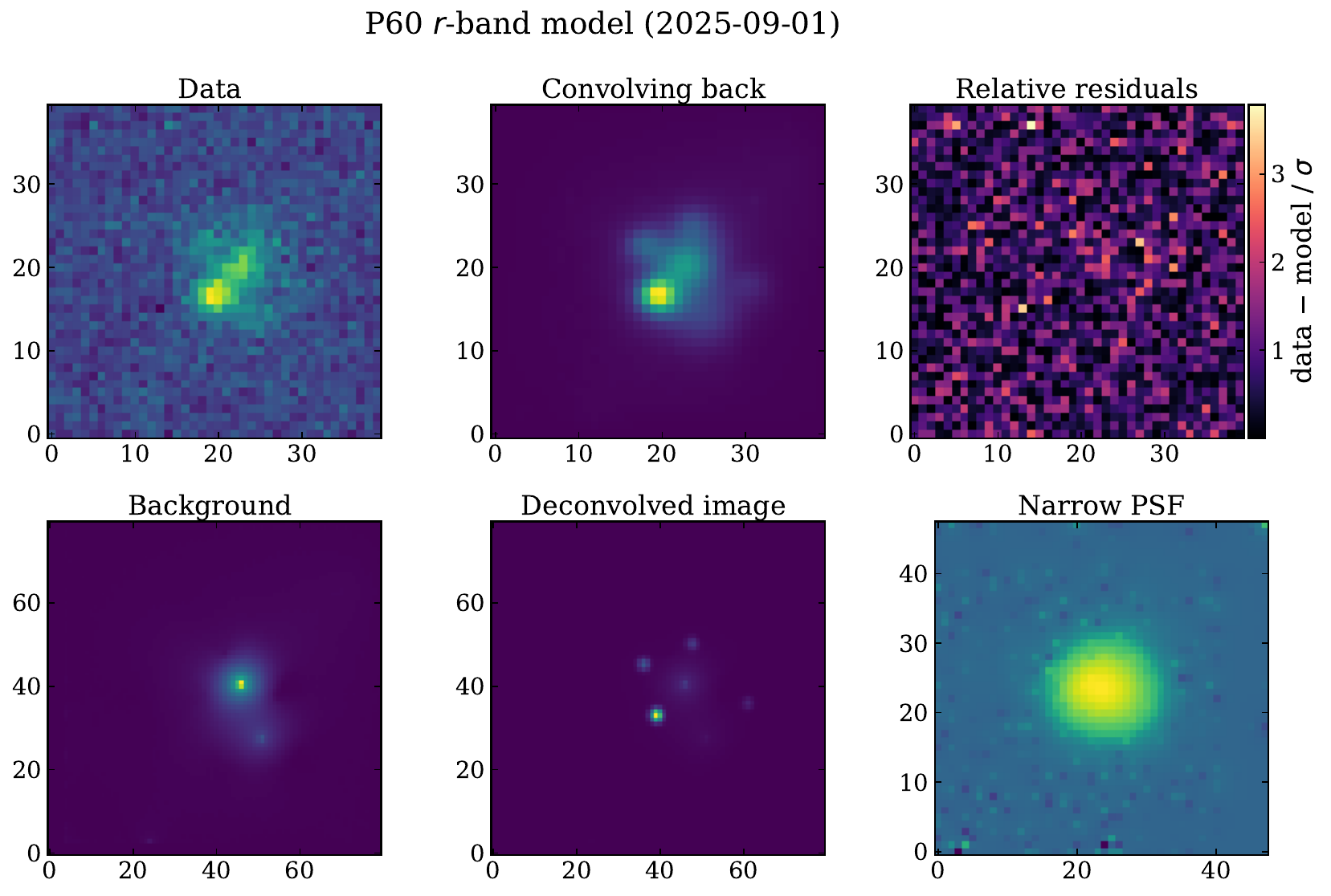}
\caption{Example of the scene reconstruction for a representative P60 $r$-band epoch on 2025-09-01. The panels show the observed image, best-fitting model convolved with the observational PSF, normalised residual map, fitted background component, recovered source model, and narrow PSF. The residual map is shown in units of the pixel noise, demonstrating that the model provides an adequate description of the observed data.}
\label{fig:smp_model_p60}
\end{figure*}

Examples of the scene reconstruction for representative $r$-band epochs are shown for Gemini (2025-12-14; Fig.~\ref{fig:smp_model_gemini}), LBC (2025-10-26; Fig.~\ref{fig:smp_model_lbc}), LT IO:O (2025-09-25; Fig.~\ref{fig:smp_model_lt}), and P60 (2025-09-01; Fig.~\ref{fig:smp_model_p60}).

\FloatBarrier

\clearpage

\section{Difference imaging photometry pipeline}
\label{sec:appendix-dia}
We also inferred the SN image fluxes using a DIA photometry pipeline. The use of two independent photometric methods allowed us to quantify one component of the systematic uncertainty budget. DIA requires an image of the transient field obtained either before the transient explosion or after it has faded below the detection threshold of the instrument used to measure the transient flux. This image, referred to as the `template', is then subtracted from the science frame before performing photometry.

For SN~2025wny, the field had been observed in the $g$ and $r$ filters by the Canada-France-Hawaii Telescope Legacy Survey \citep[CFHT;][]{cfhtls2012} and in the $griz$ filters by the Panoramic Survey Telescope and Rapid Response System \citep[Pan-STARRS or PS1;][]{Chambers2016}. Since the CFHT frames were deeper, we used the $gr$ templates from CFHT and the $iz$ templates from Pan-STARRS.

For the DIA pipeline, we began by reprojecting the science frames onto the appropriate templates in the corresponding filters. We then extracted the common cutouts around the lensing system and performed background subtraction on both the science and template frames. Calibration stars were subsequently detected and fitted to determine their centroids, fluxes, and PSF widths. We corrected the science frames for any residual astrometric shifts and matched the PSFs of the science and template frames using a 2D Gaussian kernel. We note that this follows the core idea of image subtraction introduced by \citet{Alard1998}, with a specified functional form for the kernel. Using the calibration stars, we scaled the template images to match the science frames and performed the subtraction.

Forced photometry was performed on the difference images at fixed positions for images A-- D. These positions were derived from the high-resolution HST imaging \companioncitep{Goobar2026}. The fluxes were fitted simultaneously for all images using a least-squares minimiser. In addition to the fitted flux uncertainties, we aimed to quantify the systematic uncertainties in the photometry. These included contributions from the flux scaling, zeropoint calibration, PSF matching, and potential flux leakage.

To estimate the zeropoint systematic, we generated 100 bootstrap realisations and computed the scatter of individual calibration-star zeropoints after clipping. We bootstrapped the calibration stars with a scatter $\leq 0.02$ mag, refitted the scaling and fluxes, and used the spread in the recovered SN fluxes as the zeropoint systematic uncertainty.

For the scaling systematic, we perturbed the scaling between the reference and science frames according to its calibration-star uncertainty. Specifically, we bootstrapped the stars used to determine the science/reference flux ratio, recomputed the difference images, and repeated the ABCD photometry for each bootstrap sample. The standard deviation of the recovered magnitudes was taken as the scale systematic uncertainty.

We estimated the PSF matching systematic by perturbing the fitted PSF $\sigma$ according to the scatter measured from the calibration stars, i.e., within $\pm 1\sigma$ of the estimated stellar width. Finally, the leakage systematic was quantified by injecting image A-only into the frames and recovering the ABCD fluxes. The scatter from these injections was included as an additional uncertainty term added in quadrature.

\begin{figure*}[h!]   
\centering
\includegraphics[trim=0cm 0cm 0cm 0.6cm, clip, width=.8\textwidth]{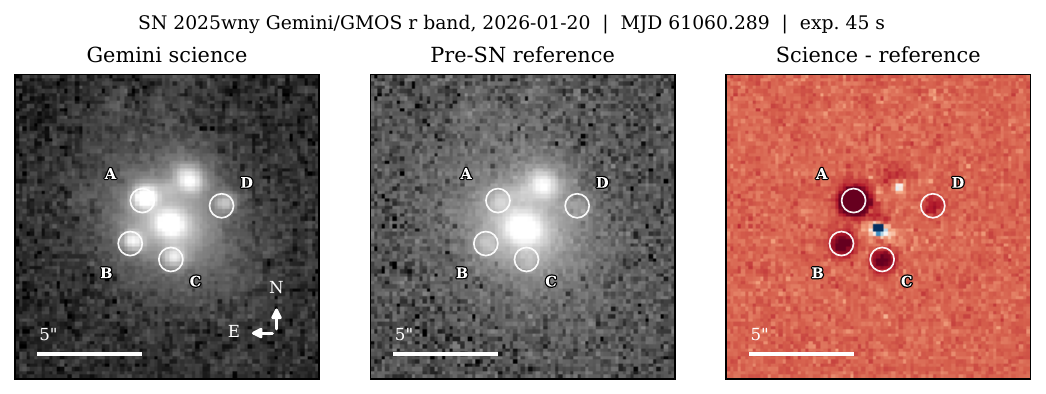}
\caption{Difference imaging analysis of an $r$-band Gemini observation (2026-01-20). The panels show the science image (left), CFHT reference image (middle), and difference image (right). Positions of individual lensed SN images are indicated.}
\label{fig:dia_diagnostic}
\end{figure*}

An example of the difference imaging process applied to an $r$-band Gemini observation (2026-01-20) is shown in Fig.~\ref{fig:dia_diagnostic}, displaying the science image, CFHT reference image, and subtracted difference frame.

\begin{figure*}
    \centering
    \includegraphics[width=0.9\linewidth]{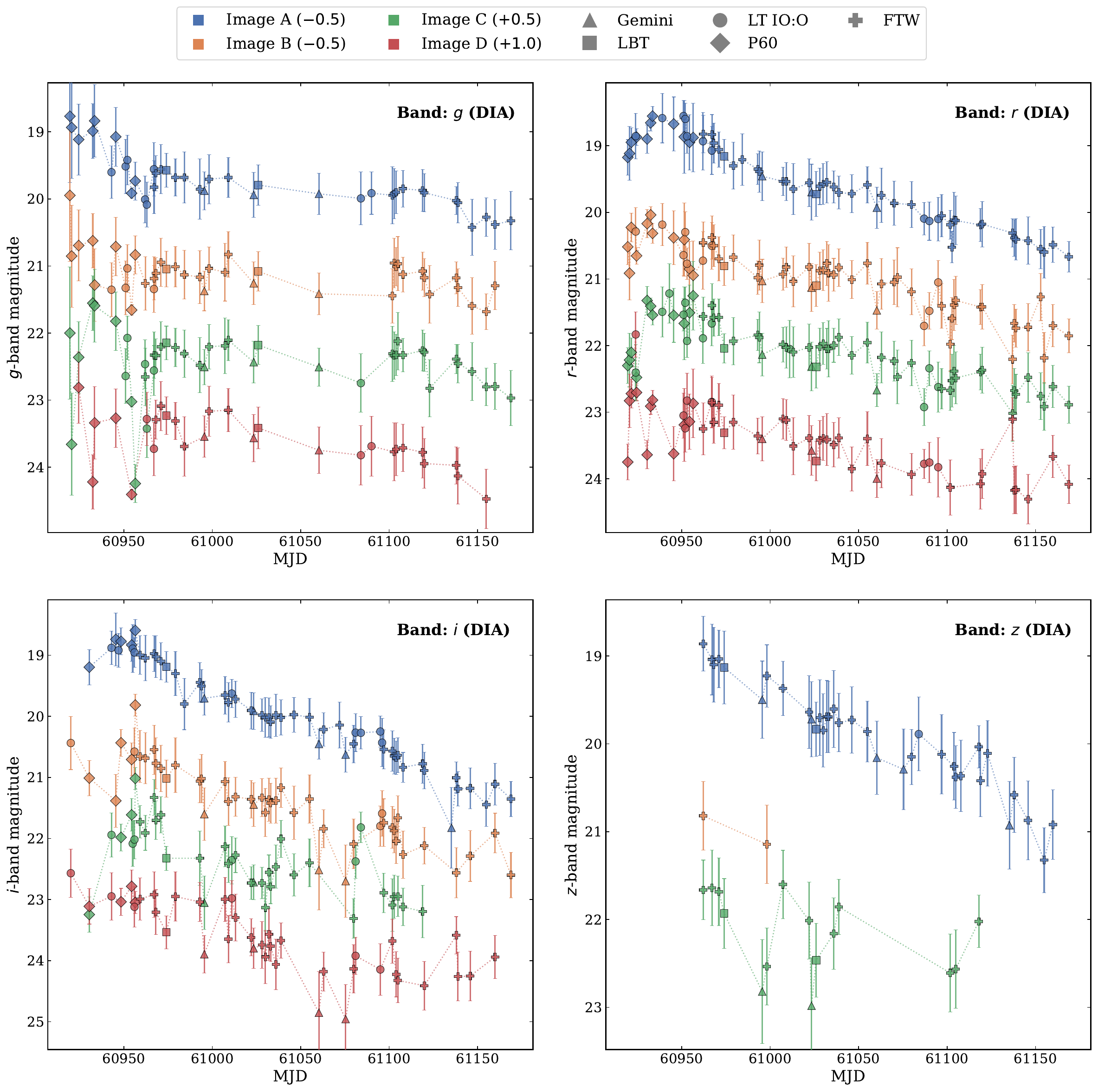}
    \caption{Multi-band $grizJ$ light curves of SN~2025wny obtained using the DIA pipeline from observations with Gemini (triangle), LBT (square), LT IO:O (circle), P60 (diamond), and FTW (cross). The resolved images A (blue), B (orange), C (green), and D (red) are shown with vertical offsets of $-0.5$, $-0.5$, $+0.5$, and $+1.0$ mag, respectively, for visual clarity. Measurements with $\mathrm{S/N}>1$ are shown; the signal-to-noise threshold is relaxed here to illustrate the full light curve evolution despite the larger photometric uncertainties at late times. The magnitudes are corrected for MW extinction. It is important to note that the uncertainties in the DIA photometry are highly correlated between epochs. This covariance should therefore be accounted for when interpreting the significance of features in the light curves.}
    \label{fig:25wny_DIA_res}
\end{figure*}

\begin{figure*}
    \centering
    \includegraphics[width=0.75\linewidth]{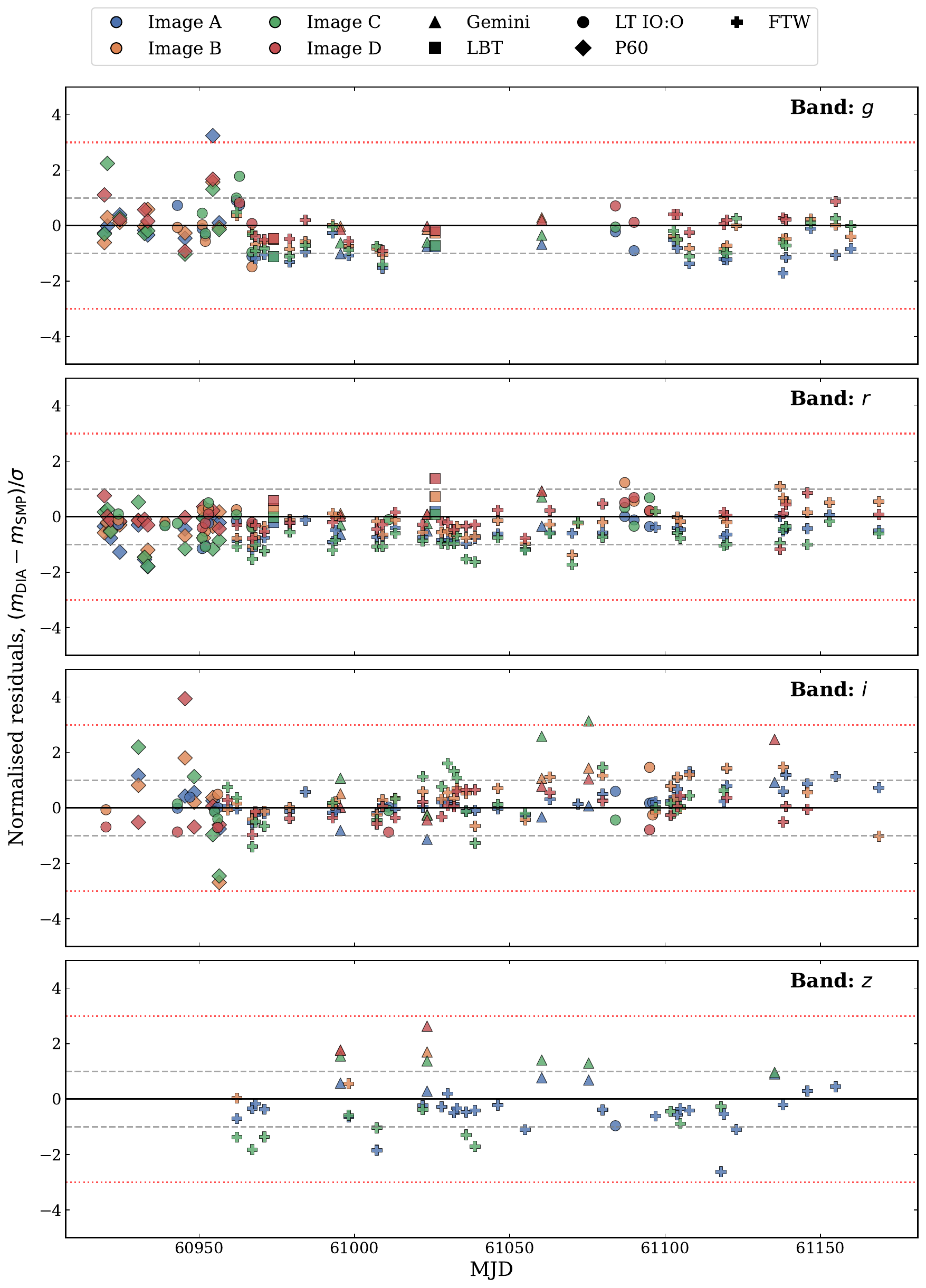}
    \caption{Normalised residuals, defined as the magnitude difference between the DIA and SMP divided by the combined uncertainty. The residuals are shown for data from Gemini (triangle), LBT (square), LT IO:O (circle), P60 (diamond), and FTW (cross) in the $griz$ bands. The images A (blue), B (orange), and C (green) are shown. The majority of data points agree within $1\sigma$, a few within $3\sigma$ and a handful of points above $3\sigma$.}
    \label{fig:25wny_SMphot_residual}
\end{figure*}

The DIA photometry for the resolved data is presented in Fig.~\ref{fig:25wny_DIA_res}, which has been corrected for MW extinction. A cut of $\mathrm{SNR}>1$ is adopted for the DIA measurements to account for their larger photometric uncertainties. A comparison of the two photometry methods is shown in Fig.~\ref{fig:25wny_SMphot_residual}, where we plot the normalised residuals, defined as the magnitude difference divided by the combined uncertainty.

\FloatBarrier

\section{Scene-modelling photometry table}
\label{sec:appendix-smpphot}
The scene-modelling photometry used in this analysis is listed in Table~\ref{tab:photometry} without extinction corrections or systematic error floors applied; bad epochs identified by our quality cuts are indicated in the mask column.

\begin{table}[htbp]
\centering
\caption{Scene-modelling Photometry Measurements and Diagnostic Metadata}
\label{tab:photometry}
\begin{tabular}{cccccccccccc}
\hline\hline
Time & Flux & Flux error & Mag$^a$ & Mag error & Image & Band & Instrument & ZP & Seeing & Mask \\
(MJD) & (counts) & (counts) & (mag) & (mag) & & & (mag) & & (\arcsec) & \\
\hline
60919.509 & 79.02 & 2.94 & 19.73 & 0.04 & A & $g$ & P60/SEDM & 24.47 &1.43 & 1 \\
60920.473 & 85.34 & 2.18 & 19.64 & 0.03 & A & $g$ & P60/SEDM & 24.47 & 1.48 & 1 \\
60924.471 & 88.48 & 5.66 & 19.60 & 0.07 & A & $g$ & P60/SEDM & 24.47 & 1.25 & 1 \\
60932.438 & 77.01 & 3.50 & 19.75 & 0.05 & A & $g$ & P60/SEDM & 24.47 & 1.27 & 1 \\
60933.481 & 79.70 & 1.99 & 19.72 & 0.03 & A & $g$ & P60/SEDM & 24.47 & 1.36 & 1 \\
60945.452 & 63.18 & 2.21 & 19.97 & 0.04 & A & $g$ & P60/SEDM & 24.47 & 1.53 & 1 \\
60951.430 & 56.53 & 1.82 & 20.09 & 0.03 & A & $g$ & P60/SEDM & 24.47 & 1.69 & 0 \\
\dots & \dots & \dots & \dots & \dots & \dots & \dots & \dots & \dots & \dots & \dots \\
\hline
\end{tabular}
\begin{minipage}{\textwidth}
\vspace{1ex}
\footnotesize
$^a$ Magnitudes are reported in the AB magnitude system.\\
\textbf{Notes.} Raw scene-modelling photometry extracted using \texttt{lightcurver}. Values are presented without MW extinction correction and without an added systematic error floor. Table \ref{tab:photometry} is published in its entirety in machine-readable format. A portion is shown here for guidance regarding its form and content. The mask column indicates epochs retained ($1$) or rejected ($0$) based on the quality cuts described in Sect.~\ref{subsec:masks}.
\end{minipage}
\end{table}

\FloatBarrier

\section{Template-based light curve fits}
\label{sec:appendix-template}

As a further consistency check, we performed light curve fits using the PLAsTiCC SLSN-I template set\footnote{\url{https://zenodo.org/records/6672739}}, generated from the MOSFiT framework \citep{Kasen2010,Nicholl2017,Guillochon2018}. This library consists of an ensemble of 999 physically motivated SLSN-I models spanning a wide range of ejecta, magnetar, and opacity parameters.

We constructed \texttt{sncosmo} \citep{sncosmo}\footnote{\url{https://sncosmo.readthedocs.io}} time-series templates from the full set of SLSN-I models and fit each template independently to the resolved photometry of image A in the $gri$ bands. For each template, we allowed the amplitude, time of peak ($t_0$), and lens extinction to vary, while fixing the redshift, Milky Way extinction, and host extinction. We adopt a Milky Way reddening of $E(B-V)=0.0529$ mag from \citet{Schlafly2011} and assume the extinction law of \citet{Fitzpatrick1999} with $R_V=3.1$. We assumed a host extinction of $E(B-V)=0.1$ mag.

\begin{figure}[h!]
    \centering
    \includegraphics[width=0.35\linewidth]{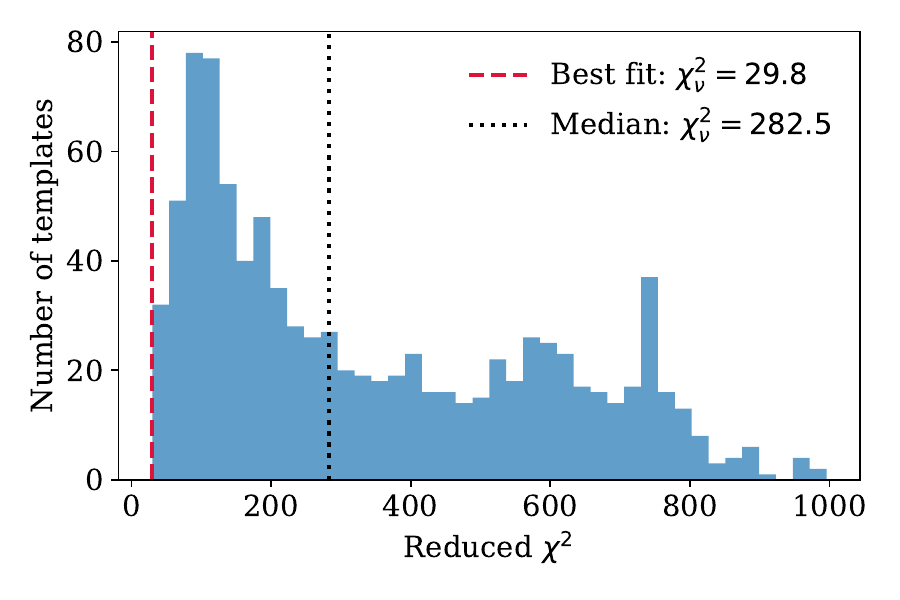}\includegraphics[width=0.65\linewidth]{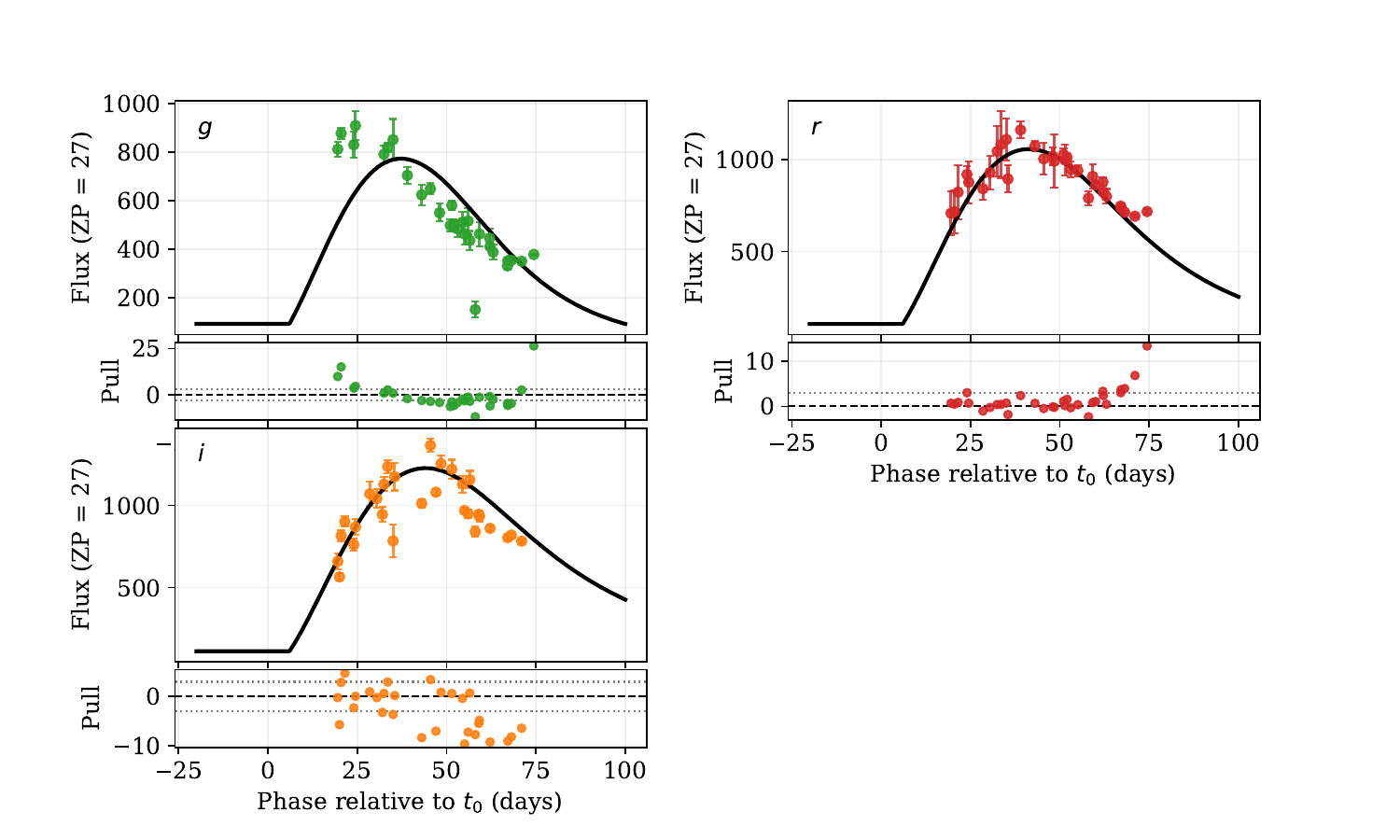}
    \caption{\textit{Left}: Distribution of reduced $\chi^2$ values for fits of 999 MOSFiT-based SLSN-I templates from the PLAsTiCC dataset to the $gri$ photometry of SN~2025wny image A. The distribution is strongly peaked at high reduced $\chi^2$, with the best-fit model still yielding $\chi^2_\nu \sim 30$, indicating that no template provides an acceptable fit to the data. \textit{Right:} Best-fitting PLAsTiCC MOSFiT SLSN-I template compared to the $gri$ light curve of SN~2025wny image A. Despite optimisation of amplitude, peak time, and extinction, the model fails to reproduce both the colour evolution and the temporal structure of the observed light curve, resulting in significant correlated residuals across all bands.}
    \label{fig:template_chi2}
\end{figure}

None of the templates provide an acceptable fit to the photometry, with the best-fit model resulting in a reduced $\chi^2 \simeq 30$ (plotted in the right panel of Fig.~\ref{fig:template_chi2}). In particular, the templates fail to simultaneously reproduce the observed multi-band colours and the temporal evolution of the light curve. The left panel of Fig.~\ref{fig:template_chi2} shows the distribution of reduced $\chi^2$ values across the 999 template models, demonstrating that the MOSFiT-based SLSN-I family does not adequately describe SN~2025wny.

We note that introducing an additional error floor does not resolve this discrepancy, as the residual structure is correlated in time and across bands. This suggests that standard SLSN-I template families are insufficient to describe SN~2025wny, reinforcing the need for a more flexible, non-parametric modelling approach in the main analysis.

\FloatBarrier

\section{Independent Gaussian process modelling}
\label{sec:appendix_indep_GP}

As an independent verification of the time-delay method with \texttt{GausSN}, we implemented a separate GP analysis of the multi-band light curves. The purpose of this analysis was twofold: first, to verify that the inferred time delays are robust to the details of the GP implementation; and second, to provide additional flexibility in modelling photometric systematics, including an explicitly fitted error floor.

We used the same dataset and raw photometric uncertainties as in the \texttt{GausSN} analysis, before application of the adopted error floor. The light curves were modelled in flux space. Image A was adopted as the reference image, and we fit the time delays and relative magnifications $(\Delta t_{AB}, \Delta t_{AC}, \beta_{B/A}, \beta_{C/A})$ simultaneously.

For a trial set of parameters, the observations from image $X$ were transformed onto the reference light curve frame according to $t' = t-t_{AX}$ and $F' = \frac{F-\delta_{X,b}}{m_X}$, where $m_X$ is the multiplicative flux scale and $\delta_{X,b}$ is an additive flux offset for image $X$ in band $b$. The corresponding statistical uncertainty was transformed as $\sigma_{F'}=\frac{\sigma_F}{m_X}$. The transformed observations from the three images were then combined within each band.

The common intrinsic light curve in each band was modelled using a GP with a squared-exponential kernel, as in Equation~\ref{equ:squared_exp_kernel}. For each trial set of time delays and nuisance parameters, the GP was evaluated on the combined, transformed light curve. The GP amplitude was estimated from the standard deviation of the transformed flux measurements. For the results presented here, a common GP correlation timescale was fitted simultaneously with the other model parameters, with a prior range of 20--50 days centred on a characteristic timescale of 30 days.

The covariance matrix used in the GP likelihood included both the quoted photometric uncertainties and an additional error floor contribution,
\begin{equation}
   K_{\rm tot} =K_{\rm GP}
+
{\rm diag}
\left(
\sigma_{\rm phot}^2+
\sigma_{\rm floor}^2
\right). 
\end{equation}
This additional term is intended to account for residual photometric systematics or underestimated uncertainties that are not captured by the reported measurement errors. The error floor was fitted independently for each photometric band. We adopted weak log-space priors centred on 0.10 mag, with the parameter constrained to lie between 0.01 and 0.30 mag. The error floor parameters are defined in magnitude units and sampled in logarithmic space; the quoted bounds and prior therefore refer to the magnitude-domain error floor before conversion to flux units. Including an explicit error floor is useful for this dataset because otherwise unmodelled photometric scatter can be absorbed by the GP as apparent short-timescale structure.

The total objective function was the sum of the GP negative log marginal likelihoods over the fitted bands, together with the contributions from the nuisance-parameter priors. For a data vector $\mathbf{y}$, the GP negative log marginal likelihood has the form

\begin{equation}
    -\ln \mathcal{L}_{\rm{GP}}
=
\frac{1}{2}
\mathbf{y}^{T}
K_{\rm{tot}}^{-1}
\mathbf{y}
+\frac{1}{2}\ln |K_{\rm{tot}}|
+
\frac{N}{2}\ln(2\pi).
\end{equation}

Before evaluating the likelihood, the transformed flux data were centred by subtracting their mean within each band; $\mathbf{y}$ therefore denotes the resulting mean-subtracted flux vector. The full objective function can be written schematically as
\begin{equation}
   \sum_b
\left(
-\ln\mathcal{L}_{{\rm GP},b}
\right)
+
\frac{1}{2}\chi^2_{\rm prior}, 
\end{equation}

where the prior term contains contributions from the flux scales, error floors, and GP correlation timescale.

To locate the region of maximum likelihood, we first evaluated the objective function on a two-dimensional grid in $(\Delta t_{AB}, \Delta t_{AC})$. This coarse grid was used to identify a suitable starting point for subsequent numerical optimisation. We then minimised the full objective function using the bounded L-BFGS-B algorithm, allowing the delays and nuisance parameters to vary simultaneously. The best-fit parameters from the L-BFGS-B optimisation were used to initialise an ensemble Markov Chain Monte Carlo analysis with \texttt{emcee}. We used $4N_{\rm dim}$ walkers (with a minimum of 32) and 1000 steps, discarding the first half of the chains as burn-in. The reported parameter values are the posterior medians, with uncertainties given by the 16th and 84th percentiles. For the relative delay $\Delta t_{BC} = \Delta t_{AC}-\Delta t_{AB}$, the difference was calculated directly from the posterior samples, thereby accounting for correlations between the individual delay parameters.

\begin{table}[t]
    \centering
    \caption{Parameter Estimates from the Independent GP Analysis.}
    \label{tab:independent_gp_parameters}
    \begin{tabular}{lc}
        \hline
        Parameter & Value \\
        \hline
        $\Delta t_{AB}$ [days] & $-12.4_{-2.7}^{ +2.4}$ \\
        $\Delta t_{AC}$ [days] & $-1.5_{-2.9}^{ +2.3}$ \\
        $\Delta t_{BC}$ [days] & $10.7_{-3.0}^{ +3.4}$ \\
        $\beta_{B/A}$ & $0.241_{-0.009}^{+0.010}$ \\
        $\beta_{C/A}$ & $0.176_{-0.007}^{+0.007}$ \\
        $\sigma_{\rm floor}$ $g$-band [mag] & $0.10_{-0.01}^ {+0.02}$ \\
        $\sigma_{\rm floor}$ $r$-band [mag] & $0.07_{-0.01}^{+0.01}$ \\
        $\sigma_{\rm floor}$ $i$-band [mag] & $0.11_{-0.01}^ {+0.02}$  \\
        \hline
    \end{tabular}
\begin{minipage}{\textwidth}
\vspace{1ex}
\footnotesize
\textbf{Notes.} The reported uncertainties correspond to the $+1\sigma$ and $-1\sigma$ credible intervals. Units are given in square brackets; parameters without units are dimensionless.
\end{minipage}
\end{table}

\begin{figure*}
    \centering
    \includegraphics[width=0.6\linewidth]{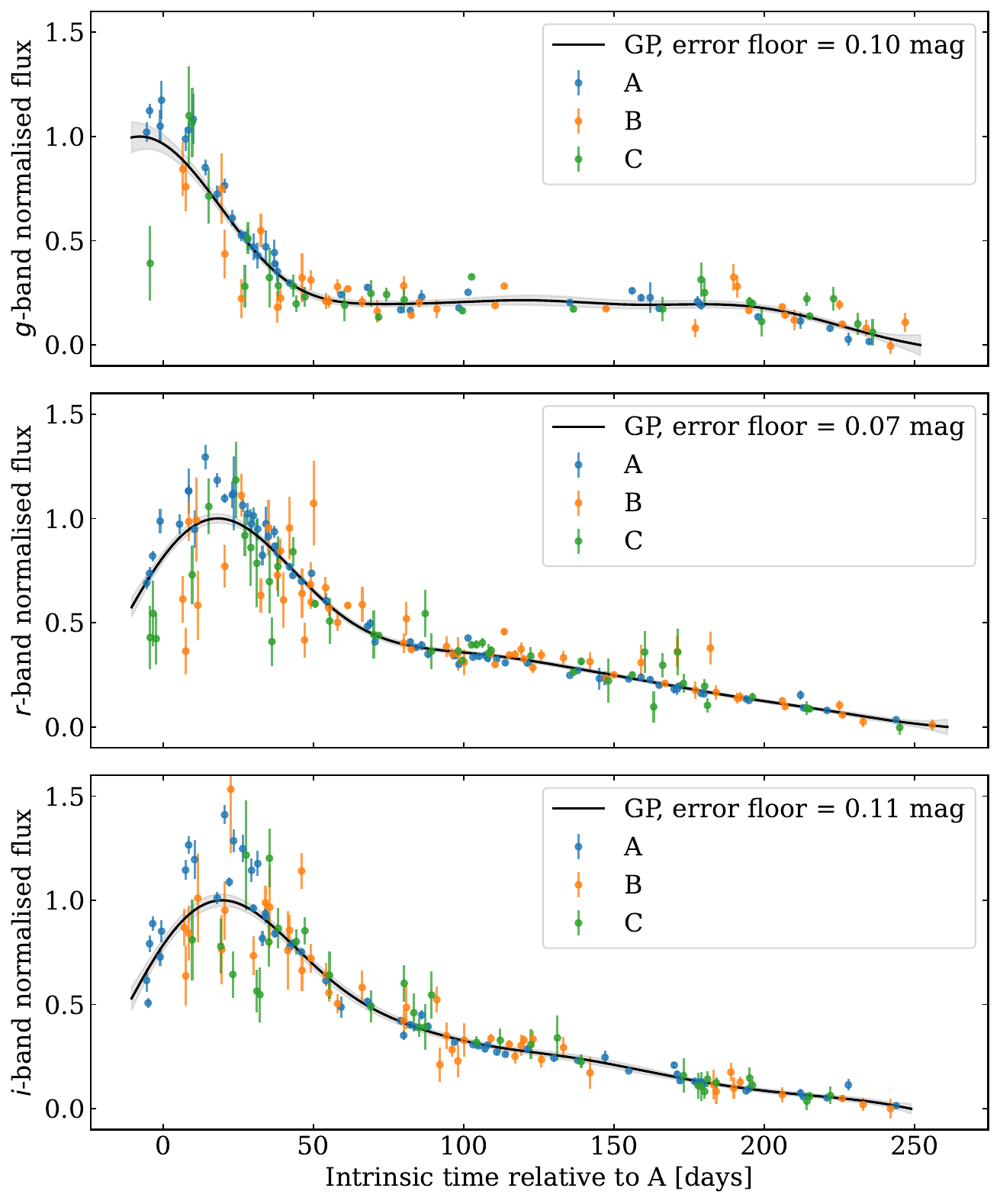}
    \caption{
    Independent GP model of the aligned light curves in the $gri$ bands. The light curves of images A, B, and C are shifted by the best-fitting relative time delays and rescaled by their fitted flux factors. Fluxes are normalised to approximately the range 0--1 in each band for visualisation. The solid black curve shows the best-fitting GP reconstruction, with the shaded region indicating the $1\sigma$ predictive uncertainty. The fitted photometric error floor is indicated in each panel.
    }
    \label{fig:indep_gp}
\end{figure*}

The best-fit parameters from the independent Gaussian-process analysis are presented in Table~\ref{tab:independent_gp_parameters}. The recovered relative time delays are consistent with those obtained from the \texttt{GausSN} analysis in Sect.~\ref{sec:time_delays_results}, providing an independent verification of the time-delay measurement. The fitted magnitude error floors are also broadly consistent across the three bands, with values of $0.10$, $0.07$, and $0.11$~mag in $g$, $r$, and $i$, respectively. To one decimal place, these values all round to a common error floor of $0.1$~mag. We therefore adopted a $0.1$~mag error floor in the \texttt{GausSN} analysis as a representative level of additional photometric uncertainty.

Overall, this analysis provides an independent GP-based cross-check of the \texttt{GausSN} results while also demonstrating that the measured delays remain stable when additional nuisance parameters, including band-dependent photometric error floors, are incorporated into the inference.

\FloatBarrier

\section{\texttt{GausSN} with constant magnification model}
\label{sec:appendix-constantgaussn}

\begin{figure*}[h!]
    \centering
    \includegraphics[width=.9\textwidth]{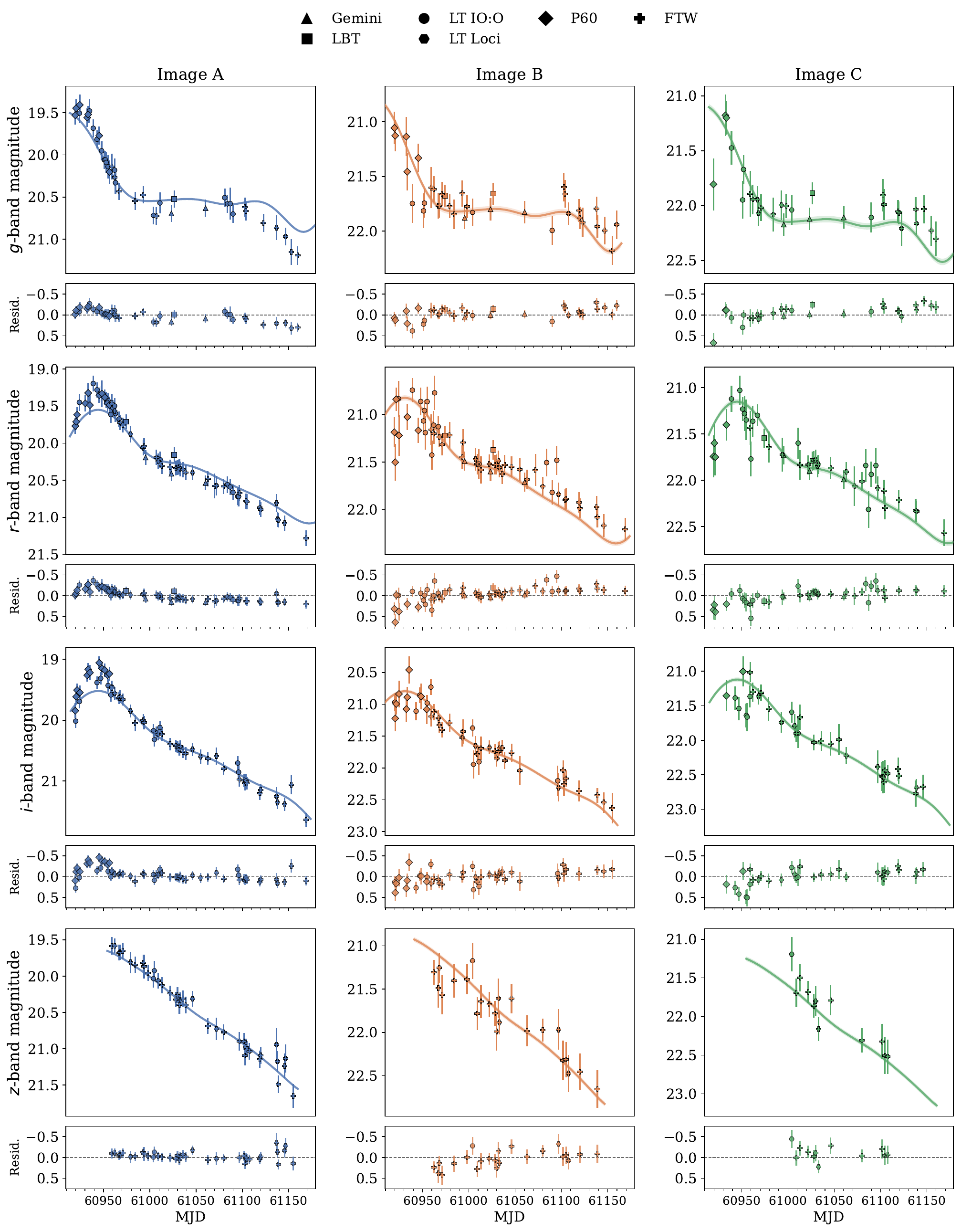}
    \caption{\texttt{GausSN} constant magnification model fit to the masked, MW-corrected $griz$ photometry. The blue, orange, and green curves show the GP model evaluated at the median posterior parameters for images A, B, and C, respectively, with the corresponding photometric observations shown as points. The lower panels show the residuals between the observed magnitudes and the posterior-median model. The inferred time delays are $\Delta t_{AB}=-13.2^{+3.1}_{-3.4}$ days and $\Delta t_{AC}=0.8^{+3.5}_{-3.8}$ days (68\% credible intervals).
    }
    \label{fig:25wny_deltat_all}
\end{figure*}

A \texttt{GausSN} constant magnification model fit to the fiducial dataset is shown in Fig.~\ref{fig:25wny_deltat_all}. The resulting photometric time delays are $\Delta t_{AB}=-13.2^{+3.1}_{-3.4}$ days and $\Delta t_{AC}=0.8^{+3.5}_{-3.8}$ days (68\% credible intervals).

The resulting fit shows noticeable tension around the peak of image A and during the decline phase of the light curves of all images. The image A data points are systematically brighter than predicted by the model, with the discrepancy increasing towards redder bands, from $g$ through $r$ to $i$. This colour-dependent behaviour could indicate that additional extinction corrections are required for image A (see Sect.~\ref{subsec:ext_corr}). Fits including an additional lens extinction correction for image A at the $\Delta E(B-V)=0.1\:\mathrm{mag}$ level, however, did not resolve this discrepancy, with the resulting time-delay measurements remaining consistent to within $\sim1.6$~days (see Sect.~\ref{subsec:sys_test} and Fig.~\ref{fig:gaussn_sys_const} in Appendix~\ref{sec:appendix-constantgaussn}).

Alternatively, this could indicate other wavelength-dependent effects, such as chromatic microlensing, which may affect image A more strongly than the other images. A time-dependent but achromatic effect could also contribute, as the peaks occur progressively later in the redder bands. This behaviour could be consistent with the effects of achromatic microlensing. \companioncite{Mortsell2026} found that image A is the most strongly affected by microlensing, with a broad magnification factor of $\mu_A=-4.7^{+2.6}_{-13.3}$ in the $F115W$-band. This makes achromatic microlensing a plausible explanation for the observed wavelength-dependent discrepancy.

\begin{figure*}[h!]
    \centering
    \includegraphics[width=.8\textwidth]{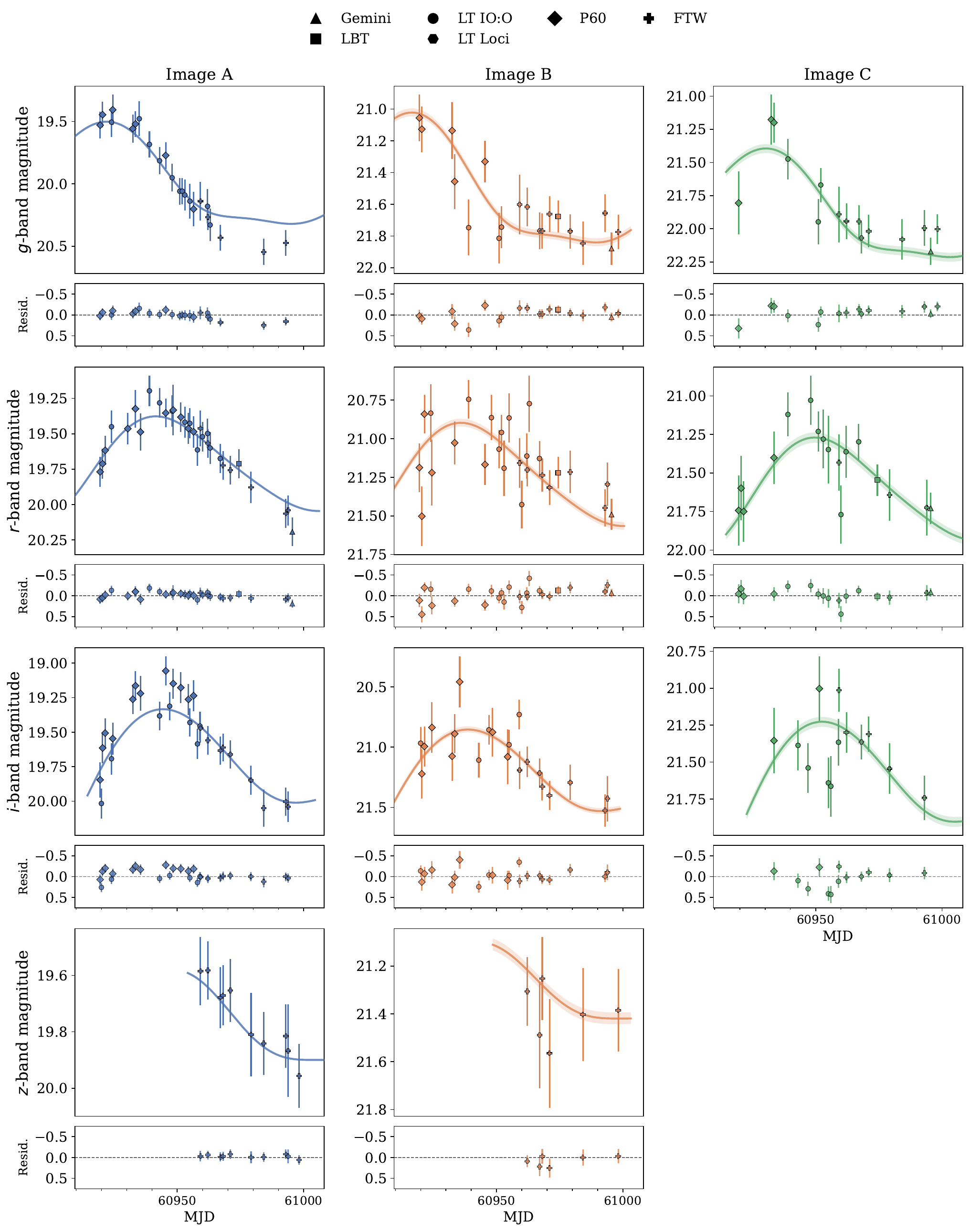}
    \caption{\texttt{GausSN} constant magnification model fit to the masked, MW-corrected $griz$ photometry with an observation date limit of $\mathrm{MJD}<61000$. The blue, orange, and green curves show the GP model evaluated at the median posterior parameters for images A, B, and C, respectively, with the corresponding photometric observations shown as points. The lower panels show the residuals between the observed magnitudes and the posterior-median model. There is no data for image C in the $z$-band at $\mathrm{MJD}<61000$. The inferred time delays are $\Delta t_{AB}-5.6^{+1.8}_{-1.9}$ days and $\Delta t_{AC}=8.2^{+3.0}_{-3.0}$ days (68\% credible intervals).
}
    \label{fig:25wny_deltat_61000}
\end{figure*}

For comparison, we also present a constant magnification model fit to a subset of data, with a cut on the observation date of $\mathrm{MJD}<61000$. This cut is chosen to isolate the rising and peak phases of the light curves, although we note that there is a lack of $z$-band for image C in this region. Table~\ref{tab:fit_comparison} compares the time delays and relative magnifications inferred from the two fits, together with the other parameters of the GP model. The resulting model is shown in Fig.~\ref{fig:25wny_deltat_61000}, and corresponds to measured photometric time delays of $\Delta t_{AB}=-5.6^{+1.8}_{-1.9}$ days and $\Delta t_{AC}=8.2^{+3.0}_{-3.0}$ days (68\% credible intervals). This is a shift of $+7.6$ and $+7.4$~days for $\Delta t_{AB}$ and $\Delta t_{AC}$, respectively, corresponding to positive shifts of approximately $2.0\,\sigma$ and $1.5\,\sigma$ of the combined statistical uncertainty, respectively. In this case, the model shows improved agreement with the peak of image A, reducing the tension seen in the fit to the full light curve.

Despite the differences in $\Delta t_{AB}$ and $\Delta t_{AC}$ between the two fits, the inferred delay between images B and C, $\Delta t_{BC}=\Delta t_{AC}-\Delta t_{AB}$, remains consistent within $\sim0.1$~days when calculated from the posterior chains. For the full light curve fit, we calculate
$\Delta t_{BC}=13.9^{+4.1}_{-4.1}$~days, while the peak-only fit results in $\Delta t_{BC}=13.8^{+3.4}_{-3.3}$~days. The consistency of $\Delta t_{BC}$, despite the substantial changes in the delays involving image A, suggests that the tension between the full and peak-only fits may be primarily associated with image A. This provides further motivation for considering time-dependent magnification effects that preferentially affect image A, such as microlensing.

The inferred relative magnifications differ between the \texttt{GausSN} fits to the full and peak-only light curves (see Table~\ref{tab:fit_comparison}). In particular, $\beta_{B/A}$ and $\beta_{C/A}$ decrease from $0.309^{+0.008}_{-0.008}$ and $0.230^{+0.006}_{-0.006}$, respectively, for the full dataset to $0.247^{+0.006}_{-0.005}$ and $0.175^{+0.005}_{-0.005}$, respectively, for the peak-only subset. These shifts are both greater than $6\,\sigma$. The fact that the inferred relative magnifications change when the fit is restricted to the early-time data suggests that the relative brightness of the images may not be well described by a single constant magnification.

Unlike the static magnification produced by the macrolens, microlensing can introduce time-dependent magnification as the expanding supernova photosphere samples different regions of the stellar microlensing magnification pattern. Consequently, the effective magnification of each image may evolve with time, leading to changes in the observed image flux ratios. We note that the evolution of the fitted magnification parameters alone does not establish microlensing as the cause. Alternative explanations, including photometric systematics (such as imperfect host galaxy subtraction), or small errors in the adopted time delays, could also contribute to apparent changes in the flux ratios. However, taken together, these results provide evidence that microlensing likely plays a significant role in the observed flux-ratio evolution of SN~2025wny. As a result, we adopt the time-delay measurements from the sigmoid magnification model fits described in Sect.~\ref{sec:time_delays_results} as our fiducial results.

\begin{table}[h!]
\centering
\caption{Comparison of \texttt{GausSN} Constant Magnification Model Parameter Estimates for the Fiducial Photometry and the Subset with $\mathrm{MJD}<61000$.}
\label{tab:fit_comparison}
\begin{tabular}{lcc}
\hline
Parameter & Full dataset & $\mathrm{MJD}<61000$ \\
\hline
Number of points, $N$ & 483 & 210 \\
Log evidence, $\log Z$ & 1023 & 372 \\
\hline
Kernel amplitude, $A$ &
$0.18^{+0.04}_{-0.03}$ &
$0.19^{+0.06}_{-0.04}$ \\

Kernel timescale, $\ell$ &
$42.5^{+4.8}_{-5.0}$ &
$24.4^{+4.8}_{-2.9}$ \\

Mean, $c$ &
$0.31^{+0.05}_{-0.05}$ &
$0.46^{+0.03}_{-0.05}$ \\

\hline

$\Delta t_{AB}$ [days] &
$-13.2^{+3.1}_{-3.4}$ &
$-5.6^{+1.8}_{-1.9}$ \\

$\Delta t_{AC}$ [days] &
$0.8^{+3.5}_{-3.8}$ &
$8.2^{+3.0}_{-3.0}$ \\

$\Delta t_{BC}$ [days] &
$13.9^{+4.1}_{-4.1}$ &
$13.8^{+3.4}_{-3.3}$ \\

$\beta_{B/A}$ &
$0.309^{+0.008}_{-0.008}$ &
$0.247^{+0.006}_{-0.005}$ \\

$\beta_{C/A}$ &
$0.230^{+0.006}_{-0.006}$ &
$ 0.175^{+0.005}_{-0.005}$ \\
\hline
\end{tabular}
\begin{minipage}{\textwidth}
\vspace{1ex}
\footnotesize
\textbf{Notes.} The reported uncertainties correspond to the $+1\sigma$ and $-1\sigma$ credible intervals. Units are given in square brackets; parameters without units are dimensionless.
\end{minipage}
\end{table}

\begin{figure*}[t!]
    \centering
    \textbf{Constant Magnification \texttt{GausSN} Model}\par\vspace{0.5em}
    \includegraphics[width=0.49\linewidth]{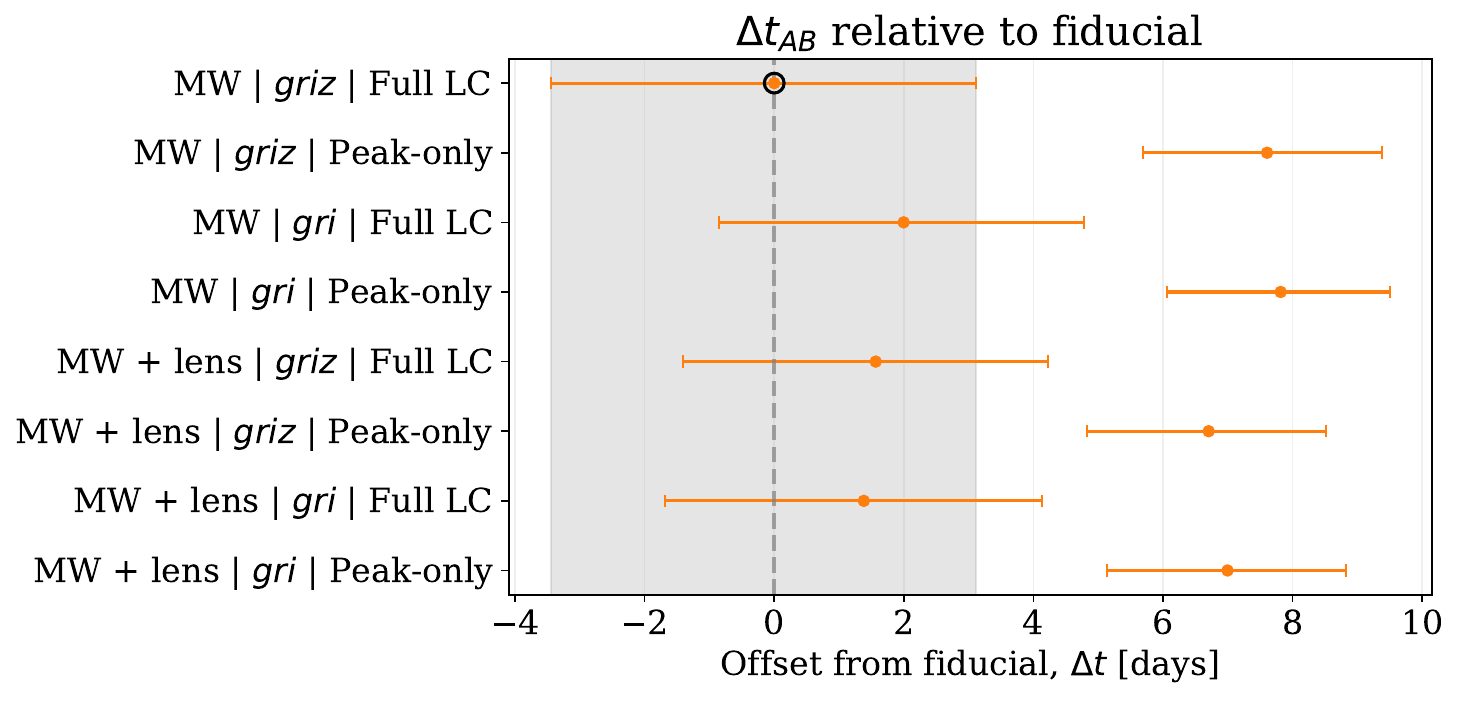}
    \includegraphics[width=0.49\linewidth]{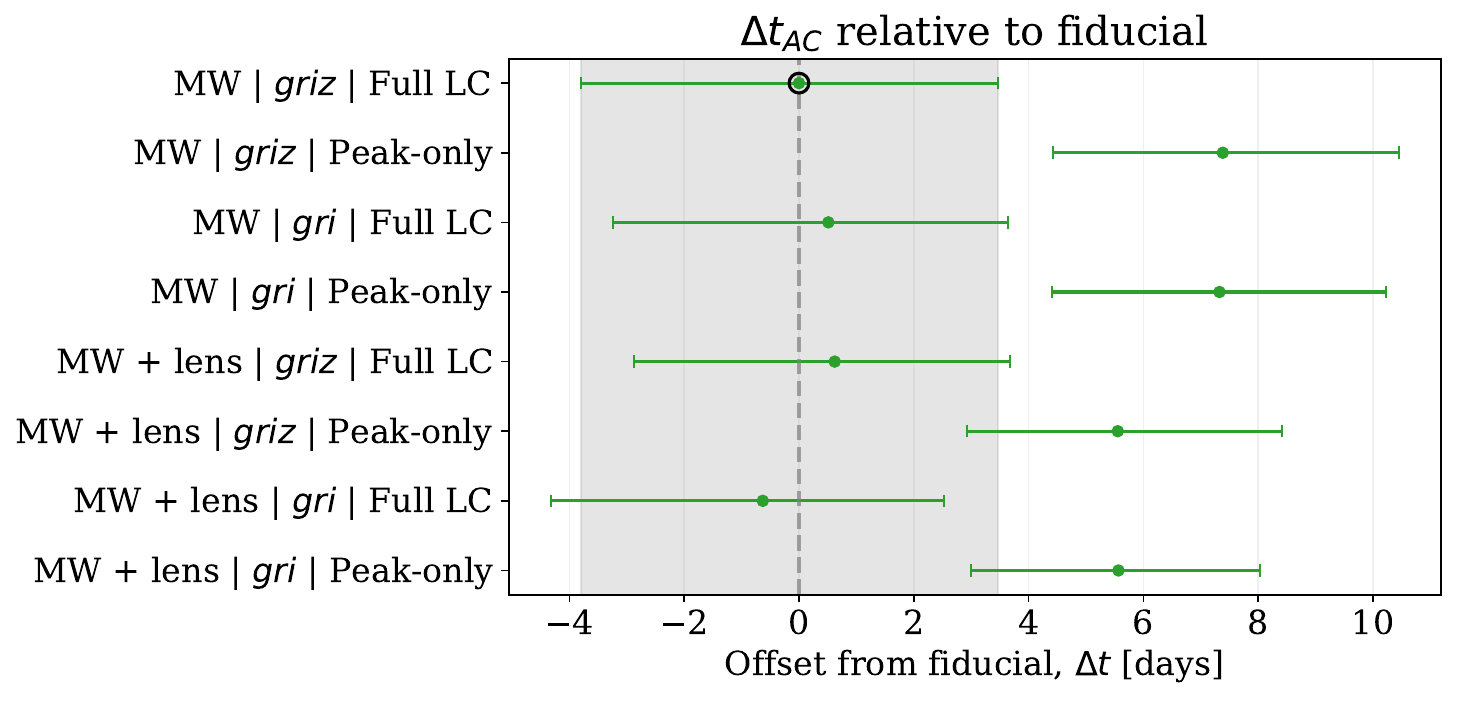}
    \caption{Forest plot showing the systematic tests of the inferred photometric time delays from the \texttt{GausSN} constant magnification model (left: $\Delta t_{AB}$; right: $\Delta t_{AC}$). The horizontal axis shows the offset in the inferred time delay relative to the fiducial analysis, defined as the MW-corrected, $griz$-band, masked, full light curve fit. Each point corresponds to one of the 8 combinations of lens galaxy extinction treatment (MW only or MW and lens correction), temporal range (full light curve or peak-only), and photometric bands ($gri$ or $griz$). Error bars show the statistical uncertainties from the corresponding individual fits. The dashed vertical line marks the fiducial result ($\Delta t=0$), and the open marker identifies the fiducial configuration.
}
    \label{fig:gaussn_sys_const}
\end{figure*}

\FloatBarrier

\section{\texttt{GausSN} with sigmoid magnification model}
\label{sec:appendix-sigmoidgaussn}

\begin{figure}[h!]
    \centering
    \includegraphics[width=0.5\linewidth]{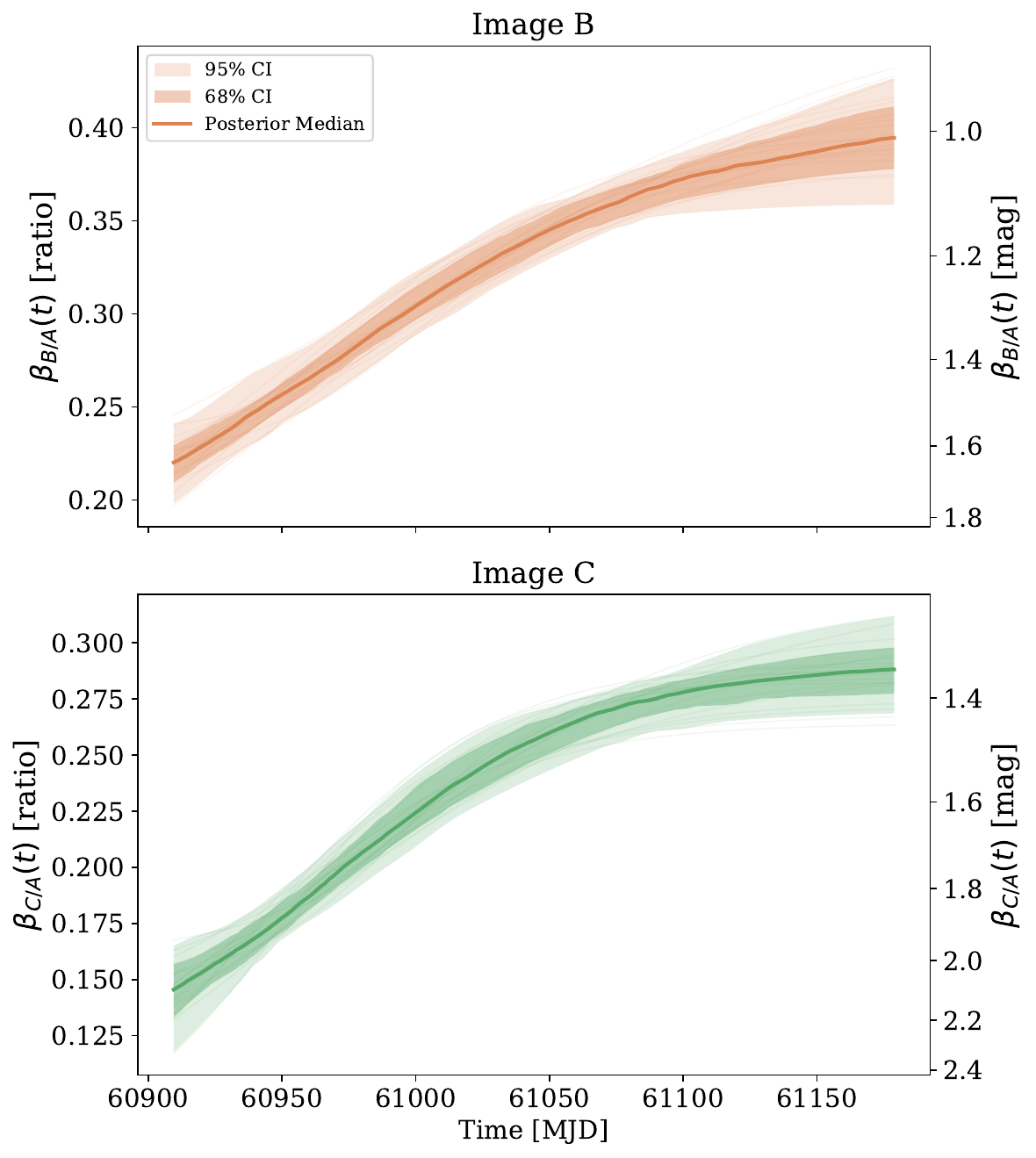}
    \caption{Time-dependent relative magnification inferred from the default \texttt{GausSN} sigmoid magnification model for images B and C relative to image A. This model was evaluated with the fiducial dataset. The solid lines show the posterior-median relative magnification, $\beta_{B/A}(t)$ and $\beta_{C/A}(t)$, while the darker and lighter shaded regions indicate the 68\% and 95\% credible intervals, respectively. The uncertainty regions are calculated from 100 equally spaced samples from the equal-weight posterior, evaluating the sigmoid magnification function for each posterior sample and taking the corresponding percentiles at each time. The right-hand axes show the equivalent differential magnification in magnitudes.
}
    \label{fig:micro_func_default}
\end{figure}

In the default implementation of \texttt{GausSN}, the magnification of each image is defined relative to the light curve of an arbitrarily chosen reference image, with the time-dependent sigmoid applied to the other images. Fitting with image A as the reference revealed nearly identical $\beta(t)$ profiles for images B and C, offset only by their initial magnification ratios. As shown in Fig.~\ref{fig:micro_func_default}, both images undergo a concurrent shift toward higher magnifications at later epochs.

This behaviour is more naturally interpreted as evidence that the time-dependent magnification is associated primarily with the reference image A, which is relatively brighter at the earlier epochs. Therefore, we modified the \texttt{GausSN} model such that images B and C are assigned constant magnifications, while image A is assigned a time-dependent sigmoid magnification. 

\begin{figure*}[h!]
    \centering
    \includegraphics[width=\textwidth]{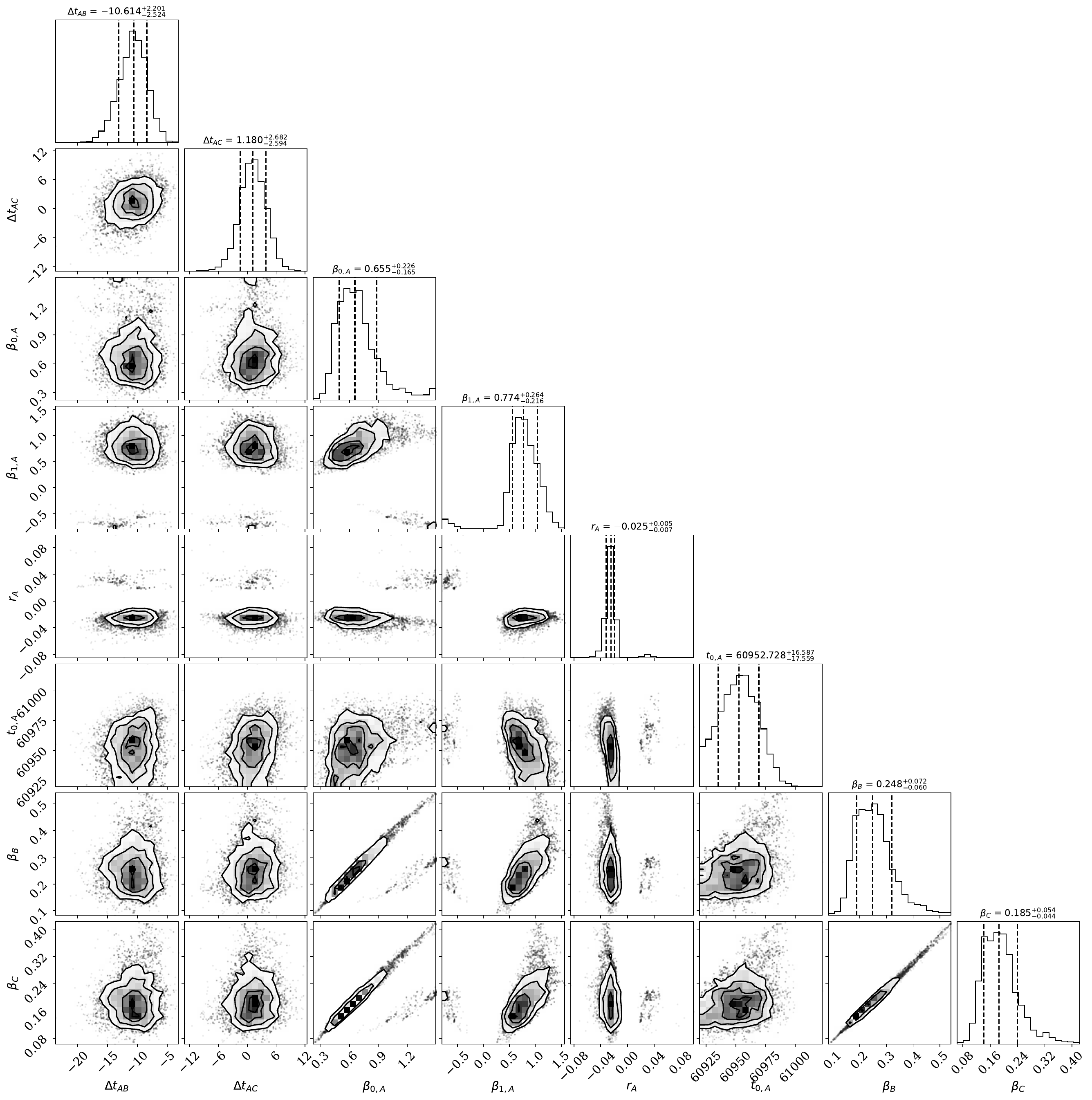}
    \caption{Posterior distributions of the sigmoid magnification parameters obtained from the \texttt{GausSN} fit to the masked, MW-corrected $griz$ dataset. The parameters shown are the relative time delays, $\Delta t_{AB}$ and $\Delta t_{AC}$; the sigmoid magnification model parameters for image A, $\beta_{0,A}$, $\beta_{1,A}$, $r_A$, and $t_{0,A}$; and the constant magnification parameters for images B and C, $\beta_B$ and $\beta_C$. The diagonal panels show the one-dimensional marginal posterior distributions, with the median and 16th and 84th percentiles indicated in the titles, while the off-diagonal panels show the corresponding two-dimensional posterior distributions and parameter correlations. The posterior samples are drawn from the equal-weighted nested sampling posterior.
}
    \label{fig:corner1}
\end{figure*}

The posterior distributions of the sigmoid magnification model parameters are presented in the corner plot shown in Fig.~\ref{fig:corner1}. The posterior distributions show a degeneracy between $\beta_{1,A}$ and $r_A$, with two distinct modes corresponding to solutions related by a simultaneous change in the signs of these parameters. These solutions produce equivalent magnification curves and therefore cannot be distinguished by the data. A strong correlation is also evident between $\beta_{0,A}$, $\beta_B$, and $\beta_C$. This arises from an overall scale degeneracy between the magnification parameters and the amplitude of the underlying GP model.

To assess the sensitivity of our time-delay measurements to analysis choices, we performed a full-factorial set of 8 analyses spanning three binary choices: lens galaxy extinction correction (MW extinction correction alone or MW correction supplemented by a correction for extinction from the lens galaxy; see Sect.~\ref{subsec:ext_corr}), temporal range (full light curve or peak-only, with $\mathrm{MJD}<61000$), and photometric bands ($gri$ or $griz$). The values of $\Delta t_{AB}$ and $\Delta t_{AC}$ from these tests are presented in Table~\ref{tab:sys_time_delays}.

\FloatBarrier

\clearpage

\section{Table of time delays from systematic tests}
\label{sec:appendix-systest}

\begin{table*}[h!]
\centering
\caption{Measured Time Delays for the Different Light Curve Analyses.}
\label{tab:sys_time_delays}
\begin{tabular}{llccccc}
\hline
Model & Corr. data & Band & Light curve &
$\Delta t_{AB}$ [days] & $\Delta t_{AC}$ [days] \\
\hline
Constant & MW-corrected & $griz$ & Full
& $-13.2^{+3.1}_{-3.4}$ & $+0.8^{+3.5}_{-3.8}$ \\
Constant & MW-corrected & $griz$ & Peak-only
& $-5.6^{+1.8}_{-1.9}$ & $+8.2^{+3.1}_{-3.0}$ \\
Constant & MW-corrected & $gri$ & Full
& $-11.2^{+2.8}_{-2.8}$ & $+1.3^{+3.1}_{-3.7}$ \\
Constant & MW-corrected & $gri$ & Peak-only
& $-5.4^{+1.7}_{-1.8}$ & $+8.1^{+2.9}_{-2.9}$ \\
Constant & MW+lens-corrected & $griz$ & Full
& $-11.7^{+2.7}_{-3.0}$ & $+1.4^{+3.0}_{-3.5}$ \\
Constant & MW+lens-corrected & $griz$ & Peak-only
& $-6.5^{+1.8}_{-1.9}$ & $+6.3^{+2.9}_{-2.6}$ \\
Constant & MW+lens-corrected & $gri$ & Full
& $-11.8^{+2.8}_{-3.1}$ & $+0.2^{+3.1}_{-3.7}$ \\
Constant & MW+lens-corrected & $gri$ & Peak-only
& $-6.2^{+1.8}_{-1.9}$ & $+6.4^{+2.5}_{-2.6}$ \\
\hline
Sigmoid & MW-corrected & $griz$ & Full
& $-10.6^{+2.2}_{-2.5}$ & $+1.2^{+2.7}_{-2.6}$ \\
Sigmoid & MW-corrected & $griz$ & Peak-only
& $-8.4^{+2.2}_{-2.6}$ & $+4.1^{+3.2}_{-3.2}$ \\
Sigmoid & MW-corrected & $gri$ & Full
& $-9.4^{+2.1}_{-2.1}$ & $+2.2^{+2.8}_{-2.9}$ \\
Sigmoid & MW-corrected & $gri$ & Peak-only
& $-9.1^{+2.3}_{-2.6}$ & $+3.8^{+3.5}_{-3.3}$ \\
Sigmoid & MW+lens-corrected & $griz$ & Full
& $-10.4^{+2.2}_{-2.3}$ & $+2.4^{+2.6}_{-2.6}$ \\
Sigmoid & MW+lens-corrected & $griz$ & Peak-only
& $-11.2^{+2.4}_{-2.6}$ & $+1.8^{+3.1}_{-3.4}$ \\
Sigmoid & MW+lens-corrected & $gri$ & Full
& $-10.1^{+2.2}_{-2.1}$ & $+1.7^{+2.7}_{-2.6}$ \\
Sigmoid & MW+lens-corrected & $gri$ & Peak-only
& $-11.6^{+2.5}_{-2.7}$ & $+1.4^{+2.9}_{-3.3}$ \\
\hline
\end{tabular}
\begin{minipage}{\textwidth}
\vspace{1ex}
\footnotesize
\textbf{Notes.} The reported uncertainties correspond to the $+1\sigma$ and $-1\sigma$ credible intervals.
\end{minipage}
\end{table*}



\end{document}